

\magnification=\magstep1

\newbox\SlashedBox
\def\slashed#1{\setbox\SlashedBox=\hbox{#1}
\hbox to 0pt{\hbox to 1\wd\SlashedBox{\hfil/\hfil}\hss}#1}
\def\hboxtosizeof#1#2{\setbox\SlashedBox=\hbox{#1}
\hbox to 1\wd\SlashedBox{#2}}

\def\mathslashed#1{\setbox\SlashedBox=\hbox{$#1$}
\hbox to 0pt{\hbox to 1\wd\SlashedBox{\hfil/\hfil}\hss}#1}

\def\ifsmall{\iffalse}  
\def\titlepagefont{}  

\def\DefineTeXgraphics{%
\special{ps::[global] /TeXgraphics { } def}}  

\def\today{\ifcase\month\or January\or February\or March\or April\or May
\or June\or July\or August\or September\or October\or November\or
December\fi\space\number\day, \number\year}
\def\eatPrefix19{}
\def\Year{\expandafter\eatPrefix\the\year}
\newcount\hours \newcount\minutes
\def\monthname{\ifcase\month\or
January\or February\or March\or April\or May\or June\or July\or
August\or September\or October\or November\or December\fi}
\def\shortmonthname{\ifcase\month\or
Jan\or Feb\or Mar\or Apr\or May\or Jun\or Jul\or
Aug\or Sep\or Oct\or Nov\or Dec\fi}

\def\TimeStamp{\hours\the\time\divide\hours by60%
\minutes -\the\time\divide\minutes by60\multiply\minutes by60%
\advance\minutes by\the\time%
${\rm \shortmonthname}\cdot\if\day<10{}0\fi\the\day\cdot\the\year%
\qquad\the\hours:\if\minutes<10{}0\fi\the\minutes$}




\def\Title#1{%
\vskip 1in{\titlefont\centerline{#1}}\vskip .5in}



\newif\ifdraftmode
\newif\ifleftlabels  

\def\nolabels{\def\wrlabeL##1{}\def\eqlabeL##1{}\def\reflabeL##1{}}
\def\writelabels{\def\wrlabeL##1{\leavevmode\vadjust{\rlap{\smash%
{\line{{\escapechar=` \hfill\rlap{\sevenrm\hskip.03in\string##1}}}}}}}%
\def\eqlabeL##1{{\escapechar-1\rlap{\sevenrm\hskip.05in\string##1}}}%
\def\reflabeL##1{\noexpand\rlap{\noexpand\sevenrm[\string##1]}}}
\def\writeleftlabels{\def\wrlabeL##1{\leavevmode\vadjust{\rlap{\smash%
{\line{{\escapechar=` \hfill\rlap{\sevenrm\hskip.03in\string##1}}}}}}}%
\def\eqlabeL##1{{\escapechar-1%
\rlap{\sixrm\hskip.05in\string##1}%
\llap{\sevenrm\string##1\hskip.03in\hbox to \hsize{}}}}%
\def\reflabeL##1{\noexpand\rlap{\noexpand\sevenrm[\string##1]}}}
\nolabels

\newdimen\fullhsize
\newdimen\hstitle
\hstitle=\hsize 
\newdimen\hsbody
\hsbody=\hsize 
\newdimen\hbodyoffset
\hbodyoffset=\hoffset 
\newbox\leftpage
\def\abstract#1{#1}
\def\rotated{\special{ps: landscape}
\magnification=1000  
\baselineskip=14pt
\global\hstitle=9truein\global\hsbody=4.75truein
\global\vsize=7truein\global\voffset=-.31truein
\global\hoffset=-0.54in\global\hbodyoffset=-.54truein
\global\fullhsize=10truein
\def\DefineTeXgraphics{%
\special{ps::[global]
/TeXgraphics {currentpoint translate 0.7 0.7 scale
              -80 0.72 mul -1000 0.72 mul translate} def}}
\let\lr=L
\def\ifsmall{\iftrue}
\def\titlepagefont{\twelvepoint}
\trueseventeenpoint
\def\almostshipout##1{\if L\lr \count1=1
      \global\setbox\leftpage=##1 \global\let\lr=R
   \else \count1=2
      \shipout\vbox{\hbox to\fullhsize{\box\leftpage\hfil##1}}
      \global\let\lr=L\fi}

\output={\ifnum\count0=1 
 \shipout\vbox{\hbox to \fullhsize{\hfill\pagebody\hfill}}\advancepageno
 \else
 \almostshipout{\leftline{\vbox{\pagebody\makefootline}}}\advancepageno
 \fi}

\def\abstract##1{{\leftskip=1.5in\rightskip=1.5in ##1\par}} }

\def\linemessage#1{\immediate\write16{#1}}

\global\newcount\secno \global\secno=0
\global\newcount\appno \global\appno=0
\global\newcount\meqno \global\meqno=1
\global\newcount\subsecno \global\subsecno=0
\global\newcount\figno \global\figno=0

\newif\ifAnyCounterChanged
\let\terminator=\relax
\def\normalize#1{\ifx#1\terminator\let\next=\relax\else%
\if#1i\aftergroup i\else\if#1v\aftergroup v\else\if#1x\aftergroup x%
\else\if#1l\aftergroup l\else\if#1c\aftergroup c\else%
\if#1m\aftergroup m\else%
\if#1I\aftergroup I\else\if#1V\aftergroup V\else\if#1X\aftergroup X%
\else\if#1L\aftergroup L\else\if#1C\aftergroup C\else%
\if#1M\aftergroup M\else\aftergroup#1\fi\fi\fi\fi\fi\fi\fi\fi\fi\fi\fi\fi%
\let\next=\normalize\fi%
\next}
\def\makeNormal#1#2{\def\doNormalDef{\edef#1}\begingroup%
\aftergroup\doNormalDef\aftergroup{\normalize#2\terminator\aftergroup}%
\endgroup}

\def\warnIfChanged#1#2{%
\ifundef#1
\else\begingroup%
\edef\oldDefinitionOfCounter{#1}\edef\newDefinitionOfCounter{#2}%
\ifx\oldDefinitionOfCounter\newDefinitionOfCounter%
\else%
\linemessage{Warning: definition of \noexpand#1 has changed.}%
\global\AnyCounterChangedtrue\fi\endgroup\fi}

\def\Section#1{\global\advance\secno by1\relax\global\meqno=1%
\global\subsecno=0%
\bigbreak\bigskip
\centerline{\twelvepoint \bf %
\the\secno. #1}%
\par\nobreak\medskip\nobreak}
\def\tagsection#1{%
\warnIfChanged#1{\the\secno}%
\xdef#1{\the\secno}%
\ifWritingAuxFile\immediate\write\auxfile{\noexpand\xdef\noexpand#1{#1}}\fi%
}
\def\section{\Section}
\def\Subsection#1{\global\advance\subsecno by1\relax\medskip %
\leftline{\bf\the\secno.\the\subsecno\ #1}%
\par\nobreak\smallskip\nobreak}
\def\tagsubsection#1{%
\warnIfChanged#1{\the\secno.\the\subsecno}%
\xdef#1{\the\secno.\the\subsecno}%
\ifWritingAuxFile\immediate\write\auxfile{\noexpand\xdef\noexpand#1{#1}}\fi%
}

\def\subsection{\Subsection}

\def\romappno{\uppercase\expandafter{\romannumeral\appno}}
\def\makeNormalizedRomappno{%
\expandafter\makeNormal\expandafter\normalizedromappno%
\expandafter{\romannumeral\appno}%
\edef\normalizedromappno{\uppercase{\normalizedromappno}}}
\def\Appendix#1{\global\advance\appno by1\relax\global\meqno=1\global\secno=0%
\global\subsecno=0%
\bigbreak\bigskip
\centerline{\twelvepoint \bf Appendix %
\romappno. #1}%
\par\nobreak\medskip\nobreak}
\def\tagappendix#1{\makeNormalizedRomappno%
\warnIfChanged#1{\normalizedromappno}%
\xdef#1{\normalizedromappno}%
\ifWritingAuxFile\immediate\write\auxfile{\noexpand\xdef\noexpand#1{#1}}\fi%
}
\def\appendix{\Appendix}
\def\Subappendix#1{\global\advance\subsecno by1\relax\medskip %
\leftline{\bf\romappno.\the\subsecno\ #1}%
\par\nobreak\smallskip\nobreak}
\def\tagsubappendix#1{\makeNormalizedRomappno%
\warnIfChanged#1{\normalizedromappno.\the\subsecno}%
\xdef#1{\normalizedromappno.\the\subsecno}%
\ifWritingAuxFile\immediate\write\auxfile{\noexpand\xdef\noexpand#1{#1}}\fi%
}

\def\subappendix{\Subappendix}

\def\eqn#1{\makeNormalizedRomappno%
\ifnum\secno>0%
  \warnIfChanged#1{\the\secno.\the\meqno}%
  \eqno(\the\secno.\the\meqno)\xdef#1{\the\secno.\the\meqno}%
     \global\advance\meqno by1
\else\ifnum\appno>0%
  \warnIfChanged#1{\normalizedromappno.\the\meqno}%
  \eqno({\rm\romappno}.\the\meqno)%
      \xdef#1{\normalizedromappno.\the\meqno}%
     \global\advance\meqno by1
\else%
  \warnIfChanged#1{\the\meqno}%
  \eqno(\the\meqno)\xdef#1{\the\meqno}%
     \global\advance\meqno by1
\fi\fi%
\eqlabeL#1%
\ifWritingAuxFile\immediate\write\auxfile{\noexpand\xdef\noexpand#1{#1}}\fi%
}
\def\defeqn#1{\makeNormalizedRomappno%
\ifnum\secno>0%
  \warnIfChanged#1{\the\secno.\the\meqno}%
  \xdef#1{\the\secno.\the\meqno}%
     \global\advance\meqno by1
\else\ifnum\appno>0%
  \warnIfChanged#1{\normalizedromappno.\the\meqno}%
  \xdef#1{\normalizedromappno.\the\meqno}%
     \global\advance\meqno by1
\else%
  \warnIfChanged#1{\the\meqno}%
  \xdef#1{\the\meqno}%
     \global\advance\meqno by1
\fi\fi%
\eqlabeL#1%
\ifWritingAuxFile\immediate\write\auxfile{\noexpand\xdef\noexpand#1{#1}}\fi%
}
\def\anoneqn{\makeNormalizedRomappno%
\ifnum\secno>0
  \eqno(\the\secno.\the\meqno)%
     \global\advance\meqno by1
\else\ifnum\appno>0
  \eqno({\rm\normalizedromappno}.\the\meqno)%
     \global\advance\meqno by1
\else
  \eqno(\the\meqno)%
     \global\advance\meqno by1
\fi\fi%
}
\def\mfig#1#2{\global\advance\figno by1%
\relax#1\the\figno%
\warnIfChanged#2{\the\figno}%
\edef#2{\the\figno}%
\reflabeL#2%
\ifWritingAuxFile\immediate\write\auxfile{\noexpand\xdef\noexpand#2{#2}}\fi%
}

\def\fig#1{\mfig{fig.\ ~}#1}

\catcode`@=11 

\newif\ifFiguresInText\FiguresInTexttrue
\newif\if@FigureFileCreated
\newwrite\capfile
\newwrite\figfile

\def\PlaceTextFigure#1#2#3#4{%
\vskip 0.5truein%
#3\hfil\epsfbox{#4}\hfil\break%
\hfil\vbox{Figure #1. #2}\hfil%
\vskip10pt}
\def\PlaceEndFigure#1#2{%
\epsfysize=\vsize\epsfbox{#2}\hfil\break\vfill\centerline{Figure #1.}\eject}

\def\LoadFigure#1#2#3#4{%
\ifundef#1{\phantom{\mfig{}#1}}\fi
\ifWritingAuxFile\immediate\write\auxfile{\noexpand\xdef\noexpand#1{#1}}\fi%
\ifFiguresInText
\PlaceTextFigure{#1}{#2}{#3}{#4}%
\else
\if@FigureFileCreated\else%
\immediate\openout\capfile=\jobname.caps%
\immediate\openout\figfile=\jobname.figs%
\fi%
\immediate\write\capfile{\noexpand\item{Figure \noexpand#1.\ }#2.}%
\immediate\write\figfile{\noexpand\PlaceEndFigure\noexpand#1{\noexpand#4}}%
\fi}

\def\listfigs{\ifFiguresInText\else%
\vfill\eject\immediate\closeout\capfile
\immediate\closeout\figfile%
\centerline{{\bf Figures}}\bigskip\frenchspacing%
\input \jobname.caps\vfill\eject\nonfrenchspacing%
\input\jobname.figs\fi}

\font\ninerm=cmr9
\font\eightrm=cmr8
\font\sixrm=cmr6

\def\loadtrueseventeenpoint{
 \font\seventeenrm=cmr10 at 17.28truept
 \font\seventeeni=cmmi10 at 17.28truept
 \font\seventeenbf=cmbx10 at 17.28truept
 \font\seventeenit=cmti10 at 17.28truept
 \font\seventeensl=cmsl10 at 17.28truept
 \font\seventeensy=cmsy10 at 17.28truept
}
\def\loadfourteenpoint{
\font\fourteenrm=cmr10 at 14.4pt
\font\fourteeni=cmmi10 at 14.4pt
\font\fourteenit=cmti10 at 14.4pt
\font\fourteensl=cmsl10 at 14.4pt
\font\fourteensy=cmsy10 at 14.4pt
\font\fourteenbf=cmbx10 at 14.4pt
}
\def\loadtruetwelvepoint{
\font\twelverm=cmr10 at 12truept
\font\twelvei=cmmi10 at 12truept
\font\twelveit=cmti10 at 12truept
\font\twelvesl=cmsl10 at 12truept
\font\twelvesy=cmsy10 at 12truept
\font\twelvebf=cmbx10 at 12truept
}

\font\ninei=cmmi9
\font\eighti=cmmi8
\font\sixi=cmmi6
\skewchar\ninei='177 \skewchar\eighti='177 \skewchar\sixi='177

\font\ninesy=cmsy9
\font\eightsy=cmsy8
\font\sixsy=cmsy6
\skewchar\ninesy='60 \skewchar\eightsy='60 \skewchar\sixsy='60

\font\ninebf=cmbx9
\font\eightbf=cmbx8
\font\sixbf=cmbx6

\font\ninett=cmtt9
\font\eighttt=cmtt8

\hyphenchar\tentt=-1 
\hyphenchar\ninett=-1
\hyphenchar\eighttt=-1

\font\ninesl=cmsl9
\font\eightsl=cmsl8

\font\nineit=cmti9
\font\eightit=cmti8


\newskip\ttglue
\def\tenpoint{\def\rm{\fam0\tenrm}%
  \textfont0=\tenrm \scriptfont0=\sevenrm \scriptscriptfont0=\fiverm
  \textfont1=\teni \scriptfont1=\seveni \scriptscriptfont1=\fivei
  \textfont2=\tensy \scriptfont2=\sevensy \scriptscriptfont2=\fivesy
  \textfont3=\tenex \scriptfont3=\tenex \scriptscriptfont3=\tenex
  \def\it{\fam\itfam\tenit}\textfont\itfam=\tenit
  \def\sl{\fam\slfam\tensl}\textfont\slfam=\tensl
  \def\bf{\fam\bffam\tenbf}\textfont\bffam=\tenbf \scriptfont\bffam=\sevenbf
  \scriptscriptfont\bffam=\fivebf
  \normalbaselineskip=12pt
  \let\sc=\eightrm
  \let\big=\tenbig
  \setbox\strutbox=\hbox{\vrule height8.5pt depth3.5pt width\z@}%
  \normalbaselines\rm}

\def\twelvepoint{\def\rm{\fam0\twelverm}%
  \textfont0=\twelverm \scriptfont0=\ninerm \scriptscriptfont0=\sevenrm
  \textfont1=\twelvei \scriptfont1=\ninei \scriptscriptfont1=\seveni
  \textfont2=\twelvesy \scriptfont2=\ninesy \scriptscriptfont2=\sevensy
  \textfont3=\tenex \scriptfont3=\tenex \scriptscriptfont3=\tenex
  \def\it{\fam\itfam\twelveit}\textfont\itfam=\twelveit
  \def\sl{\fam\slfam\twelvesl}\textfont\slfam=\twelvesl
  \def\bf{\fam\bffam\twelvebf}\textfont\bffam=\twelvebf%
  \scriptfont\bffam=\ninebf
  \scriptscriptfont\bffam=\sevenbf
  \normalbaselineskip=12pt
  \let\sc=\eightrm
  \let\big=\tenbig
  \setbox\strutbox=\hbox{\vrule height8.5pt depth3.5pt width\z@}%
  \normalbaselines\rm}

\def\fourteenpoint{\def\rm{\fam0\fourteenrm}%
  \textfont0=\fourteenrm \scriptfont0=\tenrm \scriptscriptfont0=\sevenrm
  \textfont1=\fourteeni \scriptfont1=\teni \scriptscriptfont1=\seveni
  \textfont2=\fourteensy \scriptfont2=\tensy \scriptscriptfont2=\sevensy
  \textfont3=\tenex \scriptfont3=\tenex \scriptscriptfont3=\tenex
  \def\it{\fam\itfam\fourteenit}\textfont\itfam=\fourteenit
  \def\sl{\fam\slfam\fourteensl}\textfont\slfam=\fourteensl
  \def\bf{\fam\bffam\fourteenbf}\textfont\bffam=\fourteenbf%
  \scriptfont\bffam=\tenbf
  \scriptscriptfont\bffam=\sevenbf
  \normalbaselineskip=17pt
  \let\sc=\elevenrm
  \let\big=\tenbig
  \setbox\strutbox=\hbox{\vrule height8.5pt depth3.5pt width\z@}%
  \normalbaselines\rm}

\def\seventeenpoint{\def\rm{\fam0\seventeenrm}%
  \textfont0=\seventeenrm \scriptfont0=\fourteenrm \scriptscriptfont0=\tenrm
  \textfont1=\seventeeni \scriptfont1=\fourteeni \scriptscriptfont1=\teni
  \textfont2=\seventeensy \scriptfont2=\fourteensy \scriptscriptfont2=\tensy
  \textfont3=\tenex \scriptfont3=\tenex \scriptscriptfont3=\tenex
  \def\it{\fam\itfam\seventeenit}\textfont\itfam=\seventeenit
  \def\sl{\fam\slfam\seventeensl}\textfont\slfam=\seventeensl
  \def\bf{\fam\bffam\seventeenbf}\textfont\bffam=\seventeenbf%
  \scriptfont\bffam=\fourteenbf
  \scriptscriptfont\bffam=\twelvebf
  \normalbaselineskip=21pt
  \let\sc=\fourteenrm
  \let\big=\tenbig
  \setbox\strutbox=\hbox{\vrule height 12pt depth 6pt width\z@}%
  \normalbaselines\rm}

\def\ninepoint{\def\rm{\fam0\ninerm}%
  \textfont0=\ninerm \scriptfont0=\sixrm \scriptscriptfont0=\fiverm
  \textfont1=\ninei \scriptfont1=\sixi \scriptscriptfont1=\fivei
  \textfont2=\ninesy \scriptfont2=\sixsy \scriptscriptfont2=\fivesy
  \textfont3=\tenex \scriptfont3=\tenex \scriptscriptfont3=\tenex
  \def\it{\fam\itfam\nineit}\textfont\itfam=\nineit
  \def\sl{\fam\slfam\ninesl}\textfont\slfam=\ninesl
  \def\bf{\fam\bffam\ninebf}\textfont\bffam=\ninebf \scriptfont\bffam=\sixbf
  \scriptscriptfont\bffam=\fivebf
  \normalbaselineskip=11pt
  \let\sc=\sevenrm
  \let\big=\ninebig
  \setbox\strutbox=\hbox{\vrule height8pt depth3pt width\z@}%
  \normalbaselines\rm}

\def\eightpoint{\def\rm{\fam0\eightrm}%
  \textfont0=\eightrm \scriptfont0=\sixrm \scriptscriptfont0=\fiverm%
  \textfont1=\eighti \scriptfont1=\sixi \scriptscriptfont1=\fivei%
  \textfont2=\eightsy \scriptfont2=\sixsy \scriptscriptfont2=\fivesy%
  \textfont3=\tenex \scriptfont3=\tenex \scriptscriptfont3=\tenex%
  \def\it{\fam\itfam\eightit}\textfont\itfam=\eightit%
  \def\sl{\fam\slfam\eightsl}\textfont\slfam=\eightsl%
  \def\bf{\fam\bffam\eightbf}\textfont\bffam=\eightbf \scriptfont\bffam=\sixbf%
  \scriptscriptfont\bffam=\fivebf%
  \normalbaselineskip=9pt%
  \let\sc=\sixrm%
  \let\big=\eightbig%
  \setbox\strutbox=\hbox{\vrule height7pt depth2pt width\z@}%
  \normalbaselines\rm}

\def\tenbig#1{{\hbox{$\left#1\vbox to8.5pt{}\right.\n@space$}}}
\def\ninebig#1{{\hbox{$\textfont0=\tenrm\textfont2=\tensy
  \left#1\vbox to7.25pt{}\right.\n@space$}}}
\def\eightbig#1{{\hbox{$\textfont0=\ninerm\textfont2=\ninesy
  \left#1\vbox to6.5pt{}\right.\n@space$}}}

\def\footnote#1{\edef\@sf{\spacefactor\the\spacefactor}#1\@sf
      \insert\footins\bgroup\eightpoint
      \interlinepenalty100 \let\par=\endgraf
        \leftskip=\z@skip \rightskip=\z@skip
        \splittopskip=10pt plus 1pt minus 1pt \floatingpenalty=20000
        \smallskip\item{#1}\bgroup\strut\aftergroup\@foot\let\next}
\skip\footins=12pt plus 2pt minus 4pt 
\dimen\footins=30pc 

\newinsert\margin
\dimen\margin=\maxdimen
\def\titlefont{\seventeenpoint}
\loadtruetwelvepoint 
\loadtrueseventeenpoint

\def\eatOne#1{}
\def\ifundef#1{\expandafter\ifx%
\csname\expandafter\eatOne\string#1\endcsname\relax}
\def\notTrue{\iffalse}\def\isTrue{\iftrue}
\def\ifdef#1{{\ifundef#1%
\aftergroup\notTrue\else\aftergroup\isTrue\fi}}
\def\use#1{\ifundef#1\linemessage{Warning: \string#1 is undefined.}%
{\tt \string#1}\else#1\fi}


\global\newcount\refno \global\refno=1
\newwrite\rfile
\newlinechar=`\^^J
\def\@ref#1#2{\the\refno\n@ref#1{#2}}
\def\n@ref#1#2{\xdef#1{\the\refno}%
\ifnum\refno=1\immediate\openout\rfile=\jobname.refs\fi%
\immediate\write\rfile{\noexpand\item{[\noexpand#1]\ }#2.}%
\global\advance\refno by1}
\def\nref{\n@ref} 
\def\ref{\@ref}   
\def\lref#1#2{\the\refno\xdef#1{\the\refno}%
\ifnum\refno=1\immediate\openout\rfile=\jobname.refs\fi%
\immediate\write\rfile{\noexpand\item{[\noexpand#1]\ }#2\semi}%
\global\advance\refno by1}
\def\cref#1{\immediate\write\rfile{#1\semi}}

\def\preref#1#2{\gdef#1{\@ref#1{#2}}}

\def\semi{;\hfil\noexpand\break}

\def\listrefs{\vfill\eject\immediate\closeout\rfile
\centerline{{\bf References}}\bigskip\frenchspacing%
\input \jobname.refs\vfill\eject\nonfrenchspacing}

\def\inputAuxIfPresent#1{\immediate\openin1=#1
\ifeof1\message{No file \auxfileName; I'll create one.
}\else\closein1\relax\input\auxfileName\fi%
}
\def\NPB{Nucl.\ Phys.\ B}
\def\PRL{Phys.\ Rev.\ Lett.\ }

\def\PLB{Phys.\ Lett.\ B}

\def\ZPC{Z.\ Phys.\ C}

\newif\ifWritingAuxFile
\newwrite\auxfile
\def\SetUpAuxFile{%
\xdef\auxfileName{\jobname.aux}%
\inputAuxIfPresent{\auxfileName}%
\WritingAuxFiletrue%
\immediate\openout\auxfile=\auxfileName}

\def\L{\left(}\def\R{\right)}
\def\LP{\left.}\def\RP{\right.}
\def\LB{\left[}\def\RB{\right]}


\catcode`\@=\active
\catcode`@=12  
\catcode`\"=\active



\def\Tr{\mathop{\rm Tr}\nolimits}

\def\A#1{{\cal A}_{#1}}

\def\pol{\varepsilon}

\def\ksl{\slashed{k}}

\def\L{\left(}\def\R{\right)}
\def\LP{\left.}\def\RP{\right.}
\def\spa#1.#2{\left\langle#1\,#2\right\rangle}
\def\spb#1.#2{\left[#1\,#2\right]}
\def\lor#1.#2{\left(#1\,#2\right)}
\def\sand#1.#2.#3{%
\left\langle\smash{#1}{\vphantom1}^{-}\right|{#2}%
\left|\smash{#3}{\vphantom1}^{-}\right\rangle}
\def\sandp#1.#2.#3{%
\left\langle\smash{#1}{\vphantom1}^{-}\right|{#2}%
\left|\smash{#3}{\vphantom1}^{+}\right\rangle}
\def\sandpp#1.#2.#3{%
\left\langle\smash{#1}{\vphantom1}^{+}\right|{#2}%
\left|\smash{#3}{\vphantom1}^{+}\right\rangle}
\catcode`@=11  
\def\meqalign#1{\,\vcenter{\openup1\jot\m@th
   \ialign{\strut\hfil$\displaystyle{##}$ && $\displaystyle{{}##}$\hfil
             \crcr#1\crcr}}\,}
\catcode`@=12  

\newread\epsffilein    
\newif\ifepsffileok    
\newif\ifepsfbbfound   
\newif\ifepsfverbose   
\newdimen\epsfxsize    
\newdimen\epsfysize    
\newdimen\epsftsize    
\newdimen\epsfrsize    
\newdimen\epsftmp      
\newdimen\pspoints     
\pspoints=1bp          
\epsfxsize=0pt         
\epsfysize=0pt         
\def\epsfbox#1{\global\def\epsfllx{72}\global\def\epsflly{72}%
   \global\def\epsfurx{540}\global\def\epsfury{720}%
   \def\lbracket{[}\def\testit{#1}\ifx\testit\lbracket
   \let\next=\epsfgetlitbb\else\let\next=\epsfnormal\fi\next{#1}}%
\def\epsfgetlitbb#1#2 #3 #4 #5]#6{\epsfgrab #2 #3 #4 #5 .\\%
   \epsfsetgraph{#6}}%
\def\epsfnormal#1{\epsfgetbb{#1}\epsfsetgraph{#1}}%
\def\epsfgetbb#1{%
%
%
\openin\epsffilein=#1
\ifeof\epsffilein\errmessage{I couldn't open #1, will ignore it}\else
%
%
   {\epsffileoktrue \chardef\other=12
    \def\do##1{\catcode`##1=\other}\dospecials \catcode`\ =10
    \loop
       \read\epsffilein to \epsffileline
       \ifeof\epsffilein\epsffileokfalse\else
%
%
          \expandafter\epsfaux\epsffileline:. \\%
       \fi
   \ifepsffileok\repeat
   \ifepsfbbfound\else
    \ifepsfverbose\message{No bounding box comment in #1; using defaults}\fi\fi
   }\closein\epsffilein\fi}%
%
%
\def\epsfclipstring{}
\def\epsfsetgraph#1{%
   \epsfrsize=\epsfury\pspoints
   \advance\epsfrsize by-\epsflly\pspoints
   \epsftsize=\epsfurx\pspoints
   \advance\epsftsize by-\epsfllx\pspoints
%
%
   \epsfxsize\epsfsize\epsftsize\epsfrsize
   \ifnum\epsfxsize=0 \ifnum\epsfysize=0
      \epsfxsize=\epsftsize \epsfysize=\epsfrsize
      \epsfrsize=0pt
%
%
     \else\epsftmp=\epsftsize \divide\epsftmp\epsfrsize
       \epsfxsize=\epsfysize \multiply\epsfxsize\epsftmp
       \multiply\epsftmp\epsfrsize \advance\epsftsize-\epsftmp
       \epsftmp=\epsfysize
       \loop \advance\epsftsize\epsftsize \divide\epsftmp 2
       \ifnum\epsftmp>0
          \ifnum\epsftsize<\epsfrsize\else
             \advance\epsftsize-\epsfrsize \advance\epsfxsize\epsftmp \fi
       \repeat
       \epsfrsize=0pt
     \fi
   \else \ifnum\epsfysize=0
     \epsftmp=\epsfrsize \divide\epsftmp\epsftsize
     \epsfysize=\epsfxsize \multiply\epsfysize\epsftmp
     \multiply\epsftmp\epsftsize \advance\epsfrsize-\epsftmp
     \epsftmp=\epsfxsize
     \loop \advance\epsfrsize\epsfrsize \divide\epsftmp 2
     \ifnum\epsftmp>0
        \ifnum\epsfrsize<\epsftsize\else
           \advance\epsfrsize-\epsftsize \advance\epsfysize\epsftmp \fi
     \repeat
     \epsfrsize=0pt
    \else
     \epsfrsize=\epsfysize
    \fi
   \fi
%
%
   \ifepsfverbose\message{#1: width=\the\epsfxsize, height=\the\epsfysize}\fi
   \epsftmp=10\epsfxsize \divide\epsftmp\pspoints
   \vbox to\epsfysize{\vfil\hbox to\epsfxsize{%
      \ifnum\epsfrsize=0\relax
        \includegraphics{#1}%
      \else
        \epsfrsize=10\epsfysize \divide\epsfrsize\pspoints
        \includegraphics{#1}%
      \fi
      \hfil}}%
\global\epsfxsize=0pt\global\epsfysize=0pt}%
%
%
{\catcode`\%=12 \global\let\epsfpercent=
%
%
\long\def\epsfaux#1#2:#3\\{\ifx#1\epsfpercent
   \def\testit{#2}\ifx\testit\epsfbblit
      \epsfgrab #3 . . . \\%
      \epsffileokfalse
      \global\epsfbbfoundtrue
   \fi\else\ifx#1\par\else\epsffileokfalse\fi\fi}%
%
%
\def\epsfempty{}%
\def\epsfgrab #1 #2 #3 #4 #5\\{%
\global\def\epsfllx{#1}\ifx\epsfllx\epsfempty
      \epsfgrab #2 #3 #4 #5 .\\\else
   \global\def\epsflly{#2}%
   \global\def\epsfurx{#3}\global\def\epsfury{#4}\fi}%
%
%
\def\epsfsize#1#2{\epsfxsize}
%
%


\SetUpAuxFile

\loadfourteenpoint
\hfuzz 60 pt


\def\Gr{{\rm Gr}}
\def\cg{c_\Gamma}
\def\rg{r_\Gamma}
\def\Split{\mathop{\rm Split}\nolimits}
\def\tn#1#2{t^{[#1]}_{#2}}

\def\tree{{\rm tree}}

\def\eps{\epsilon}
\def\e{\epsilon}

\def\Atree{A^{\rm tree}}
\def\dlips{d^D{\rm LIPS}}
\def\Slash#1{\slash\hskip -0.17 cm #1}
\def\Li{\mathop{\hbox{\rm Li}}\nolimits}

\def\Lz{\mathop{\hbox{\rm L}}\nolimits_0}
\def\Kz{\mathop{\hbox{\rm K}}\nolimits_0}
\def\Mz{\mathop{\hbox{\rm M}}\nolimits_0}
\def\tr{\mathop{\hbox{\rm tr}}\nolimits}

\def\S{{\cal C}}
\def\Ord{{\cal O}}
\def\hf{{\textstyle {1\over 2}}}
\def\Det{\hat\Delta}
\def\lr{\leftrightarrow}
\def\lsl{\not{\hbox{\kern-2.3pt $\ell$}}}
\def\ksl{\not{\hbox{\kern-2.3pt $k$}}}
\def\soft#1#2#3{{\cal S}_{#2}(#1,#3)}

\def\Fn{n}
\def\Fs#1#2{F^{{#1}}_{\Fn:#2}}
\def\Fone{\Fs{\rm 1m}}
\def\Feasy{\Fs{{\rm 2m}\,e}}
\def\Fhard{\Fs{{\rm 2m}\,h}}
\def\Fthree{\Fs{\rm 3m}}
\def\Ffour{\Fs{\rm 4m}}
\def\Wsix#1{W_6^{(#1)}}


\preref\SpinorHelicity{
F.A.\ Berends, R.\ Kleiss, P.\ De Causmaecker, R.\ Gastmans and T.T.\ Wu,
        \PLB 103:124 (1981)\semi
P.\ De Causmaeker, R.\ Gastmans,  W.\ Troost and  T.T.\ Wu,
        \NPB206:53 (1982)\semi
R.\ Kleiss and W.J.\ Stirling, \NPB 262:235 (1985)\semi
J. F.\ Gunion and Z.\ Kunszt, Phys.\ Lett.\ 161B:333 (1985)\semi
R.\ Gastmans and T.T.\ Wu,
        {\it The Ubiquitous Photon: Helicity Method for QED and QCD}
        (Clarendon Press, 1990)\semi
Z. Xu, D.-H.\ Zhang and L. Chang, \NPB 291:392 (1987)}

\preref\RecursiveA{F.A.\ Berends and W.T.\ Giele, \NPB 306:759 (1988)}

\preref\RecursiveB{D.A.\ Kosower, \NPB335:23 (1990)}

\preref\MahlonA{G.D.\ Mahlon, Phys.\ Rev.\ D49:2197 (1994)}

\preref\MahlonB{G.D.\ Mahlon, Phys.\ Rev.\ D49:4438 (1994)}

\preref\Long{Z. Bern and D.A.\ Kosower, \NPB 379:451 (1992)}

\preref\StringBased{
Z. Bern and D.A.\ Kosower, \PRL 66:1669 (1991)\semi
Z. Bern and D.A.\ Kosower, in {\it Proceedings of the PASCOS-91
Symposium}, eds.\ P. Nath and S. Reucroft (World Scientific, 1992)\semi
Z. Bern, Phys.\ Lett.\ 296B:85 (1992)\semi
K. Roland, Phys.\ Lett.\ 289B:148 (1992)\semi
Z. Bern, D.C.\ Dunbar and T. Shimada, Phys.\ Lett.\ 312B:277 (1993)\semi
G. Cristofano, R. Marotta and K. Roland, Nucl.\ Phys.\ B392:345 (1993)\semi
D.C.\ Dunbar and P.S.\ Norridge, hep-th/9408014}

\preref\FiveGluon{Z. Bern, L. Dixon and D.A.\ Kosower, Phys.\ Rev.\ Lett.\
70:2677 (1993)}

\preref\ParkeTaylor{S.J.\ Parke and T.R.\ Taylor, \PRL 56:2459
(1986)}

\preref\TreeCollinear{F.A.\ Berends and W.T.\ Giele, Nucl.\ Phys.\
B313:595 (1989)}

\preref\AllPlus{Z. Bern, G. Chalmers, L. Dixon and D.A.\ Kosower,
Phys.\ Rev.\ Lett.\ 72:2134 (1994)}

\preref\SusyFour{Z. Bern, L. Dixon, D.C.\ Dunbar and D. Kosower, preprint
SLAC-PUB-6415,
hep-ph/9403226, to appear in Nucl.\ Phys.\ B;
SLAC-PUB-6490, hep-ph/9405248}

\preref\QCDConf{Z. Bern, L. Dixon, D.C.\ Dunbar and D. Kosower,
preprint SLAC-PUB-6490, hep-ph/9405248}

\preref\GG{W.T.\ Giele and E.W.N.\ Glover,
Phys.\ Rev.\ D46:1980 (1992)\semi
W.T.\ Giele, E. W. N.\ Glover and D. A. Kosower,
Nucl.\ Phys.\ B403:633 (1993)}

\preref\KunsztSingular{Z. Kunszt and D. Soper, Phys.\ Rev.\ D46:192 (1992)\semi
 Z. Kunszt, A. Signer and Z. Tr\'ocs\'anyi,
Nucl.\ Phys.\ B420:550 (1994)}

\preref\Superspace{S.J.\ Gates, M.T.\ Grisaru, M. Rocek and W. Siegel,
 {\it Superspace}, (Benjamin/Cummings, 1983)}

\preref\BDKconf{Z. Bern, L. Dixon and D.A.\ Kosower, Proceedings of
Strings 1993, eds. M.B.\ Halpern, A. Sevrin and G. Rivlis
(World Scientific, Singapore, 1994), hep-th/9311026}

\preref\TreeColor{J.E.\ Paton and H.M.\ Chan, Nucl.\ Phys.\ B10:516 (1969)\semi
F.A.\ Berends and W.T.\ Giele, \NPB 294:700 (1987)\semi
M.\ Mangano, \NPB 309:461 (1988)}

\preref\MPX{M.\ Mangano, S.J.\ Parke and Z.\ Xu,
Nucl.\ Phys.\ B298:653 (1988)}

\preref\Color{Z. Bern and D.A.\ Kosower, Nucl.\ Phys.\ B362:389 (1991)}

\preref\Mapping{Z. Bern and D.C.\ Dunbar,  \NPB 379:562 (1992)}

\preref\Lam{C.S.\ Lam, Nucl.\ Phys.\ B397:143 (1993);
                       Phys.\ Rev.\ D48:873 (1993)}

\preref\Background{G. 't Hooft,
in Acta Universitatis Wratislavensis no.\
38, 12th Winter School of Theoretical Physics in Karpacz, {\it
Functional and Probabilistic Methods in Quantum Field Theory},
Vol. 1 (1975)\semi
B.S.\ DeWitt, in {\it Quantum gravity II}, eds. C. Isham, R.\ Penrose and
D.\ Sciama (Oxford, 1981)\semi
L.F.\ Abbott, Nucl.\ Phys.\ B185:189 (1981)\semi
L.F. Abbott, M.T. Grisaru and R.K. Schaefer,
Nucl.\ Phys.\ B229:372 (1983)}

\preref\Subsequent{M.J.\ Strassler,  Nucl.\ Phys.\ B385:145 (1992)\semi
M.G.\ Schmidt and C. Schubert, Phys.\ Lett.\ 318B:438 (1993);
 Phys.\ Lett.\ 331B:69 (1994)\semi
D. Fliegner, M.G.\ Schmidt and C. Schubert, HD-THEP-93-44, hep-ph/9401221}

\preref\Tasi{
Z. Bern, hep-ph/9304249, in {\it Proceedings of Theoretical
Advanced Study Institute in High Energy Physics (TASI 92)},
eds.\ J. Harvey and J. Polchinski (World Scientific, 1993)\semi
Z.\ Bern and A.\ Morgan, Phys.\ Rev.\ D49:6155 (1994)}

\preref\Siegel{W. Siegel, Phys.\ Lett.\ 84B:193 (1979)\semi
D.M.\ Capper, D.R.T.\ Jones and P. van Nieuwenhuizen, Nucl.\ Phys.\
B167:479 (1980)\semi
L.V.\ Avdeev and A.A.\ Vladimirov, Nucl.\ Phys.\ B219:262 (1983)}

\preref\DimensionalRegularization{G. 't\ Hooft and M. Veltman,
Nucl.\ Phys.\ B44:189 (1972)}

\preref\CollinsBook{J.C.\ Collins, {\it Renormalization}
(Cambridge University Press, 1984)}

\preref\EpsHelicity{D.A. Kosower, Phys.\ Lett.\ 254B:439 (1991)}

\preref\PV{L.M.\ Brown and R.P.\ Feynman, Phys.\ Rev.\ 85:231 (1952)\semi
G.\ Passarino and M.\ Veltman, Nucl.\ Phys.\ {B160:151} (1979)\semi
G. 't Hooft and M. Veltman, \NPB{153:365 (1979)}\semi
R. G. Stuart, Comp.\ Phys.\ Comm.\ 48:367 (1988)\semi
R. G. Stuart and A. Gongora, Comp.\ Phys.\ Comm.\ 56:337 (1990)}

\preref\VNV{
W. van Neerven and J.A.M. Vermaseren, Phys.\ Lett.\ 137B:241 (1984)}

\preref\OtherMPoint{
D. B. Melrose, Il Nuovo Cimento 40A:181 (1965)\semi
W. van Neerven and J.A.M. Vermaseren, Phys.\ Lett.\ 137B:241 (1984)\semi
G. J. van Oldenborgh and J.A.M. Vermaseren, \ZPC{46:425 (1990)}\semi
G. J. van Oldenborgh, PhD thesis, University of Amsterdam (1990)\semi
A. Aeppli, PhD thesis, University of Zurich (1992)}

\preref\GramDet{V.E.\ Asribekov, Sov.\ Phys.\ -- JETP 15:394 (1962)\semi
N. Byers and C.N.\ Yang, Rev.\ Mod.\ Phys.\ 36:595 (1964)}

\preref\Lewin{L.\ Lewin, {\it Dilogarithms and Associated Functions\/}
(Macdonald, 1958)}

\preref\Cutting{L.D.\ Landau, Nucl.\ Phys.\ 13:181 (1959)\semi
 S. Mandelstam, Phys.\ Rev.\ 112:1344 (1958), 115:1741 (1959)\semi
 R.E.\ Cutkosky, J.\ Math.\ Phys.\ 1:429 (1960)}

\preref\GielePrivate{W. Giele, private communication}

\preref\Susy{M.T.\ Grisaru, H.N.\ Pendleton and P.\ van Nieuwenhuizen,
Phys. Rev. {D15}:996 (1977)\semi
M.T.\ Grisaru and H.N. Pendleton, Nucl.\ Phys.\ B124:81 (1977)\semi
S.J.\ Parke and T. Taylor, Phys.\ Lett.\ 157B:81 (1985)\semi
Z. Kunszt, Nucl.\ Phys.\ B271:333 (1986)}

\preref\ManganoParke{M. Mangano and S.J.\ Parke, \NPB 299:673 (1988)}

\preref\GSB{M.B.\ Green, J.H.\ Schwarz and L.\ Brink,
 Nucl.\ Phys.\ B198:472 (1982)}

\preref\IntegralsLong{Z. Bern, L. Dixon and D.A.\ Kosower,
\NPB 412:751 (1994)}

\preref\ManganoReview{M. Mangano and S.J.\ Parke, Phys.\ Rep.\ 200:301 (1991)}

\preref\QCDReview{J.C.\ Collins, D.E.\ Soper,
and G.\ Sterman, in {\it Perturbative Quantum Chromodynamics},
ed.\ A.H.\ Mueller (World Scientific, 1989)}

\preref\ThreeMassTriangle{H.-J. Lu and C. Perez, SLAC--PUB--5809}

\preref\IntegralsShort{Z. Bern, L. Dixon and D.A.\ Kosower,
Phys.\ Lett.\ 302B:299 (1993); erratum {\it ibid.} 318:649 (1993)}

\preref\qqggg{Z. Bern, L. Dixon and D.A.\ Kosower, in preparation}

\preref\FourMassBox{A. Denner, U. Nierste and R. Scharf,
  \NPB{367:637 (1991)}\semi
N.I.\ Usyukina and A.I.\ Davydychev, Phys.\ Lett.\ {298B:363 (1993)};
Phys.\ Lett.\ {305B:136 (1993)}}


\nopagenumbers

\noindent

$\null$

\vskip -1.6 cm

hep-ph/9409265
\hfill SLAC-PUB-6563\hfil\break
\rightline{Saclay/SPhT--T94/95}
\rightline{UCLA/TEP/94/29}
\rightline{SWAT-94-36}

\vskip -2.4 cm

\baselineskip 12 pt
\Title{\bf Fusing Gauge Theory Tree Amplitudes Into Loop Amplitudes}

\vskip -.7 cm
\centerline{\ninerm ZVI BERN${}^{\sharp}$}
\baselineskip=13pt
\centerline{\nineit Department of Physics, UCLA, Los Angeles, CA 90024, USA}
\vglue 0.3cm

\centerline{\ninerm LANCE DIXON${}^{\star}$}
\centerline{\nineit Stanford Linear Accelerator Center, Stanford University,
Stanford, CA 94309, USA}
\vglue 0.3cm

\centerline{\ninerm DAVID C. DUNBAR${}^{\dagger}$ }
\centerline{\nineit Department of Physics, UCLA, Los Angeles, CA 90024, USA}

\vglue 0.2cm
\centerline{\ninerm and}
\vglue 0.2cm
\centerline{\ninerm DAVID A. KOSOWER${}^{\ddagger}$}
\baselineskip12truept
\centerline{\nineit Service de Physique Th\'eorique,
 Centre d'Etudes de Saclay}
\centerline{\nineit F-91191 Gif-sur-Yvette cedex, France}

\vglue 0.7cm
\centerline{\tenrm ABSTRACT}
\vglue 0.3cm
{\rightskip=3pc
\leftskip=3pc
\tenrm\baselineskip=12pt
\noindent
We identify a large class of one-loop amplitudes for massless
particles that can be constructed via unitarity from tree amplitudes,
without any ambiguities.  One-loop amplitudes for massless
supersymmetric gauge theories fall into this class; in addition, many
non-supersymmetric amplitudes can be rearranged to take advantage of
the result.  As applications, we construct the one-loop amplitudes for
$n$-gluon scattering in $N=1$ supersymmetric theories with the
helicity configuration of the Parke-Taylor tree amplitudes, and for
six-gluon scattering in $N=4$ super-Yang-Mills theory for all helicity
configurations.  }
\baselineskip 15 pt

\vglue 0.3cm

\vfil\vskip .2 cm
\noindent\hrule width 3.6in\hfil\break
${}^{\sharp}$Research supported in part by the US Department of Energy
under grant DE-FG03-91ER40662 and in part by the
Alfred P. Sloan Foundation under grant BR-3222. \hfil\break
${}^{\star}$Research supported by the Department of
Energy under grant DE-AC03-76SF00515.\hfil\break
${}^{\dagger}$Address after Sept. 1, 1994: University
College of Swansea,
UK. Research supported in part by the NSF under grant
PHY-9218990 and in part by the Department of Energy under grant
DE-FG03-91ER40662. \hfil\break
${}^{\ddagger}$Laboratory of the {\it Direction des Sciences de la Mati\`ere\/}
of the {\it Commissariat \`a l'Energie Atomique\/} of France.\hfil\break
\eject

\footline={\hss\tenrm\folio\hss}


\section{Introduction}
\tagsection\IntroSection

While the computation of amplitudes in perturbative QCD is central to
the study of jet and jet-associated physics at current and future
hadron colliders, and thereby important to the prospects of
discovering new physics there, such computations are not easy.  Recent
years have nonetheless seen a number of improvements in techniques
both at tree level and at the one-loop level, which have allowed the
computation of a more extensive set of matrix elements than previously
possible.  These techniques include the spinor helicity basis
[\use\SpinorHelicity]; recurrence relations for amplitudes
[\use\RecursiveA,\use\RecursiveB,\use\MahlonA,\use\MahlonB]; and
string-based techniques for one-loop amplitudes
[\use\Long,\use\StringBased,\use\FiveGluon].  Collinear limits are useful
in constructing ans\"atze for amplitudes with an arbitrary number of
gluons, both at tree level [\use\ParkeTaylor,\use\TreeCollinear] and at loop
level~[\use\AllPlus].  They also provide strong checks on results
obtained by other means.

Recently we gave a formula [\use\SusyFour,\use\QCDConf] for all-multiplicity
one-loop amplitudes in an $N=4$ supersymmetric gauge theory, with the
helicity configuration of the Parke-Taylor [\use\ParkeTaylor]
tree-level amplitudes (which have maximal helicity violation).
The computation of this formula utilized a new
technique, based on the observation that for an $N=4$ supersymmetric
amplitude with all external gluons, all integrals that enter are
determined by their absorptive parts (that is, their cuts), and thus
the amplitude can be computed from the interference of appropriate
tree amplitudes.  The choice of external helicities in turn restricts
the set of tree helicity amplitudes that appear, and thereby renders the
computation tractable even for an arbitrary number of external legs.
These amplitudes constitute part of the computation of the
corresponding one-loop amplitudes in QCD: the latter amplitudes have a
natural decomposition~[\use\FiveGluon,\use\Tasi]
into three pieces: an $N=4$ supersymmetric amplitude,
the contribution of an internal $N=1$ supersymmetric chiral
matter multiplet, and the contribution of an internal scalar.

In this paper, we show that a larger class of amplitudes can be
calculated solely from the knowledge of their cuts.  In particular,
the collection of amplitudes that can be calculated in this manner ---
amplitudes which we shall term {\it cut-constructible} --- includes all
purely massless one-loop amplitudes for which the loop integrals
satisfy certain power-counting criteria, given in
section~\use\UnitarityProofSection.  As we shall show, all
color-ordered amplitudes in massless supersymmetric gauge theories
with trivial superpotential are cut-constructible.  (We expect that
the result to carry over to all amplitudes in massless supersymmetric
theories.)  This observation is also useful in non-supersymmetric
theories, where sums or differences of certain contributions to
amplitudes satisfy the criterion, allowing one to trade a harder
calculation for an easier one.

As applications of the method, we compute the one-loop maximally
helicity-violating $n$-gluon amplitudes in $N=1$ supersymmetric gauge
theory, and those one-loop six-gluon amplitudes in $N=4$
supersymmetric gauge theory that were not previously computed in
ref.~[\use\SusyFour].

For amplitudes that are not cut-constructible, such as scalar-loop
contributions to $n$-gluon amplitudes, one may still use the cuts to
determine efficiently all terms with logarithms and dilogarithms.  We
will present an explicit scalar-loop example with an arbitrary number
of external legs.  In this case, there are additional polynomial terms
which cannot be directly determined from the cuts, although the form of the
missing polynomials is restricted by the simple factorization
properties of color-ordered amplitudes in the soft and collinear
limits. (The sorts of universal functions that enter into these
limits also provide a powerful means of summarizing the integrations
over singular regions of phase space in the construction of numerical
programs~[\use\GG].) A technique utilizing collinear factorization
has been used to
construct an ansatz for the all-plus helicity configuration with an
arbitrary number of external legs~[\use\BDKconf,\use\AllPlus], which
was subsequently proven by Mahlon via recursive
techniques~[\use\MahlonB].  Recursive techniques have also been used
to calculate a variety of other non-cut-constructible
amplitudes~[\use\MahlonA,\use\MahlonB].  In this way the collinear and
recursive techniques complement the cutting method discussed in this
paper.

In section~\use\ReviewSection\ we briefly review salient features of
color and supersymmetry decompositions of QCD amplitudes.
Section~\use\UnitarityProofSection\ contains a ``uniqueness'' result
which shows that a one-loop amplitude whose diagrams all satisfy a certain
power-counting criterion can be completely determined from its cuts.
A general discussion of the applicability of this result to
massless supersymmetric gauge theories is given in
section~\use\SusySection.
A concrete application follows in section~\use\MHVSection:  by
systematically determining all the cuts, we construct the maximally
helicity-violating one-loop amplitudes in $N=1$ supersymmetric gauge
theory, for an arbitrary number of external gluons.
In section~\use\NonMHVSection\ we present those
six-gluon amplitudes in $N=4$ supersymmetric gauge theory that have
not appeared previously.
In section~\use\NonSusySection\ we show how one can apply
cutting methods to non-supersymmetric theories.
Section~\use\ConclusionSection\ contains our conclusions.


\section{Review}
\tagsection\ReviewSection

At tree level, amplitudes in an $U(N_c)$ or $SU(N_c)$ gauge theory
with $n$ external gluons can be written in terms of
color-ordered partial amplitudes $A_n^\tree$, multiplied by
an associated color trace [\use\TreeColor,\use\MPX].
In order to obtain the full amplitude,
one must sum over all non-cyclic permutations,
$$
\A{n}^\tree(\{k_i,\lambda_i,a_i\}) =
g^{n-2} \sum_{\sigma\in S_n/Z_n} \Tr(T^{a_{\sigma(1)}}
\cdots T^{a_{\sigma(n)}})
\ A_n^\tree(k_{\sigma(1)}^{\lambda_{\sigma(1)}},\ldots,
            k_{\sigma(n)}^{\lambda_{\sigma(n)}})\ ,
\eqn\TreeAmplitudeDecomposition
$$
where $k_i$, $\lambda_i$, and $a_i$ are respectively the momentum,
helicity ($\pm$), and color index of the $i$-th external
gluon, $g$ is the coupling constant, and $S_n/Z_n$ is the set of
non-cyclic permutations of $\{1,\ldots, n\}$.
The $U(N_c)$ ($SU(N_c)$) generators $T^a$ are the set of hermitian
(traceless hermitian) $N_c\times N_c$ matrices,
normalized so that $\Tr\L T^a T^b\R = \delta^{ab}$.
The color decomposition~(\use\TreeAmplitudeDecomposition)
can be derived in conventional field theory by using
$$
f^{abc} = -{i\over\sqrt2} \Tr\L \LB T^a, T^b\RB T^c\R,
\eqn\struct
$$
where the $T^a$ may by either $SU(N_c)$ matrices or $U(N_c)$ matrices.
The structure constants $f^{abc}$ vanish when any index belongs to the
$U(1)$, which is generated by the matrix
$T^{a_{U(1)}} \equiv {\bf 1}/\sqrt{N_c}$;
therefore the partial amplitudes satisfy the $U(1)$ decoupling
identities~[\use\MPX,\use\RecursiveA]
$$
\A{n}^{\rm tree}(\{k_i,\pol_i,a_i\}_{i=1}^{n-1};k_n,\pol_n,a_{U(1)}) = 0 \; .
\eqn\DecouplingTreeBase
$$

One may perform a similar color decomposition for
one-loop amplitudes;
in this case, there are up to two traces
over color matrices [\use\Color],
and one must also sum over the different spins $J$ of the internal
particles circulating in the loop.
When all internal particles transform as color adjoints,
the result takes the form
$$
{\cal A}_n\L \{k_i,\lambda_i,a_i\}\R =
  g^n \sum_{J=0,{1\over2},1} n_J\,\sum_{c=1}^{\lfloor{n/2}\rfloor+1}
      \sum_{\sigma \in S_n/S_{n;c}}
     \Gr_{n;c}\L \sigma \R\,A_{n;c}^{[J]}(\sigma),
\eqn\ColorDecomposition$$
where ${\lfloor{x}\rfloor}$ is the largest integer less than or equal to $x$,
and $n_J$ is the number of particles of spin $J$.
The leading color-structure factor
$$
\Gr_{n;1}(1) = N_c\ \Tr\L T^{a_1}\cdots T^{a_n}\R
\anoneqn
$$
is just $N_c$ times the tree color factor, and the subleading color
structures are given by
$$
\Gr_{n;c}(1) = \Tr\L T^{a_1}\cdots T^{a_{c-1}}\R\,
\Tr\L T^{a_c}\cdots T^{a_n}\R.
\anoneqn
$$
$S_n$ is the set of all permutations of $n$ objects,
and $S_{n;c}$ is the subset leaving $\Gr_{n;c}$ invariant.
Once again it is convenient to use $U(N_c)$ matrices; the extra $U(1)$
decouples from all final results [\use\Color].
(For internal particles in the fundamental ($N_c+\bar{N_c}$) representation,
only the single-trace color structure ($c=1$) would be present,
and the corresponding color factor would be smaller by a factor of $N_c$.
In this case the $U(1)$ gauge boson will {\it not} decouple from
the partial amplitude, so one should only sum over $SU(N_c)$ indices
when color-summing the cross-section.)
In each case the massless spin-$J$ particle is taken to have two
helicity states whether it is a gauge boson, a Weyl fermion, or a
complex scalar.

In the next-to-leading-order correction to the cross-section, summed
over colors, the leading contribution for large $N_c$ comes from
$A_{n;1}^{[J]}$; the remaining partial amplitudes $A^{[J]}_{n;c}$
($c>1$) produce corrections which
are down by a factor of $1/N_c^2$~[\use\Color].  For
amplitudes whose external legs are purely in the adjoint representation,
we showed in
section~7 of ref.~[\use\SusyFour] how to obtain
$A^{[J]}_{n;c}$ as a sum over permutations of the leading contribution
$A^{[J]}_{n;1}$.  Therefore, it is sufficient to
calculate the $A^{[J]}_{n;1}$.
These color-ordered partial amplitudes involve
a limited class of loop integrals, those where the
external legs (including attached trees we amputate to leave off-shell
legs) are ordered sequentially.

We also draw lessons from string-based techniques for computing
one-loop amplitudes in gauge theories~[\use\Long,\use\StringBased], used
previously to calculate all one-loop five-gluon helicity
amplitudes~[\use\FiveGluon].  The techniques, though independent of the
conventional Feynman-diagram expansion, can be understood using a
reorganization of a conventional approach [\use\Mapping,\use\Lam ]
utilizing the background-field method~[\use\Background] and a
second-order formalism for internal fermions.
Such an approach has also proven useful for the
computation of effective actions~[\use\Subsequent].
The string-based techniques further reveal that gluon amplitudes
are most naturally written in a form~[\use\FiveGluon,\use\Tasi] that
reflects the simplicity of contributions from different supersymmetry
multiplets,
$$
\eqalign{
A_{n;1}^{[0]} &= c_\Gamma  \L V^s_n \Atree_n + i F^s_n \R\,,\cr
A_{n;1}^{[1/2]} &= -c_\Gamma \L (V^f_n+V^s_n) \Atree_n
                             + i(F^f_n+F^s_n) \R\,
,\cr
A_{n;1}^{[1]} &= c_\Gamma \L (V^g_n+4V^f_n+V^s_n) \Atree_n
+ i (F^g_n + 4F^f_n + F^s_n) \R
  \,,\cr
}
\eqn\Totalamp
$$
where, with $\eps= (4-D)/2$ the dimensional regularization parameter,
$$
\rg = {\Gamma(1+\e)\Gamma^2(1-\e) \over \Gamma(1-2\e) },\hskip 1cm
\cg = {\rg\over(4\pi)^{2-\eps}}\; ,
\eqn\Prefactor
$$
and we have assumed use of a supersymmetry preserving regulator
[\use\Siegel,\use\Long].
In equation~(\Totalamp) the superscripts $g$, $f$ and $s$ reflect the
supersymmetric decomposition,
whereas the split into $V$ and $F$ type pieces indicates the singularities
as $\eps\to0$.

All singular pieces (poles in $\e$) are proportional to the tree
amplitude, and hence are contained in the $V_n$ factors; the $F_n$
terms, in contrast, are finite as $\e\rightarrow0$ and may involve
different spinor-product forms [\use\SpinorHelicity] than the tree.
While $V_n$ contains
only momentum invariants but no spinor products, the ratio of $F_n$
to the tree amplitude will in general contain not just momentum invariants
but spinor products as well.  (The latter could be replaced in such a
phase-free ratio by combinations of momentum invariants and
contracted antisymmetric tensors such as
$\varepsilon(1,2,3,4) \equiv  4i \varepsilon_{\mu\nu\rho\sigma}
 k_1^\mu k_2^\nu k_3^\rho k_4^\sigma$, but the latter combinations
cannot be eliminated from the $F_n$.)
There is of course some freedom in shifting terms between $V_n$
and $F_n$.

The organization~(\use\Totalamp) in terms of $g$, $f$ and $s$ pieces
is equivalent to calculating the fermion and gluon loop
contributions in terms of the scalar loop contributions plus the
contributions from supersymmetric multiplets.
In an $N=4$ super-Yang-Mills theory, summing over the contributions
from one gluon, four Weyl fermions and three complex (or six real)
scalars, all functions except $V_n^g$ and $F_n^g$ cancel from
eq.~(\use\Totalamp) and the amplitudes are
$$
A_{n;1}^{N=4}\ \equiv\
A_{n;1}^{[1]} + 4A_{n;1}^{[1/2]}+3 A_{n;1}^{[0]}\ =\
c_\Gamma \,  ( V^g_n \; A_{n}^{\rm tree}   + i F^g_n).
\eqn\Neqfoursum
$$
For an $N=1$ chiral multiplet, containing one scalar and one Weyl fermion,
only the functions $V^f_n$ and $F^f_n$ survive,
$$
A_{n;1}^{N=1\ {\rm chiral}}\ \equiv\
A_{n;1}^{[1/2]}\
+A_{n;1}^{[0]}
=\
-c_\Gamma\, \L V^f_n \Atree_n + i F^f_n \R.
\eqn\Neqonesum
$$

We can thus rewrite equation~(\use\Totalamp) in terms of the $N=4$
supersymmetric amplitude, the contribution of an $N=1$ supersymmetric
chiral multiplet, and the contribution of a complex scalar,
$$
\eqalign{
A_{n;1}^{[1/2]} &= A_{n;1}^{N=1\ {\rm chiral}} - A_{n;1}^{[0]}\,,\cr
A_{n;1}^{[1]} &= A_{n;1}^{N=4}-4 A_{n;1}^{N=1\,{\rm chiral}} + A_{n;1}^{[0]}
       \,.\cr
}\eqn\TotalampB
$$

These equations, which hold for supersymmetry preserving regulators
[\use\Siegel,\use\Long], may
be converted to the 't Hooft-Veltman (HV) [\use\DimensionalRegularization]
or conventional [\use\CollinsBook] schemes by
accounting for the $[\eps]$-scalar differences between the schemes.
In the 't~Hooft-Veltman or conventional scheme
we must remove $[\eps]$-scalars propagating in the gluon loop.
For the amplitudes with external gluon helicities in four dimensions
$(\pm)$ we modify the above relation to
$$
\eqalign{
A_{n;1}^{[1/2]}\Bigr|_{\rm HV,\ conventional}
&= A_{n;1}^{N=1\ {\rm chiral}} - A_{n;1}^{[0]}\,,\cr
A_{n;1}^{[1]}\Bigr|_{\rm HV,\ conventional}
&= A_{n;1}^{N=4}-4 A_{n;1}^{N=1\,{\rm chiral}} +
     (1-\eps) A_{n;1}^{[0]}
       \, ,\cr
}\eqn\TotalampC
$$
where the quantities on the right-hand-side are ones computed in this
paper.  In the conventional scheme there are additional
amplitudes with external $[\eps]$-helicities
[\use\EpsHelicity,\use\Long]. For the case of $n$ external gluons
these are given by the corresponding tree amplitudes with
$[\eps]$-helicities (which are proportional to $\eps$) multiplied by
the universal singularities [\use\GG,\use\KunsztSingular] contained in
the $V_n$.

Given the explicit results of section~\use\MHVSection\ for
$A_{n;1}^{N=1\ {\rm chiral}}$ in the maximally helicity-violating
(MHV) configurations, and of ref.~[\SusyFour] for the corresponding
$N=4$ amplitudes, only the computation of the scalar contribution
$A_{n;1}^{[0]}$ remains in order to obtain the full QCD amplitude
for this choice of helicities.


\section{A Uniqueness Result}
\tagsection\UnitarityProofSection

In this section, we shall prove that a certain class of
amplitudes can be determined entirely from the
knowledge of their cuts; we call such amplitudes {\it
cut-constructible}.  The class consists of color-ordered one-loop
amplitudes in massless theories for which a diagrammatic
representation exists where all the loop-momentum integrals satisfy
the following power-counting criterion: the $m$-point integrals have
at most $m-2$ powers of the loop momentum in the numerator of the
integrand, and the two-point (bubble) integrals have at most one power
of the loop momentum.  In other words, in the dimensionally-regulated
integral
$$
I_m[ P(p^\mu)]\ =\ (-1)^{m+1} i \L4\pi\R^{2-\e} \,\int
{d^{4-2\e}p\over \L2\pi\R^{4-2\e}}
\;{P(p^\mu)\over p^2 \L p-K_1\R^2 \L p-K_1-K_2\R^2 \cdots \L p+K_m\R^2}\;,
\eqn\GeneralMPoint
$$
where $K_i$ are (sums of) external momenta $k_i$ for the amplitude,
the loop-momentum polynomial $P(p^\mu)$ should have degree
at most $m-2$, except for $m=2$ when it should be at most linear.

We refer to this result as a ``uniqueness'' result because the
issue is whether the cuts uniquely determine an amplitude.
An equivalent question is whether one can rule out terms
in an amplitude that are pure ``polynomials''
--- actually, rational functions --- in the kinematic invariants,
which cannot be detected by any cut.
By inspecting the integrals appearing in cut-constructible amplitudes,
we shall show that they have no such additive polynomial ambiguity.

In general, the power counting associated with a given amplitude
depends on the specific gauge choice, regulator, and diagrammatic
organization.  In order to apply the uniqueness result, we need only
find one organization of the diagrams satisfying the power-counting
criterion.  It is often convenient to use string-based diagrams or
background-field-gauge (superspace) diagrams~[\use\Superspace],
because they can satisfy the power-counting criterion even when the
corresponding diagrams in conventional Feynman gauge do not.
Furthermore, amplitudes which satisfy the power-counting criterion
with a supersymmetry preserving regulator
[\use\Siegel,\use\Long] will generally not do so in the
't~Hooft-Veltman [\use\DimensionalRegularization] or conventional
[\use\CollinsBook] schemes.

\def\IntSet{{\cal F}}
In a previous paper~[\use\SusyFour] we considered
the case of $N=4$ supersymmetric amplitudes, where the loop integrals
satisfy an even more stringent power-counting criterion
($m$-point integrals have at most $m-4$ powers of the loop-momentum).
The proof of uniqueness presented here is an extension of the proof
given in ref.~[\use\SusyFour].
An outline of the proof follows.
We first note that any one-loop $m$-point tensor integral in
$4-2\eps$ dimensions with $m\geq4$ can be reduced to a combination of
scalar box integrals, tensor triangle integrals, and tensor bubble integrals,
using standard integral reduction
formulae~[\use\PV,\use\VNV,\use\OtherMPoint].
(A tensor integral means any integral with non-constant $P(p^\mu)$,
while a scalar integral has $P(p^\mu)=1$.)
When applied to integrals obeying the power-counting criterion,
the reduction formulae maintain the discrepancy of two units between the
number of legs in the integral and the degree of the loop-momentum polynomial,
so that only scalar box integrals and {\it linear} triangle and bubble
integrals appear.
(In the $N=4$ case the triangle and bubble integrals were absent.)
The linear bubble integrals arise directly from internal propagator
diagrams, in amplitudes with external fermions.
We define a set of integral functions $\IntSet_n$ related to these
integrals, such that any cut-constructible $n$-point amplitude $A_n$
can be written as a linear combination of the elements of $\IntSet_n$,
$$
  A_n =\  \sum_{i|I_i\in \IntSet_n} c_i I_i\ ,
\eqn\AnDecomp
$$
with coefficients $c_i$ allowed to be arbitrary {\it rational\/} functions
of the momentum invariants.
Finally, direct inspection of the integrals in $\IntSet_n$
shows that any cut-free linear combination
of these integrals must vanish, and therefore
the cuts contain sufficient information to
completely determine $A_n$.

We do not review in detail the general techniques~[\use\PV,\VNV,\OtherMPoint]
for reducing dimensionally-regulated $(m\geq4)$-point tensor integrals
to lower-point and lower-degree integrals,
because we need only two particular features.
First, scalar $(m>4)$-point integrals can be reduced to a linear
combination of scalar box integrals~[\use\VNV,\OtherMPoint], provided that
at least four of the external momentum vectors are kept in four
dimensions.%
\footnote{${}^\dagger$}{
There is a technical complication having to do with the number
of independent kinematic variables for $n$-point processes
in four dimensions.  Momentum conservation leaves $n(n-3)/2$ kinematic
invariants, e.g. the $t_i^{[r]}$ defined below,
but for $n\geq6$ only $3n-10$ of these are actually independent in
four dimensions,
due to constraints imposed by the vanishing of various Gram
determinants~[\use\GramDet].
In explicit loop calculations we wish to restrict
all external momenta to four dimensions, so that we can use
four-dimensional tree amplitudes to evaluate cuts of the
loop amplitudes.
However, then the Gram-determinantal constraints
would complicate our argument that the cut-constructible amplitudes
are uniquely determined, which implicitly assumes that the $t_i^{[r]}$
are all independent.
A solution is to restrict exactly four momenta to $D=4$ --- this is
the minimum number needed for the $(m>4)$-point reduction formulae
to hold~[\use\VNV] --- but leave the rest free in $D=4-2\e$,
thus removing the Gram-determinantal constraints and validating
the uniqueness argument.}
Second, in the reduction of $(m>4)$-point tensor integrals,
the degree of the loop-momentum polynomial
always shrinks along with the number of legs.
The reduction procedure involves replacing a loop momentum in the
numerator by a linear combination of inverse propagators $(p-p_i)^2$,
which are cancelled against factors in the denominator to reduce the
number of legs of the integral, plus a constant (in the loop momentum),
$p^\mu\ = \ \sum_i \alpha_i^\mu (p-p_i)^2 + \alpha_0^\mu$.
The reduction thus maintains
the criterion of having two fewer powers of loop momentum in the integrand
than the number of external legs.
Reducing the (quadratic and linear) tensor box integrals
in the same way~[\use\PV],
one arrives finally at the aforementioned linear combination
of scalar box integrals, linear triangle integrals and linear
bubble integrals.

As shown in appendix~\use\IntegralsAppendix, the linear bubble integrals
can be reduced to the corresponding scalar bubble integrals, and the
linear triangle integrals can be reduced to scalar triangle integrals plus
scalar bubbles.
The momenta flowing out of the legs of the integrals, $K_i$ are sums
of external momenta $k_i$ of the original $n$-point integral.
Assuming that the amplitude is color-ordered (as is the case for
all leading-color amplitudes), the momentum invariants
encountered as arguments of the integrals all involve sums of
color-adjacent momenta,
$$
  t_i^{[r]}\ \equiv\ (k_i+k_{i+1}+\cdots + k_{i+r-1})^2,
\eqn\tirdef
$$
where momentum labels are taken mod $n$.
In particular, the external legs of the box, triangle, and bubble integrals
may be off-shell, or massive ($K_i^2\neq0$).  Thus the set $\IntSet_n$ of
functions that can appear in color-ordered cut-constructible $n$-point
amplitudes $(n>4)$ is
$$
  \IntSet_n\ \equiv\ \{
  I^{\rm 4m}_{4:r,r',r'';i}, \, I^{\rm 3m}_{4:r,r';i}, \,
  I^{{\rm 2m}\,h}_{4:r;i}, \, I^{{\rm 2m}\,e}_{4:r;i}, \,
  I^{\rm 1m}_{4:i}, \,
  I^{\rm 3m}_{3:r,r';i}, \, I^{\rm 2m}_{3:r;i}, \,
  I^{\rm 1m}_{3:i}, \,
  I_{2:r;i} \}.
\eqn\Fndefn
$$
The elements of $\IntSet_n$ are given in eqs.~(\use\Fboxes),
(\use\OneMassTriangle), (\use\TwoMassTriangle), (\use\deqfourtrians)  and
(\use\bubbles) and are depicted in \fig\BoxesFigure,
\fig\TrianglesFigure, and \fig\BubblesFigure.  The indices labeling the
ordered external momenta $k_i$ increase in the clockwise direction
in the figure.
Note that these functions
are not linearly independent in general.

\vskip -.7 truecm

\LoadFigure\BoxesFigure{\baselineskip 13 pt
\noindent\narrower\ninerm The box integrals that may appear.}
{\epsfysize 4.0truein}
{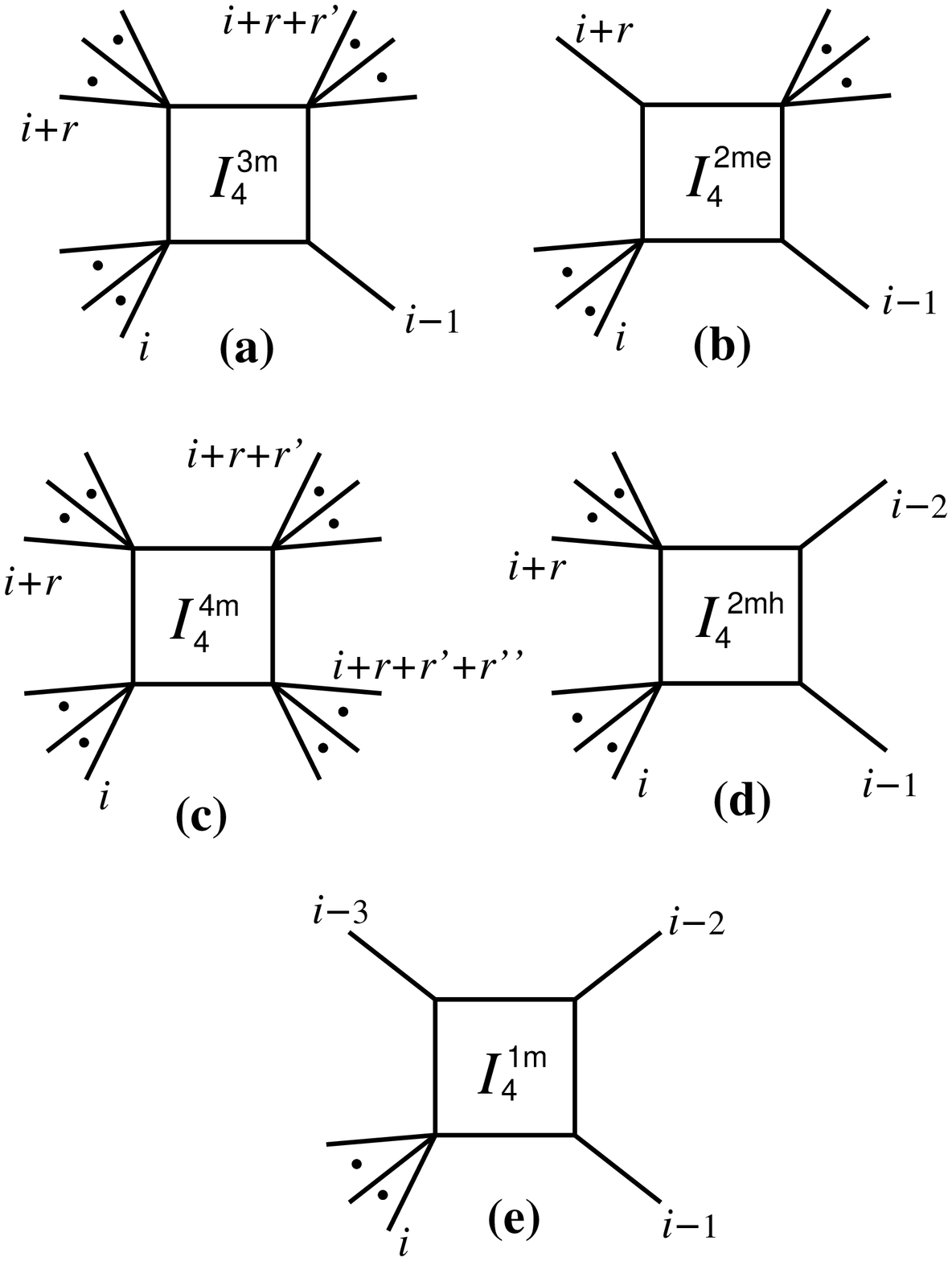}

\LoadFigure\TrianglesFigure{\baselineskip 13 pt
\narrower\ninerm The triangle integrals that may appear.}
{\epsfysize 2.0truein}{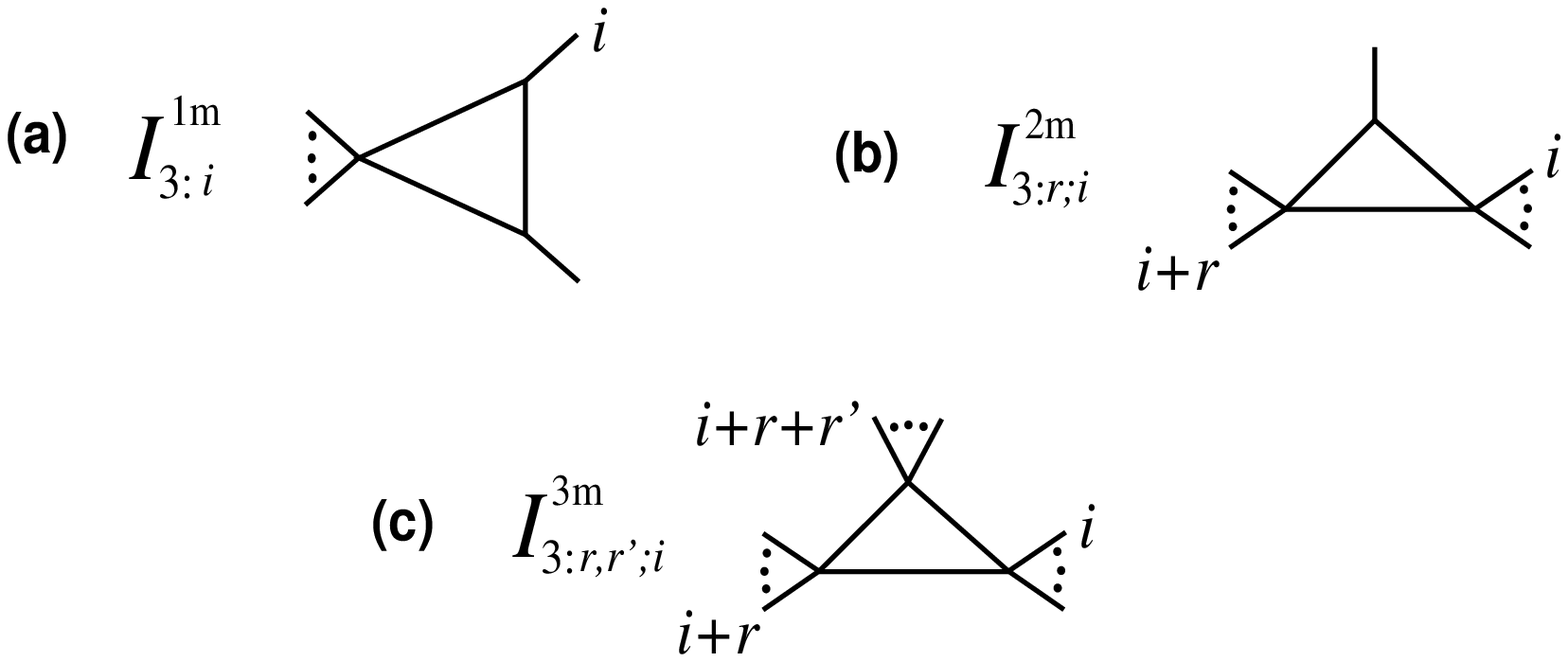}

\vskip -1. truecm
\LoadFigure\BubblesFigure{\baselineskip 13 pt
\narrower\ninerm The bubble integrals that may appear.}
{\epsfysize .75truein}{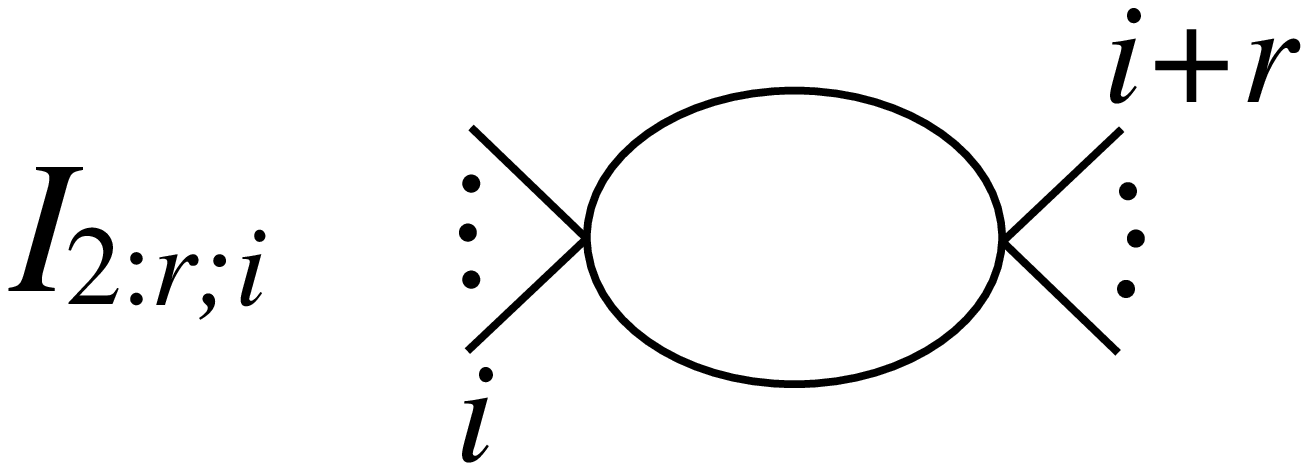}

Any cut-constructible amplitude $A_n$
is a linear combination of functions in $\IntSet_n$,
as in equation~(\use\AnDecomp), where the coefficients $c_i$
are rational functions of the kinematic invariants, arising
in part from the diagrammatic rules and in part from
the integral reduction procedure.   (`Kinematic invariants' here
includes the spinor products~[\use\SpinorHelicity] used to give compact
forms for helicity amplitudes.)
The functions in $\IntSet_n$ contain logarithms and dilogarithms
whose arguments depend only on the kinematic invariants $t_i^{[r]}$.

We wish to show that any such amplitude $A_n$ is uniquely
determined by its cuts.  To be more precise, given
two linear combinations of integral functions
possessing the same cuts,
$$
\sum_{i|I_i\in \IntSet_n} c_i  I_i \biggr|_{\rm cuts} =
 \sum_{i|I_i\in \IntSet_n} c_i' I_i \biggr|_{\rm cuts}\ ,
\anoneqn
$$
their difference must be a rational function of the momentum invariants,
$$
\sum_{i|I_i\in \IntSet_n} \L c_i-c'_i\R\, I_i = \; {\rm rational},
\eqn\EQindependent
$$
and we wish to show that the rational function appearing on the
right-hand side of~(\EQindependent) must vanish.

A cursory examination of the set
of functions shows why one might expect this to be true.
The cuts uniquely determine the coefficients of the logarithms and
dilogarithms, and the set of functions is dominated in content by these
functions.  At $\Ord(\eps^0)$, rational terms only appear in a small number of
places --- in the bubbles and in the $\pi^2$ terms in the boxes ---
and they are always associated with logarithms and dilogarithms.

We shall prove the result by inspecting the different channels that
can possess cuts, working always at $\Ord(\eps^0)$,
and demonstrating that the different functional
dependence in the different channels either uniquely
singles out one of the functions, or else singles out a set of functions
lacking cut-free parts.
We first illustrate the method using the four-point case.
In this case,
$$
\IntSet_4 = \{I_4^{0\rm m},
  I^{\rm 1m}_{3:3},\,
  I^{\rm 1m}_{3:2},\,
  I_{2:2;1},\,
  I_{2:2;4}\} \ ,
\anoneqn$$
is the set of scalar integrals depicted in \fig\FourBasisFigure.
(Other scalar integral functions are equivalent to these:
$I^{\rm 1m}_{3:1} = I^{\rm 1m}_{3:3}$, etc.)
As discussed in appendix~\use\IntegralsAppendix,
the tensor integrals that appear in an explicit
evaluation of a cut-constructible four-point amplitude,
such as triangle integrals with one power of
loop momentum in the numerator of their integrands,
can be written as a linear combination of this set of integrals.
Because we are only considering color-ordered partial
amplitudes, the legs of all diagrams follow the same ordering,
say, $1,2,3,4$.
For this simple case, we make the kinematic variable
dependence explicit with the notation
$$
\eqalign{
&I_4^{0\rm m} (s,t) \equiv  I_4^{0\rm m} \ ,  \hskip 1 cm
I_3^{\rm 1m}(s) \equiv I^{\rm 1m}_{3:3} \ , \hskip 1 cm
I_3^{\rm 1m}(t) \equiv I^{\rm 1m}_{3:2} \ , \cr
& \hskip 1 cm I_2(s) \equiv I_{2:2;1} \ , \hskip 2.5 cm
I_2(t) \equiv  I_{2:2;4}\ ,  \cr }
\anoneqn$$
where $s = (k_1+k_2)^2$ and $t = (k_2 + k_3)^2$ are
the usual Mandelstam variables.

\vskip -1. truecm
\LoadFigure\FourBasisFigure{\baselineskip 13 pt
\narrower\ninerm The integral functions that may appear in a
four-point calculation.}{\epsfysize 3.0truein}
{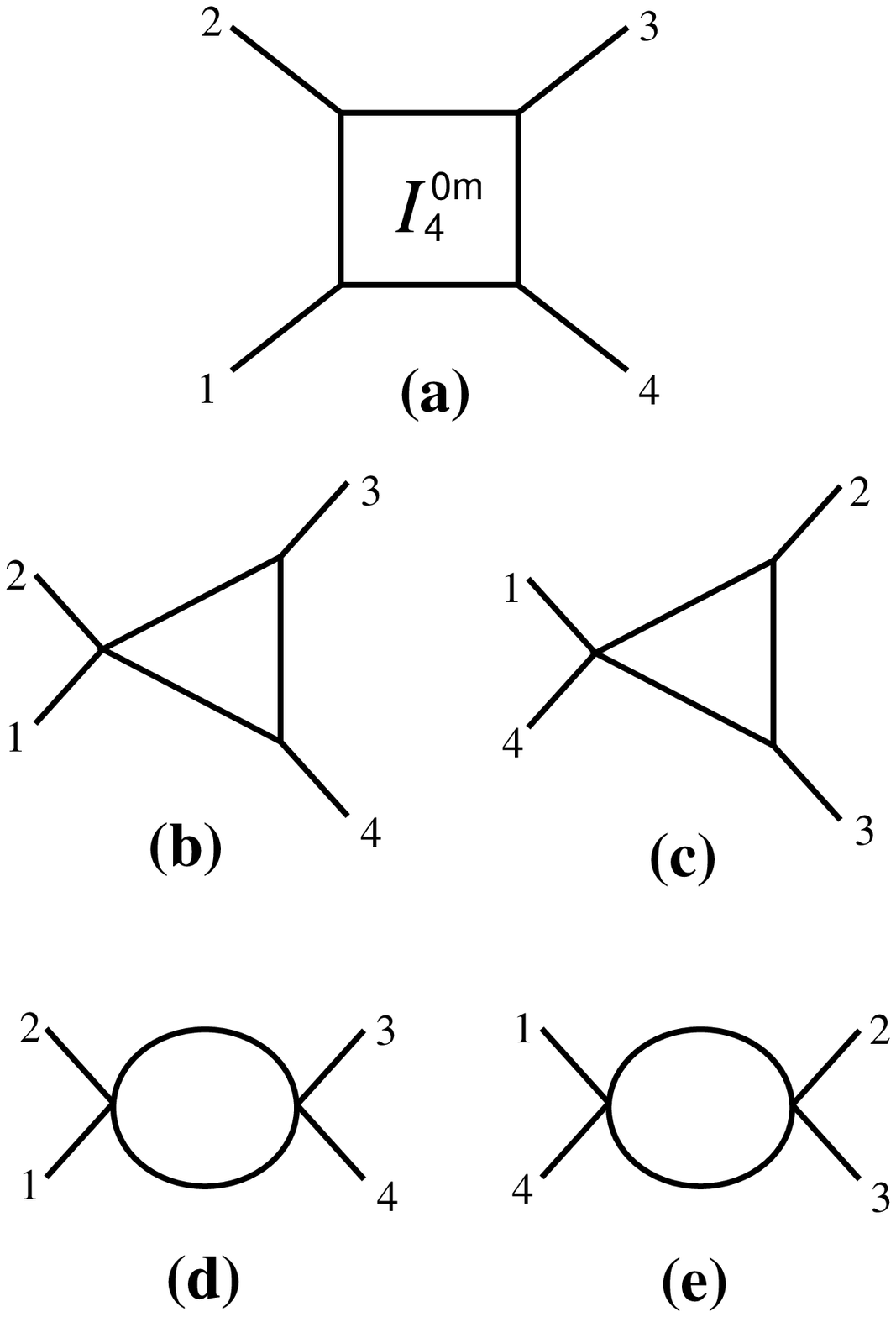}

The five integral functions in $\IntSet_4$ are given explicitly
in equations~(\use\ZeroMassBox),~(\use\OneMassTriangle)
and~(\use\bubbles) of appendix~\use\IntegralsAppendix.
Each one contains at $\Ord(\eps^0)$ a logarithm or a product of
logarithms unique to that function, as shown in Table~1.
Therefore the only way to satisfy eq.~(\use\EQindependent)
is to choose the coefficient of every integral to vanish, $c_i-c'_i=0$.
In other words, one cannot
construct a non-vanishing cut-free massless four-point integral,
and hence amplitude, that satisfies the power-counting criterion.
This result applies, for example, to one-loop four-gluon amplitudes
in a supersymmetric gauge theory.

\vskip .5 cm
\hskip 3.8 truecm
\hbox{
\def\tend{\cr \noalign{\hrule}}

\vbox{\offinterlineskip
{
\hrule
\halign{
        &\vrule#
        &\quad\vrule#
        &\strut\quad#\hfil\quad\vrule
        & \quad\hfil\strut # \hfil
        \cr
height13pt & &{\bf Integral}  &{\bf Unique Function}      &\tend
height12pt $\;\;$a &&  $I^{\rm 0m}_4(s,t)$
& $\ln(-s)\ln(-t)$  &\tend
height12pt $\;\;$b && $I^{\rm 1m}_3(s)$
& $\ln(-s)^2$ &\tend
height12pt $\;\;$c && $I^{\rm 1m}_3(t)$
& $\ln(-t)^2$ &\tend
height12pt $\;\;$d  &&  $I_2(s)$
& $\ln(-s)$  &\tend
height12pt $\;\;$e  &&  $I_2(t)$
& $\ln(-t)$  &\tend
}
}
}
}
\vskip .2 cm
\nobreak
{\baselineskip 13 pt
\narrower\smallskip\noindent\ninerm
{\ninebf Table 1:}  The set of integral functions that may appear
in a cut-constructible massless four-point amplitude,
together with the independent logarithms.
\smallskip}

The proof for the $n$-point case is similar, although more involved.
We will again show, by inspecting the integral functions in $\IntSet_n$,
that the only solutions to (\use\EQindependent) have zero right-hand side.
(Note that, as in the four-point case, some of the integral functions
have multiple names, so we need not count these twice;
$I^{{\rm 2m}\, e}_{4:r;i} = I^{{\rm 2m}\, e}_{4:n-r-2;i+r+1}$,
while $I_{2:r;i}$, $I^{\rm 3m}_{3:r,r';i}$ and $I^{\rm 4m}_{4:r,r',r'';i}$
respectively have two-, three- and four-fold naming degeneracies.)
We successively inspect the integrals in kinematic
regimes where one of the kinematic invariants, say $s$,
is taken to be large, while the others remain fixed.  In this
regime we isolate terms of the form $\ln(-s) \ln(-s')$.
The region of large kinematic invariants is useful because it simplifies
the functional form of the integrals, particularly the three-mass
triangle and the four-mass box. For example,
$$
\Li_2 \L 1 - {s s' \over t t'} \R \rightarrow - \ln(-s) \ln(-s') + \cdots
\anoneqn
$$
where $\Li_2$ is the dilogarithm [\use\Lewin].
In these kinematic regimes the terms involving
square-root arguments also reduce to simple products
of logarithms.
In taking certain kinematic invariants to be large,
we do not take the limit of infinite
$s$, but keep the full functional form in the coefficients $c_i-c'_i$.

With successive judicious choices of $s$ one can isolate a single
coefficient at a time from the sum, for certain of the functions in
$\IntSet_n$; that is only a single integral contains the chosen
$\ln(-s) \ln(-s')$, after noting the vanishing of prior coefficients.
The coefficient of the isolated function must therefore be zero as
well, in order to obtain a cut-free result~(\use\EQindependent).  We
display in table~2 the order in which we examine the integrals for
large $\tn{r}{i}$ and the product of logarithms thereby isolated.

Consider the first entry, the three-mass box $I^{\rm 3m}_{4:r,r';i}$.
For a color-ordered amplitude with the ordering of legs shown in
fig.~\use\BoxesFigure{a} we examine the regime where the kinematic
variable $\tn{r}{i}$ is large.  The combination
$\ln(-\tn{r}{i})\ln(-\tn{n-r-r'-1}{i+r+r'})$ appears only in the
single integral function $I^{\rm 3m}_{4:r,r';i}$; in fact, these two
kinematic variables appear together only in this function.  (For
non-color ordered amplitudes more functions can contain this
combination, complicating the situation.)  One then deduces that the
coefficients of the three-mass boxes in eq.~(\use\EQindependent) are
zero.  The ordering of the table is such that at each stage only a
single function contains the pairs $s$ and $s'$ together. In checking
the behavior for large kinematics one merely checks that the
coefficient of $\ln(-s) \ln(-s')$ is nonzero, thereby requiring the
corresponding difference of coefficients $c_i-c'_i$ to be set to zero.
We can then proceed through the table, at each turn forcing additional
differences of coefficients to vanish.

Upon reaching the end of the table, we are left with three kinds of
integrals: one- and two-mass triangle integrals, and bubble integrals.
The former two contain no rational pieces at all at ${\cal O}(\e^0)$, but
only logarithms squared; whatever linear combination makes the logarithms
vanish eliminates these integrals entirely.  This leaves us only with
bubble integrals; while these do contain rational pieces at ${\cal O}(\e^0)$,
requiring the
coefficient of each of the single powers of logarithms of independent
invariants to vanish eliminates the bubble integrals, leaving us with the
desired result, that the rational function on the right-hand side of
equation~(\use\EQindependent) indeed vanishes.
\vskip .5 cm
\hskip 1.3 truecm
\hbox{
\def\tend{\cr \noalign{\hrule}}

\vbox{\offinterlineskip
{
\hrule
\halign{
        &\vrule#
        &\quad\vrule#
        &\strut\quad#\hfil\quad\vrule
        & \quad\hfil\strut # \hfil
        \cr
height13pt & &{\bf Integral}
&{\bf Unique Function}      &\tend
height12pt $\;\;$a &&  $I^{\rm 3m}_{4:r,r';i}$
& $\ln(-\tn{r}{i})\ln(-\tn{n-r-r'-1}{i+r+r'})$  &\tend
height12pt $\;\;$b && $I^{{\rm 2m}\, e}_{4:r;i}$
& $\ln(-\tn{r}{i})\ln(-\tn{n-r-2}{i+r+1})$ &\tend
height12pt $\;\;$c && $I^{\rm 4m}_{4:r,r',r'';i}$
& $\ln(-t_{i}^{[r]})\ln(-t_{i+r+r'}^{[r'']})$ &\tend
height12pt $\;\;$d && $I^{{\rm 2m}\, h}_{4:r;i}$
& $\ln(-t_{i}^{[r]})\ln(-t_{i+r}^{[n-r-1]})$ &\tend
height12pt $\;\;$e && $I^{\rm 1m}_{4;i}$
& $\ln(-t_{i}^{[r]})\ln(-t_{i}^{[r+1]})$ &\tend
height12pt $\;\;$f && $I^{\rm 3m}_{3:r,r';i}$
& $\ln(-t_{i}^{[r]})\ln(-t_{i+r}^{[r']})$ &\tend
}
}
}
}
\nobreak
{\baselineskip 13 pt
\narrower\smallskip\noindent\ninerm
{\ninebf Table 2:}
Following the ordering shown and taking large $t_{i}^{[r]}$
makes the proof of uniqueness of the cuts straightforward.
\smallskip}

\vskip .2 cm

For integrals and thus amplitudes that fail to satisfy the power-counting
criterion given at the beginning of the section,
for example $m$-point loop diagrams containing terms with $m$ powers
of the loop momentum, the result should not be expected to hold,
and in fact can be shown explicitly to be violated.
Here, after a Passarino-Veltman reduction,
one has additional functions beyond those mentioned previously,
for example bubble integrals with up to {\it two\/} powers of
the loop momentum
in the numerator of the integrand.  Denote the
loop momentum by $p^\mu$;
the bubble integrals with loop-momentum
polynomial $1$ and $p^{\mu} p^{\nu}$, defined in equation~(\use\GeneralFdefnB),
are respectively
$$
\eqalign{
I_2[1](K)&= {\rg \over \eps (1-2\eps)} (-K^2)^{-\eps} \, , \cr
I_2[p^{\mu} p^{\nu}](K)&=
{\rg  (-K^2)^{-\eps} \over \eps (1-2\eps)}
\biggl[
{K^{\mu}K^{\nu} \over 3} \Bigl( 1+{\eps\over 6} \Bigr)
-{\eta^{\mu\nu} K^2 \over 12 } \Bigl( 1+{2\eps\over 3} \Bigr) \biggr]\,,
\cr}
\anoneqn
$$
where $K^\mu$ is the momentum flowing out one side of the bubble.
The linear combination
$$
\Bigl( {K^{\mu}K^{\nu} \over 3} -{\eta^{\mu\nu} K^2 \over 12 } \Bigr)
I_2[1](K) -I_2[p^{\mu} p^{\nu}](K)
=-{ \rg \over 18} ( {K^{\mu}K^{\nu} }-\eta^{\mu\nu} K^2 )
\anoneqn
$$
provides a solution to eq.~(\EQindependent).
The presence of such a combination within the amplitude thus cannot be
determined from the cuts.
Such combinations of integrals do occur in QCD calculations.
Simple examples are the four-gluon amplitudes [\use\Long]
where all external helicities are the same or where one leg is of
opposite helicity.  In these two cases the cuts all vanish but the
amplitudes do not; furthermore, they are not equal,
showing that the cuts in this case do
not uniquely determine the amplitude.  (The integrals in these amplitudes
do violate the required power-counting criterion.)


\section{Supersymmetric Theories}
\tagsection\SusySection

In this section, we show that the power-counting criterion of
section~\UnitarityProofSection\ is satisfied by all color-ordered
one-loop amplitudes in massless supersymmetric gauge theories
with no superpotential.
Thus such amplitudes can be computed solely
from the knowledge of their cuts.
Color-ordering is irrelevant to the power-counting,
but it is required for the uniqueness argument of
section~\UnitarityProofSection\ to be valid.
The leading-color contributions to a given amplitude in $SU(N_c)$
(super-) gauge theory are always color-ordered, and in many cases
the subleading-color contributions can be expressed as sums over
permutations of color-ordered objects~[\use\SusyFour,\use\QCDConf,\use\qqggg].
While we expect that the result extends to the full amplitude, as well
as to amplitudes in all massless supersymmetric theories (i.e. including
nonvanishing terms in the superpotential), we shall not present an
argument for such an extension here.

We shall use ordinary (as opposed to superspace) diagrams,
and background-field gauge. We always assume the use of a
supersymmetric regulator~[\use\Siegel,\use\Long].
Use of ordinary diagrams reveals that
the power-counting criterion does not always require
supersymmetric cancellations, and one can see how to apply it to
certain nonsupersymmetric calculations as well.
(At the end of the section we sketch how a superspace argument
would proceed.)
The presence of trees attached to the loop does not change the
power-counting of the loop integrand;
hence we amputate external trees in favor of taking
legs off-shell, that is we restrict our attention to the computation
of the one-loop effective action.

If all external legs are gluons, we can write a compact determinantal
formula for the color-ordered (and color-stripped) effective action
(one should think of this object as the generating functional for
the amputated color-ordered diagrams),
$$\eqalign{
\Gamma^{\rm SUSY} &=
\ln \det{}^{-1/2}_{[1]}  [D^2 \eta_{\alpha\beta}
- g (\Sigma_{\mu\nu})_{\alpha\beta} F^{\mu\nu}]
+ \ln \det{}_{[0]} [D^2]
+{1\over 2}\ln\det{}^{1/2}_{[1/2]}
[D^2 -  {g\over 2} \sigma_{\mu\nu} F^{\mu\nu}] \cr
&+ n_m \L {1\over 2}\ln\det{}^{1/2}_{[1/2]}
    [D^2 -  {g\over 2} \sigma_{\mu\nu} F^{\mu\nu}]
+\ln \det{}^{-1}_{[0]} [D^2]\R\, ,  \cr
}\eqn\SUSYEffectiveAction$$
where $\det{}_{[J]}$ is the one-loop determinant for a particle of
spin $J$ in the loop.
This formula assumes two-component fields, the fermions being taken to
be Majorana to allow a definition of the determinant;
$D$ is the covariant derivative, $n_m$ is the number of $N=1$ chiral
matter multiplets,  ${1\over2}\sigma_{\mu\nu}$
($\Sigma_{\mu\nu}$) are the spin-${1\over2}$ (spin-1) Lorentz
generators, and we have used the fact that the contribution of a Weyl
fermion in a non-chiral theory is half that of a Dirac fermion.
The fermion determinant is written in a second-order
form [\use\Mapping], making the similarity of the fermion and
gluon loop contributions clear.
Note that the gluon determinant contains Lorentz indices and the fermion
determinant spinor indices, and the effective action implicitly includes
traces over these indices.

We now extract the $m$-point contribution
by differentiating $m$ times with respect to the background field,
thereby expanding the logarithms in equation~(\use\SUSYEffectiveAction).
It is convenient to think of performing this expansion in two stages:
first, expanding with respect to the field strength $F^{\mu\nu}$,
and only then expanding with respect to the background field.
Expanding the determinants,
$$
\eqalign{
\ln \det{}^{-1/2}_{[1]}  [D^2 \eta_{\alpha\beta}
- g (\Sigma_{\mu\nu})_{\alpha\beta} F^{\mu\nu}]
&= -2 \ln \det{}_{[0]} [D^2] + \cdots \, ,
\cr
{1\over 2}\ln\det{}^{1/2}_{[1/2]}
[D^2 -  {g\over 2} \sigma_{\mu\nu} F^{\mu\nu}]
&= +\ln \det{}_{[0]} [D^2] + \cdots \, ,
\cr}
\anoneqn
$$
in the first stage of the expansion it is clear
that terms with no $F$s cancel within each supermultiplet.  Furthermore,
terms with a lone $F$ also vanish, because the trace (over Lorentz indices)
of the accompanying Lorentz generator ($\sigma$ or $\Sigma$) vanishes.
Thus each term must contain at least two $F$s.
Whereas the terms generated by expanding $D^2$ ($A\cdot\partial$, etc.)
may have up to one power of the loop momentum per background field,
$F$ does not contain any powers of the loop momentum, but rather
only powers of external momenta.  Thus each insertion of $F$ reduces
by one the number of powers of loop momenta in the numerator of the loop
integrand, from a maximum of $m$ to a maximum of $m-2$.
We conclude that the effective action for external gluons,
and hence the amplitudes whose external legs are all gluons,
satisfies the desired power-counting criterion in a supersymmetric
theory.  (For amplitudes containing only external gluons, formula~(7.2)
of ref.~[\SusyFour] ensures that the entire amplitude --- subleading as
well as leading-color terms --- may be determined in this fashion.)

Next consider one-particle irreducible graphs with
multiple pairs of external fermion lines.
For each pair of external fermions there is one fermion line running
along the loop.  Using the standard first-order formalism for the
fermions, the interaction vertex for the fermion line with a gauge boson
or a scalar does not contain the loop momentum $\ell^\mu$,
but the rationalized Dirac propagator ($\sim \Slash{\ell}/\ell^2$)
contributes one power of the loop momentum to the numerator.
Thus each fermion line running along the loop leads to a reduction
of the maximum degree of the loop-momentum polynomial by one.
Graphs with two or more pairs of external fermions automatically
satisfy the power-counting criterion, without requiring any
supersymmetric cancellation.

Graphs with only one pair of external fermion lines
require us to find a reduction of the degree of the loop-momentum polynomial
by one more unit.
If there is also (at least) one external scalar line, this can be
achieved within a single diagram.
For example, if the scalar couples directly to the fermion line running
around the loop, then there is a factor of the form
$$
  \Slash{\ell} (a+b\gamma_5) (\Slash{\ell}-\Slash{k_s})
  \ =\ (a-b\gamma_5) \ell^2\ +\ \Ord(\ell)\, ,
\eqn\yukterm
$$
from the Yukawa coupling and adjacent propagators.
We can use the factor of $\ell^2$ to cancel one of the propagators in
the denominator of the $m$-point integral,
thus converting it to an $(m-1)$-point integral
with a loop polynomial of maximum degree $m-3 = (m-1)-2$.
(One power is lost due to the external fermion pair, and one
due to the $\ell^2$ cancellation.)
The power-counting criterion is then satisfied.
If the scalar does not couple directly to the fermion line,
then in the leading loop-momentum terms it is still possible to find a
factor of $\ell^2$ to cancel against the propagator,
again establishing the criterion.

If there are no external scalars, but just two external fermions and
$m-2$ gluons, then supersymmetric cancellations are required.
For example in the pure super-Yang-Mills case with only gluinos and
gluons, the graphs that cancel are shown in \fig\GenSusyFigure.
Since the maximum degree of the loop momentum polynomial is $m-1$ for
each graph, we only need to show that they cancel to leading order
in $\ell$.  It is straightforward to show this by repeated use of the
identity
$$
  \Slash{\ell}\gamma_\mu\Slash{\ell}
  \ =\ 2\ell_\mu \Slash{\ell} - \gamma_\mu \ell^2.
\eqn\gammaid
$$
The second term in~(\use\gammaid) can be ignored by the
cancelled-propagator argument used above,
while the first term has converted a
gluon emission off a fermion line into the form of a gluon emission
off a scalar line, or off a gluon line.  (In background-field gauge,
the latter two vertices have the same form to leading order in $\ell$.)
In this way the two graphs where the fermion line runs around
the loop two different ways can be made to look identical at
leading order, apart from an overall minus sign, so that the sum cancels.
This argument breaks down if there are no external gluons, i.e. just a
fermion propagator bubble, because there is no second diagram to
cancel against.  However such a linear bubble is permitted
by the uniqueness result.

\vskip -.7 truecm
\LoadFigure\GenSusyFigure{\baselineskip 13 pt
\narrower\ninerm Two graphs with external fermions
between which cancellations occur
in satisfying the power-counting criterion.}{\epsfysize
1.5truein}
{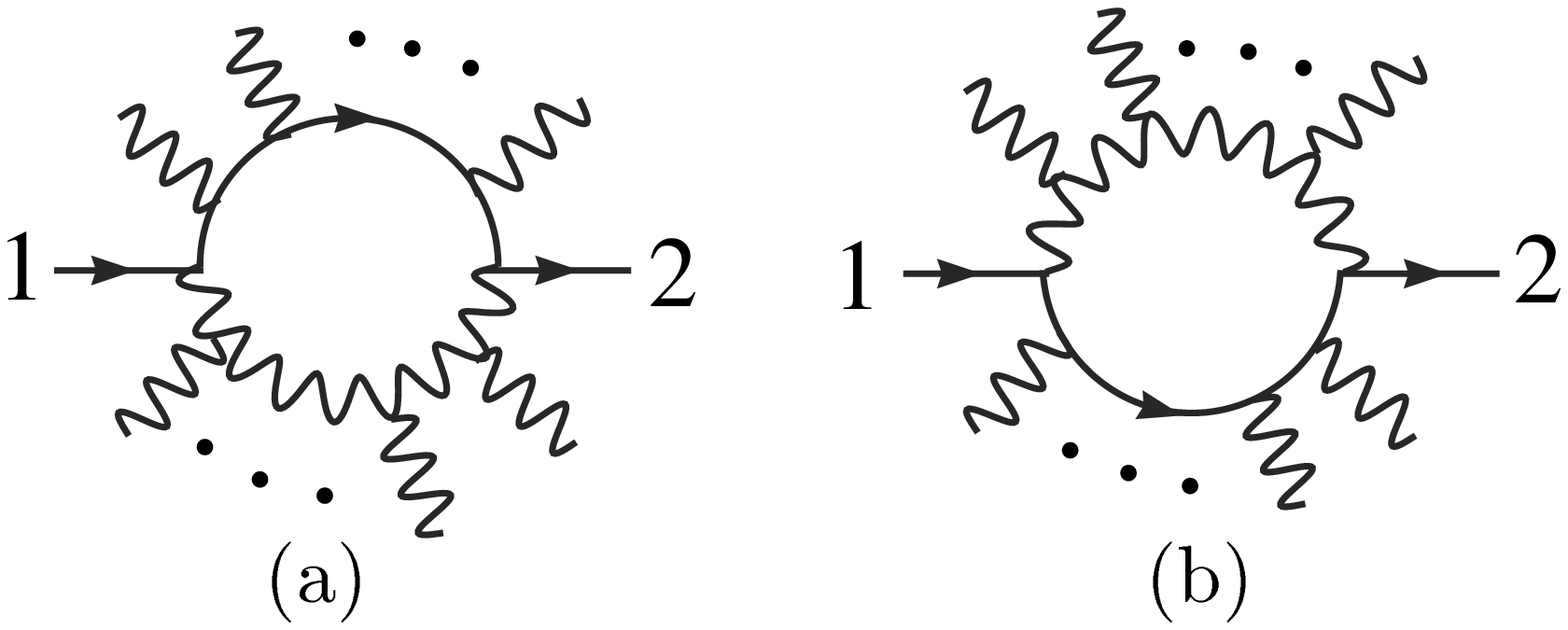}

Finally there are the one-particle irreducible graphs with
no external fermions, but including (two or more) external scalars.
For a fermion line running all the way around the loop,
the power-counting criterion is satisfied (with one exception)
by virtue of equation~(\use\yukterm) applied twice,
which reduces a degree-$m$ $m$-point integral to a degree
$m-4$ $(m-2)$-point integral.
The exception is for exactly two {\it adjacent} external scalars;
in this case a supersymmetric cancellation occurs
against corresponding graphs with gluons and scalars in the loop.
The graphs with gluons and scalars in the loop can also be shown to
satisfy the criterion without invoking supersymmetry (with the same exception);
however, here a few graphs with four-point vertices must be added to the
``parent diagram'' with only three-point vertices.

Supersymmetric cancellations of the type described above should be
manifest in superspace.  For $n$-gluon amplitudes in an $N=4$
supersymmetric theory, where only a vector superfield $V$ circulates
in the loop, a superspace argument is straightforward
to construct~[\use\Superspace,\use\SusyFour].
With chiral superfields, the chiral constraints lead to large numbers
of supercovariant derivatives $D_\alpha$ and $\overline{D}_{\dot\alpha}$
in the supergraphs, which should be removed by $D$-algebra, because
they correspond to high-degree loop-momentum polynomials.
In brief, the commutator $[D^2,\overline{D}^2] \propto \partial^2$
can be used to cancel a propagator just as in the ordinary diagram
analysis performed above.
Combining this fact with $D$-integration-by-parts, in order
to remove $D$'s on the loop in favor of $D$'s on external legs,
one can drastically reduce the degree of the loop-momentum
polynomials, and it should be possible to verify the power-counting
criterion in this way too.

In conclusion we have shown that in a supersymmetric theory
the ``cut-constructible'' conditions may be met. We have
avoided the use of elegant superspace arguments in order to illustrate
how components of non-supersymmetric amplitudes can satisfy the
conditions also.


\section{All-Multiplicity $N=1$ Supersymmetric MHV Amplitudes}
\tagsection\MHVSection

In this section we apply the unitarity result of the previous
section to obtain
explicit formulas for ``maximally helicity-violating'' (MHV)
gluon amplitudes
in an $N=1$ supersymmetric gauge theory.  Recall that, in the convention
where all momenta are taken as outgoing, supersymmetric all-gluon amplitudes
where all the gluons, or all but one, have identical helicities,
vanish on account of a supersymmetry Ward identity~[\use\Susy],
$$
A_n^{\rm SUSY}(1^\pm,2^+,\ldots,n^+) = 0 \; .
\eqn\susyward
$$
  The MHV amplitudes --- with two opposite-helicity gluons --- are thus
the simplest non-vanishing amplitudes we can consider.
We shall denote these MHV $n$-gluon amplitudes by
$$
\eqalign{
  A_{ij}(1,2,\ldots,n)\ &\equiv\
  A_{n;1} (1^+,\ldots,i^-,\ldots,j^-,\ldots,n^+) \; .
     \cr}
\eqn\mhvdef
$$
The corresponding amplitudes with predominantly negative helicities
may of course be obtained by complex conjugation.

In previous work~[\use\SusyFour], we computed the amplitudes
with this helicity configuration in an $N=4$ supersymmetric theory.
In this section, we present the result for the $N=1$ supersymmetric
chiral matter multiplet contribution to the $n$-gluon MHV amplitudes.
The contribution to an $n$-gluon amplitude from any supersymmetric multiplet
is a linear combination of the $N=4$ amplitude
and $N=1$ chiral multiplet contributions.
For example, using eq.~(\TotalampB),
the amplitude in a pure $N=1$ supersymmetric nonabelian
gauge theory (a theory with just the vector supermultiplet,
containing a gluon and a gluino) is given by the linear combination
$$
A_{n;1}^{N=1\ \rm vector} (1,2,\ldots,n)
= A_{n;1}^{N=4} (1,2,\ldots,n)
-3A_{n;1}^{N=1\ \rm chiral} (1,2,\ldots,n).
\anoneqn
$$
As discussed in section~\use\ReviewSection,
the contribution of an $N=1$ supersymmetric chiral multiplet contribution
is also one of the three components of a QCD $n$-gluon amplitude.
To obtain the gluon scattering amplitude in QCD we need
one further contribution, essentially the contribution of an
internal scalar loop;
this remaining component contains rational functions of the invariants
that cannot be fixed by the cut techniques discussed in this paper.

Consider the cut in the channel
$\tn{m_2-m_1+1}{m_1} \equiv
 (k_{m_1}+k_{m_1+1}+\cdots+k_{m_2-1}+k_{m_2})^2$ for the
loop amplitude $A_{n;1}^{N=1\ \rm chiral}(1,2,\ldots,n)$,
depicted in \fig\CutsFigure\ and given by [\use\Cutting]
$$
\eqalign{
  &i\int \dlips(-\ell_1,\ell_2)
  \ A^{\rm tree}(-\ell_1,m_1,\ldots,m_2,\ell_2)
  \ A^{\rm tree}(-\ell_2,m_2+1,\ldots,m_1-1,\ell_1)\, , \cr
}
\eqn\CutEquation
$$
where the integration is over $D$-dimensional Lorentz-invariant phase-space.
The intermediate states with momenta $\ell_1$ and $\ell_2$
are fermions or scalars, the remaining external states are gluons.
The overall sign is unimportant, since it is a simple matter
to fix the overall
sign in final expressions from the known universal ultraviolet
and infrared singularities [\use\KunsztSingular].
Instead of evaluating the phase-space integrals we evaluate
the off-shell integral
$$
\eqalign{
  &\int {d^{D} \ell_1 \over (2\pi)^D}
  \ A^{\rm tree}(-\ell_1,m_1,\ldots,m_2,\ell_2) {1\over \ell_2^2}
  \ A^{\rm tree}(-\ell_2,m_2+1,\ldots,m_1-1,\ell_1) {1\over \ell_1^2}
   \biggr|_{\rm cut} . \cr
}\eqn\caseacut
$$
whose cut in this channel is (\use\CutEquation).
This replacement is valid only in this channel. In evaluating this
off-shell integral, we may substitute
$\ell_1^2 = \ell_2^2 = 0$ in the numerator; terms with $\ell_1^2$
or $\ell_2^2$ in the numerator do not produce a cut in this
channel because the $\ell_1^2$ or $\ell_2^2$ cancels a cut propagator.
We emphasize that the cuts are evaluated
not for a lone diagram at a time, but for the whole amplitude.

\LoadFigure\CutsFigure{\baselineskip 13 pt
\narrower\ninerm The cuts needed to obtain the MHV amplitudes.}
{\epsfysize 4.2truein}
{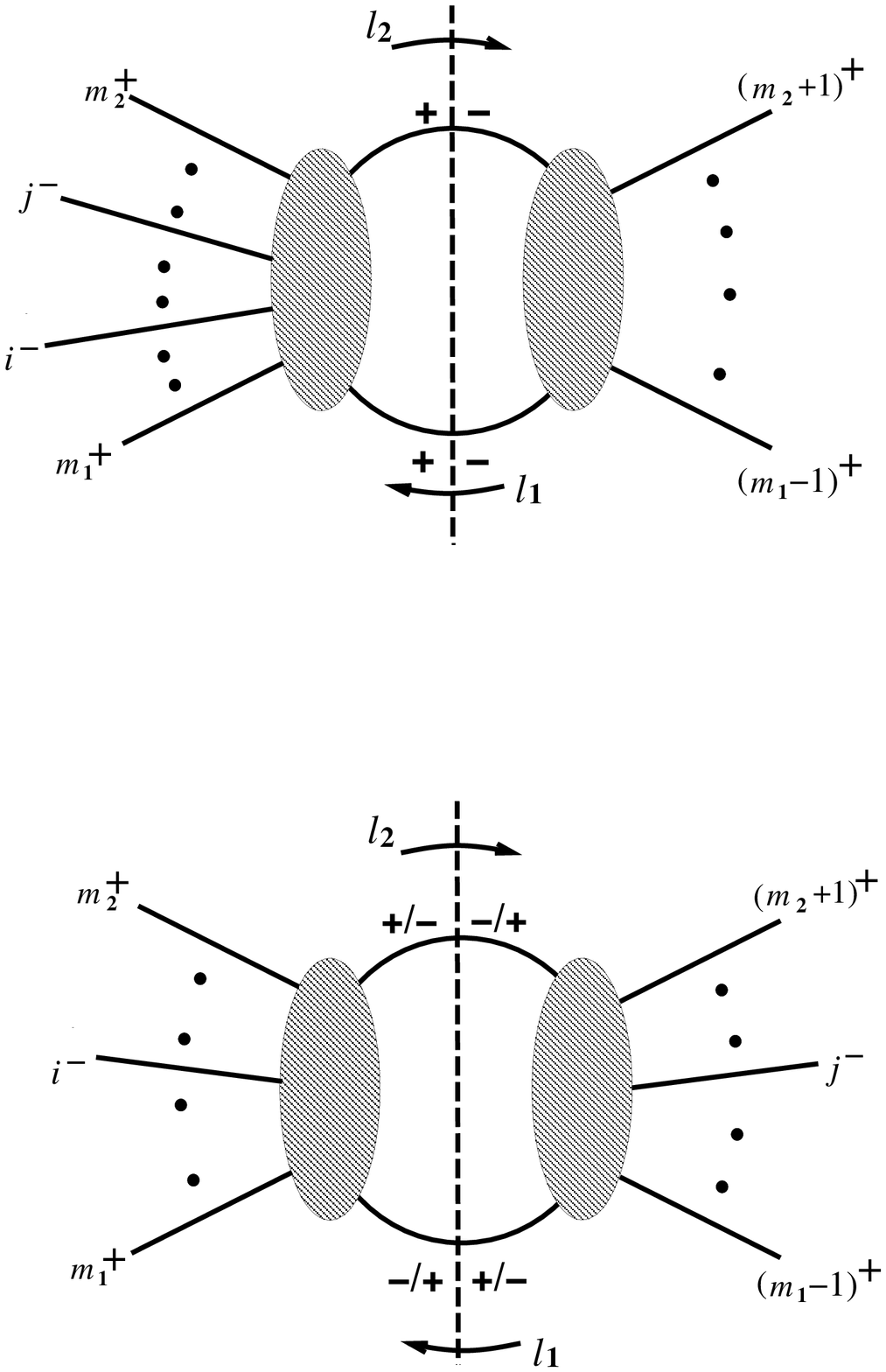}

In quoting final results for the amplitudes we do so in the
dimensional reduction [\use\Siegel] or `four-dimensional
helicity' [\use\Long] schemes (which for practical purposes are
equivalent at one-loop).  In these schemes, after using
Passarino-Veltman reduction [\PV], the coefficients of all integral
functions do not depend on $\eps$ since all tensor and spinor manipulations
are performed in four dimensions.  In reconstructing the full
amplitudes from cuts we therefore take the coefficients to be free of
$\eps$, and automatically obtain results in this scheme.  This scheme
choice has the additional advantage of preserving supersymmetry, which
is necessary for the power-counting criterion to be satisfied.  At the
end of the calculation, one can convert to more conventional schemes
by accounting for the $[\eps]$-scalars as discussed near
eq.~(\use\TotalampC).

There is one subtlety we must address.
We perform the cut integration in $4-2\e$ dimensions, yet the
tree amplitudes on either side are evaluated in four dimensions.  Since
the calculation contains divergences, one might worry that this
$\Ord(\eps)$ discrepancy
in the integrand might lead to errors in the final result.  In fact,
this is not the case, and in performing the cut
calculations through $\Ord(\eps^0)$
there is no need to track which momenta in the numerator are
four-dimensional and which are $(4-2\eps)$-dimensional.
(Momenta in the denominator must of course be $(4-2\eps)$-dimensional
to regulate singularities.)
Ultraviolet
and non-overlapping infrared singularities lead only to single
poles in $\eps$; in this case an
$\Ord(\eps)$ error in the numerator of eq.~(\use\caseacut) could only
change the final result by a `polynomial' at $\Ord(\eps^0)$.
Such singularities thus do not affect
the cuts.
The case of overlapping soft and collinear singularities is a bit more subtle,
since here there is a double pole in $\eps$, and correspondingly there are
$\ln(s)/\eps$-type terms.  One might worry that these can `feed down' and
affect the $\Ord(\eps^0)$ cuts.  To see that this is not the case,
one may employ the same prescription
for handling dimensional regularization with spinor
helicity advocated by Mahlon~[\use\GielePrivate,\use\MahlonA].
This prescription separates the integration in (\use\caseacut) as follows,
$$
\int {d^{4-2\eps} \ell \over (2 \pi)^{4-2\eps} }
= -{\eps (4 \pi)^\eps \over \Gamma(1-\eps)}
\int {d^4 \ell \over (2 \pi)^{4} }
\int_0^\infty d \ell_\eps^2 (\ell_\eps^2)^{-1-\eps} \ .
\anoneqn
$$
A discrepancy caused by cavalier handling of $\ell_\eps$
contributions leads to a discrepancy containing an explicit
$\ell_\eps$ in the numerator.  A lone power of $\ell_\eps$ will lead
to a vanishing integral, because all external momenta with which
$\ell_\eps$ can be contracted may be taken to be four-dimensional.
Integrals with two or more extra powers of $\ell_\eps$ in the
numerator have only ultraviolet, not infrared, singularities (and can
be interpreted as integrals in $D>4$ [\use\IntegralsLong]).  Thus, the
`error' induced by incorrect $\ell_\e$ terms in the numerator is of
the form $\e \times (1/\e)\times$~polynomial, which we can again
neglect since we are extracting the cuts alone to order $\e^0$.
(Errors in the spinor helicity manipulations due to a mismatch of
dimensions do introduce errors in the finite cut-free polynomials;
explicit examples of this phenomenon have been provided by Mahlon
[\use\MahlonA,\use\MahlonB].)

To evaluate the integral (\use\caseacut) we need explicit formulae for
the tree amplitudes, preferably in a compact form.  To obtain the
$N=1$ chiral multiplet contributions we need formulae for tree
amplitudes with two external fermions or scalars and up to $n-2$
external gluons.  For the helicity configuration (\use\mhvdef) the
necessary tree-level results exist.
There are two cases to consider, as shown in fig.~\use\CutsFigure .
In the first case, the two negative helicities are on the same side
of the cut. Here the cut vanishes since there is no assignment of
helicity for the intermediate scalars or fermions where the tree on
the right-side does not vanish. Tree amplitudes with all positive-helicity
gluons and two
external fermions or scalars vanish by helicity conservation
if the two (outgoing) fermions or
scalars have the same helicity, and vanish by a supersymmetry
Ward identity [\use\Susy] if they have opposite helicity.
(For complex scalars the two
`helicities' mean particles and antiparticles respectively.)  The
second case, where the two external negative helicities are on opposite
sides of the cut, has two possible intermediate helicity
configurations as shown in fig.~\use\CutsFigure{b}.  The tree
amplitudes on either side of the cut are the MHV amplitudes with two
external fermions ($\Lambda$) or scalars ($\phi$). These can be
obtained using the supersymmetry Ward identities~[\use\Susy]
$$
\eqalign{
\Atree(\Lambda_1^-, g_2^+, \ldots , g_j^-, \ldots, \Lambda_n^+)
\ &=\ { \spa{j}.{n} \over \spa{j}.{1} }
\ \Atree( g_1^-, g_2^+ , \ldots, g_j^-, \ldots, g_n^+),
\cr
\Atree(\phi_1^-, g_2^+, \ldots ,g_j^-, \ldots, \phi_n^+)
\ &=\ { {\spa{j}.{n}}^2 \over {\spa{j}.{1}}^2 }
\ \Atree( g_1^-, g_2^+ , \ldots, g_j^-, \ldots, g_n^+),
\cr}
\eqn\scalarContrib
$$
where `$\ldots$' denotes positive-helicity gluons
and the Parke-Taylor $n$-gluon MHV tree amplitudes
[\use\ParkeTaylor,\use\ManganoParke] are
$$
\eqalign{
  A_{ij}^{\rm tree} (1,2,\ldots,n)\
&=\ i\, { {\spa{i}.{j}}^4 \over \spa1.2\spa2.3\cdots\spa{n}.1 }\,.
  \cr}
\eqn\PT
$$
The result~(\PT) is written in terms of spinor inner-products,
$\spa{j}.{l} = \langle j^- | l^+ \rangle = \bar{u}_-(k_j) u_+(k_l)$ and
$\spb{j}.{l} = \langle j^+ | l^- \rangle = \bar{u}_+(k_j) u_-(k_l)$,
where $u_\pm(k)$ is a massless Weyl spinor with momentum $k$ and
chirality $\pm$~[\use\SpinorHelicity,\use\ManganoReview].
We thus have explicit formulae for all the required tree amplitudes.

The evaluation of the cuts is similar to (although somewhat more
complicated than) the evaluation of the cuts for the $N=4$ MHV
amplitudes presented in ref.~[\use\SusyFour].  In performing the
calculation there is no need to track the overall sign, which is fixed
by the pole terms in $\eps$. The off-shell integral reduces to a sum of
scalar box and triangle integrals; we shall not present the details of
this reduction.

Let us consider first the special case where the two negative-helicity gluons
are adjacent.  In this case the calculation leads to a result which
does not contain box integral functions and is particularly simple.
For convenience we take the negative-helicity legs to be the first
and second ($i=1$ and $j=2$), obtaining for the unrenormalized amplitude
[\use\QCDConf],
$$
\eqalign{
A_{1\, 2}^{N=1\ \rm chiral} & (1, \ldots, n) =
{\cg (\mu^2)^\e A_{1 \, 2}^{\rm tree}(1,\ldots, n)  \over2}
\biggl\{
\biggl(
\Kz ( t_2^{[2]} ) + \Kz( t_n^{[2]} )
\biggr)
\cr
&
 - {1 \over  \tn{2}1 }
   \sum_{m=4}^{n-1} {\Lz \L -\tn{m-2}2/(-\tn{m-1}2)\R\over \tn{m-1}2}
 \Bigl( \tr_+[\Slash{k_1}\Slash{k_2}\Slash{k_{m}}\Slash{q_{m,1}} ]
-\tr_+[\Slash{k_1}\Slash{k_2}\Slash{q_{m,1}}\Slash{k_{m}} ] \Bigr)
\biggr\}
\cr}
\eqn\AdjacentMinusAmplitude
$$
where $\Kz$ and $\Lz$ are integral functions defined in
appendix~\use\FunctionAppendix ,
$$q_{m,l}=\left\{\eqalign{\sum_{i=m}^{l} k_i,&\qquad m \leq l ,\cr
                   \sum_{i=m}^{n} k_i + \sum_{i=1}^l k_i,&\qquad m>l,\cr}\RP
\anoneqn$$
 $\mu$ is the renormalization scale,
$$
\tr_+[\rho] \equiv {1\over 2} \tr[(1+\gamma_5) \rho] \; ;
\anoneqn
$$
and $\cg$ was defined in equation~(\use\Prefactor).

The expression~(\use\AdjacentMinusAmplitude) has the correct cuts in
all channels, as does the more general
expression~(\use\NonadjacentMinusAmplitude) we present below.  Since it is
written in terms of the integral functions, $\Kz$ and $\Lz$,
the unitarity result of section~\use\UnitarityProofSection\ implies it
is the complete answer including all rational, cut-free parts.
Note that the $\Lz$'s appearing in equation~(\use\AdjacentMinusAmplitude)
do not form an independent set, and therefore the result may be expressed in
a number of seemingly inequivalent ways.  We have chosen the form in
equation~(\use\AdjacentMinusAmplitude) in
order to make the reflection (anti-)symmetry of this color-ordered
amplitude for $n$ even (odd) manifest.
We give a schematic representation of this amplitude in
\fig\AdjacentFigure, where the coefficients $c_m$ are the ones appearing
in eq.~(\use\AdjacentMinusAmplitude).
(As discussed in appendix~\FunctionAppendix\
the function $\Lz$ can be regarded as arising from a two mass
triangle integral.)

\vskip -1.truecm
\LoadFigure\AdjacentFigure {\baselineskip 13 pt
\narrower\ninerm Schematic representation of the $N=1$ chiral
amplitude with legs 1 and 2 negative helicities and the rest positive
helicities. For the boundary terms $(m=3,n)$,
$\Lz$ reduces to $\Kz$. }
{\epsfysize 1.4truein}{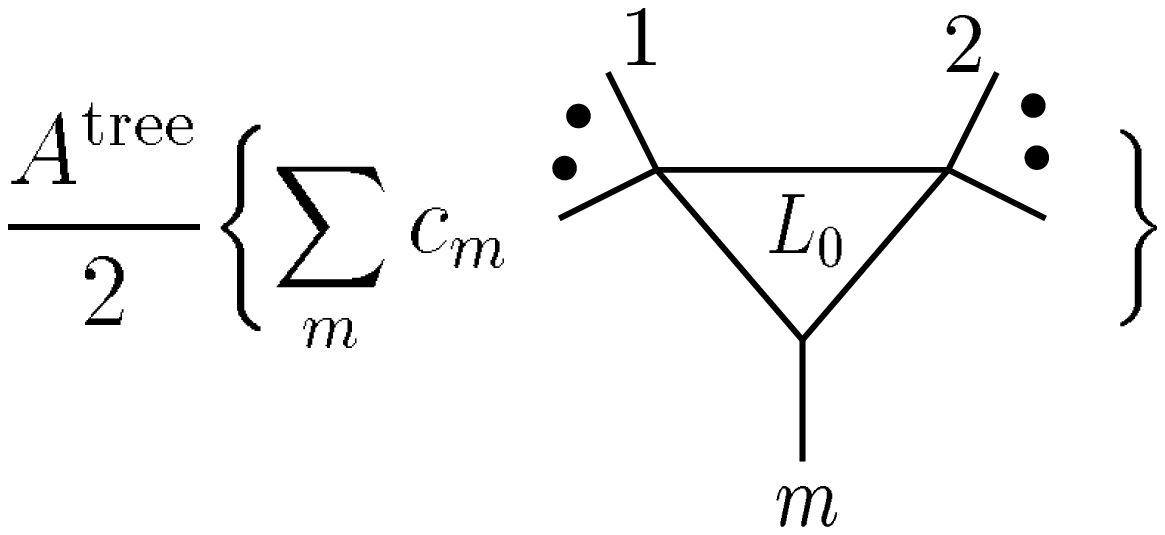}

The general case in which the two negative-helicity legs need
not be adjacent yields a more complicated expression.  Choose the first leg
to be a
negative-helicity leg ($i=1$) and label the second negative-helicity
leg by $j$; the
color-ordered amplitude is then
$$
\eqalign{
A_{1j }^{N=1\ \rm chiral} &   (1,\ldots,n)
= { {  \cg (\mu^2)^\e { A_{1j}^{\rm tree}   (1,\ldots,n)} \over
2} }
\cr
& \times\biggl\{\sum_{m_1=2\;}^{j-1} \quad \sum_{\; m_2= j+1}^n  b^j_{m_1, m_2}
\Mz\L \tn{m_2-m_1}{m_1+1}, \tn{m_2-m_1}{m_1};
     \tn{m_2-m_1-1}{m_1+1}, \tn{n+m_1-m_2-1}{m_2+1}\R\cr
&
+ \sum_{ 2 \le m < j }\;
\sum_{ a \in \hat\S_m}
c^j_{m,a}  { \Lz (  (-\tn{a-m}{m+1})/(-\tn{a-m+1}{m}) ) \over \tn{a-m+1}{m} }
  \cr
&
+ \sum_{ j < m \leq n } \;
\sum_{a \in \S_m}
c^j_{m,a}  { \Lz (  (-\tn{m-a}{a+1})/(-\tn{m-a-1}{a+1}) )
                   \over\tn{m-a-1}{a+1} }
\cr
&
+{c^j_{2,n}  \over  t_{1}^{[2]} }\Kz ( t_{1}^{[2]} )
+{c^j_{n,1}  \over t_{n}^{[2]} }  \Kz ( t_{n}^{[2]} )
+{c^j_{j+1,j-1} \over  t_{j}^{[2]}  }  \Kz ( t_{j}^{[2]} )
+{c^j_{j-1,j}  \over   t_{j-1}^{[2]}  }  \Kz ( t_{j-1}^{[2]} ) \biggr\}
\cr}
\eqn\NonadjacentMinusAmplitude
$$
where the various integral functions are defined in
appendix~\use\FunctionAppendix\ and
$$
\eqalign{
b^j_{m_1, m_2} & =
-2{\tr_+[\Slash{k_1}\Slash{k_j}\Slash{k_{m_1}}\Slash{k_{m_2}} ]
\tr_+ [\Slash{k_1}\Slash{k_j} \Slash{k_{m_2}}\Slash{k_{m_1}}]
\over \LB (k_1 + k_j )^2\RB^2 \;\LB (k_{m_1}+k_{m_2} )^2\RB^2 } \cr
& =
2 { \spa{1}.{m_1} \spa{1}.{m_2} \spa{j}.{m_1} \spa{j}.{m_2}
   \over {\spa{1}.{j}}^2 {\spa{m_1}.{m_2}}^2 }\; ,  \cr }
\anoneqn
$$
$$
\eqalign{
 c^j_{m,a} &=
{(
{\tr_{+}} [ \Slash{k_1}\Slash{k_j} \Slash{k_m}\Slash{q_{m,a}}]
-{\tr_{+}} [ \Slash{k_1}\Slash{k_j} \Slash{q_{m,a}}\Slash{k_m}]
)\over \LB (k_1 + k_j )^2\RB^2 }
\biggl(
{
{\tr_{+}} [ \Slash{k_m}\Slash{k_{a+1}}\Slash{k_j}\Slash{k_1}]
\over ( k_m+k_{a+1} )^2 }
-
{ {\tr_{+}} [ \Slash{k_m}\Slash{k_a}\Slash{k_j}\Slash{k_1}]
\over ( k_m+k_{a} )^2 }
\biggr) \; , \cr
&= {(
{\tr_{+}} [ \Slash{k_1}\Slash{k_j} \Slash{k_m}\Slash{q_{m,a}}]
-{\tr_{+}} [ \Slash{k_1}\Slash{k_j} \Slash{q_{m,a}}\Slash{k_m}]
) \over \LB (k_1 + k_j )^2\RB^2 }
\spa{m}.1\spb1.j\spa{j}.m\,\soft{a}m{a+1}\;,
}
\anoneqn
$$

$$
\soft{i}{k}{j}\ =\
{{\spa{i}.{j}\over \spa{i}.{k}\spa{k}.{j}}}\,,
\anoneqn$$

$$
\hskip - .4 cm
\S_m = \left\{ \eqalign{
& \{1,2, \ldots ,j-2 \}, \hskip .5cm   m =j+1 , \cr
& \{1,2, \ldots ,j-1 \} , \hskip .5cm  j+1 <m < n , \cr
& \{2, \ldots ,j-1 \} , \hskip .8cm m =n , \cr } \right.
\hskip .5 cm
\hat\S_m = \left\{ \eqalign{
& \{j, j+1, \ldots, n-1 \} , \hskip .4cm  m = 2 , \cr
&\{j, j+1, \ldots, n \} , \hskip 1.0 cm 2 <m < j-1 , \cr
& \{j+1, j+2, \ldots, n \} , \hskip .4cm m = j-1 . \cr }
\right.
\anoneqn
$$
The reader may verify that equation~(\use\NonadjacentMinusAmplitude)
reduces to equation~(\use\AdjacentMinusAmplitude) for $j=2$.
These equations agree for $n=5$ with previously published
results~[\use\FiveGluon]; note that in equations~(\use\AdjacentMinusAmplitude)
and~(\use\NonadjacentMinusAmplitude), we have effectively combined $V_n$
and $F_n/A^{\rm tree}_n$.

We may observe that
$$\eqalign{
{c^j_{2,n}\over\tn21}+{c^j_{n,1}\over\tn2n} &= 1\,,\cr
{c^j_{j+1,j-1}\over \tn2j}+{c^j_{j-1,j}\over\tn2{j-1}} &= 1\,,\cr
}\eqn\cRelations$$
and that as a result, equation~(\use\NonadjacentMinusAmplitude)
has the correct $1/\eps$ singularity since
the pole terms, contained in the $\Kz$ functions,
are
$$
\eqalign{
   \cg { A_{1j}^{\rm tree }   (1,\ldots,n) \over 2 }
& \biggl(
{  c^j_{2,n} \over t_{1}^{[2]} }
+{ c^j_{n,1} \over  t_{n}^{[2]} }
+{ c^j_{j+1,j-1} \over t_{j}^{[2]}   }
+{ c^j_{j-1,j}  \over t_{j-1}^{[2]}  } \biggr){1\over\eps} \cr
&=   \cg  {1\over\eps} A_{ij}^{\rm tree}   (1,\ldots,n)
\;.\cr}
\anoneqn
$$
The amplitude may be described in terms of a set
of $(D=6)$ scalar two-mass box integrals and linear two mass triangle
integrals (plus the boundary terms of single mass integrals) shown
schematically in \fig\SusyResultFigure .
In appendix~\use\CollinearLimitCheck,
we check the collinear limits of these expressions (indeed,
equation~(\use\AdjacentMinusAmplitude) was first constructed from
a knowledge of these limits).

\LoadFigure\SusyResultFigure{\baselineskip 13 pt
\narrower\ninerm Schematic representation of the general
$N=1$ chiral loop amplitude with two negative helicity external legs.}
{\epsfysize 2.5truein}
{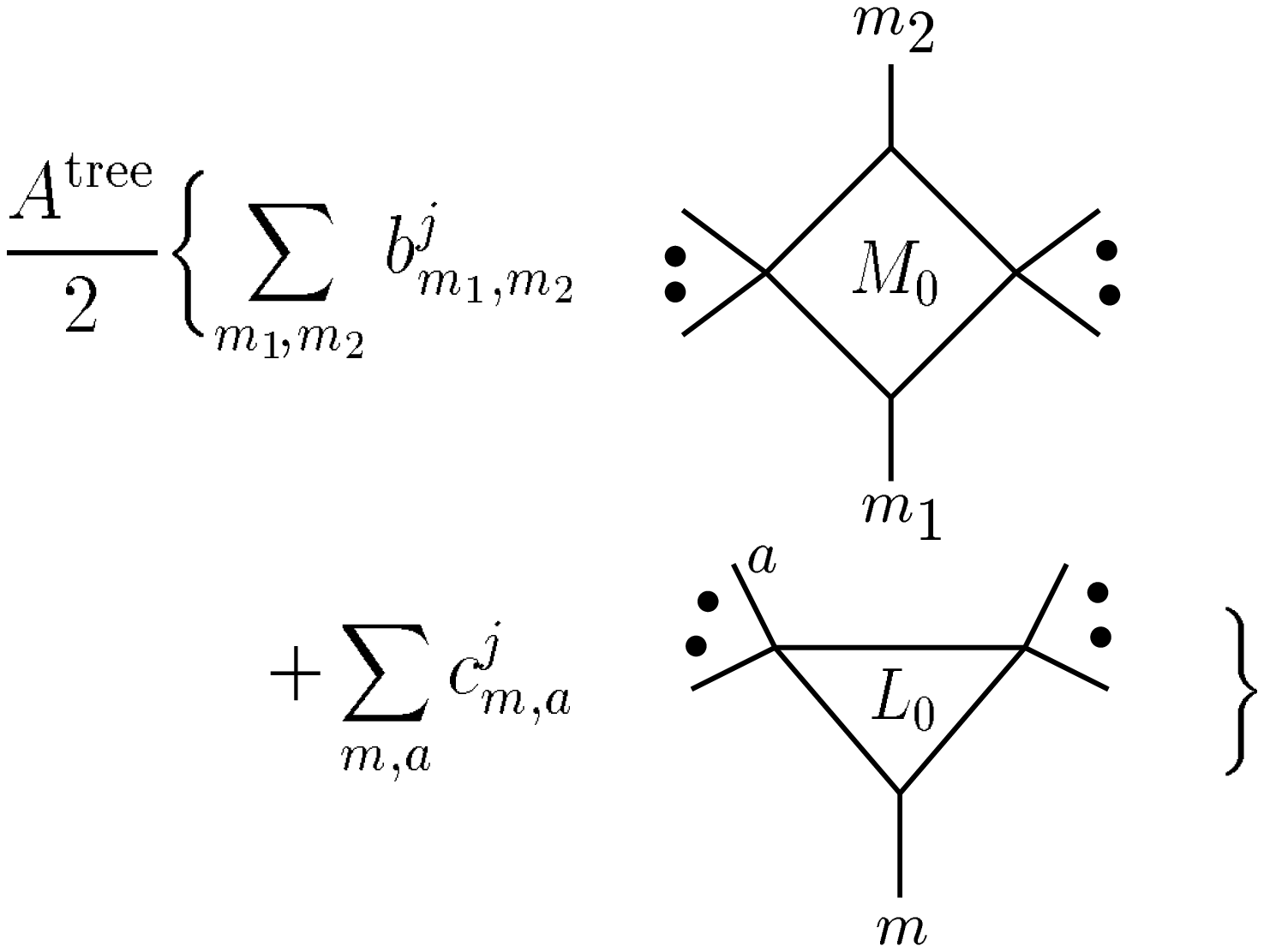}


\section{$N=4$ Supersymmetric Non-MHV Six-Gluon Amplitudes}
\tagsection\NonMHVSection
\def\Fn{6}

We calculated the $N=4$ supersymmetric MHV one-loop $n$-gluon amplitudes
$A_{jk}^{N=4}(1,\ldots,n)$ in
a previous paper~[\use\SusyFour].
They are proportional to the tree amplitude,
$$
  A_{jk}^{N=4}(1,\ldots,n)\ =\
  \cg \times A_{jk}^\tree(1,\ldots,n) \times V_n^g(t_i^{[r]})\ ,
\eqn\NeqFourMHV
$$
where the ``universal'', cyclically symmetric function $V_n^g$
is independent of the locations $j,k$ of the negative helicities,
and contains no spinor inner products, but only the momentum invariants
$t_i^{[r]}$.  (For the $N=4$ four-point amplitude this simple structure was
first observed by Green, Schwarz and Brink, who obtained the amplitude
as the low-energy limit of a superstring amplitude~[\use\GSB].)
One might wonder whether the {\it non}-MHV $N=4$ supersymmetric
amplitudes also have this structure.
Non-MHV helicity configurations first appear in six-point amplitudes,
for which there are three distinct configurations,
$(+{}+{}+{}-{}-{}-)$, $(+{}+{}-{}+{}-{}-)$, and $(+{}-{}+{}-{}+{}-)$.
In this section we calculate these three $N=4$ supersymmetric
six-gluon amplitudes from their unitarity cuts.
We will find that the structure of non-MHV amplitudes
 is more elaborate than that in eq.~(\use\NeqFourMHV).

{}From our previous work~[\use\SusyFour] we know that $N=4$ amplitudes
can be
expressed as linear combinations of scalar box integrals only.
It will turn out that the coefficient of each box integral
in each of the three non-MHV helicity configurations can be
obtained (using various symmetries) from the cut in the
$t_{123}$ channel [$t_{ijl} = \L k_i+k_j+k_l\R^2$] of the first configuration,
$A_{6;1}^{N=4}(1^+,2^+,3^+,4^-,5^-,6^-)$.
We thus begin by evaluating this cut.


\subsection{ The $t_{123}$ cut of
  $A_{6;1}^{N=4}(1^+,2^+,3^+,4^-,5^-,6^-)$}

Using the supersymmetry Ward identities at tree-level, it is easy to
see that only the gluon loop contributes to this cut, and in fact the
cut is given by a product of two five-point, pure-glue,
MHV tree amplitudes.
However, unlike the cuts in MHV loop amplitudes evaluated in
ref.~[\use\SusyFour] and in section~\use\MHVSection,
in this case one of the MHV amplitudes is complex conjugated,
since it is for the helicity configuration $(+{}+{}-{}-{}-)$
instead of $(-{}-{}+{}+{}+)$.
Thus the $t_{123}$ cut is given by the cut hexagon integral
$$
\eqalign{
 C_{123}\ &\equiv\ i\int \dlips(-\ell_1,\ell_2)
   \ A^\tree_5((-\ell_1)^-,1^+,2^+,3^+,\ell_2^-)
   \ A^{\tree}_5((-\ell_2)^+,4^-,5^-,6^-,\ell_1^+) \cr
 &=\ - { i \over \spa1.2\spa2.3\spb4.5\spb5.6 }
 \int \dlips(-\ell_1,\ell_2)
   { {\spa{\ell_1}.{\ell_2}}^3 {\spb{\ell_1}.{\ell_2}}^3
    \over \spa{\ell_1}.{1} \spa{3}.{\ell_2}
          \spb{\ell_2}.{4} \spb{6}.{\ell_1} } \cr
  &=\ - { i(t_{123})^3 \over \spa1.2\spa2.3\spb4.5\spb5.6 }
  \int \dlips(-\ell_1,\ell_2)
   \ { \spb{3}.{\ell_2} \spa{\ell_2}.{4}
     \spb{1}.{\ell_1} \spa{\ell_1}.{6} \over
    (\ell_1-k_1)^2 (\ell_2+k_3)^2 (\ell_2-k_4)^2 (\ell_1+k_6)^2 }
 \ , \cr
}\eqn\nonmhvcut
$$
using $\spa{\ell_1}.{\ell_2} \spb{\ell_1}.{\ell_2}
= (\ell_1-\ell_2)^2 = t_{123}$.
(We suppress an overall factor of $\mu^{2\e}$.)

There are several ways to reduce this cut hexagon integral
to a linear combination of cut scalar box integrals.
We choose to Feynman parametrize the integral, letting
$$
\eqalign{
  \ell_1\ &=\ q + a_2 k_1 + a_3 (k_1+k_2) + a_4 (k_1+k_2+k_3)
                - a_5 (k_5+k_6) - a_6 k_6 \, , \cr
  \ell_2\ &=\ \ell_1 - k_1 - k_2 - k_3 \, , \cr}
\eqn\Feyn
$$
with $\sum_{i=1}^6 a_i = 1$.
The numerator polynomial in eq.~(\use\nonmhvcut) becomes, after a
little spinor simplification,
$$
\eqalign{
 &\spb{3}.{\ell_2} \spa{\ell_2}.{4} \spb{1}.{\ell_1} \spa{\ell_1}.{6}
 \cr
 &\to
   \langle3^+|\gamma_\mu|4^+\rangle \langle1^+|\gamma_\nu|6^+\rangle
     \ q^\mu q^\nu \cr
 &\quad\ +\ \hbox{linear term in $q^\mu$} \cr
 &\quad\ +\ \Bigl( \spb3.5\spa5.4 a_6 + (\spb3.5\spa5.4+\spb3.6\spa6.4) a_1
                           - \spb3.2\spa2.4 a_2 \Bigr)  \cr
 &\quad\quad\times
            \Bigl( \spb1.2\spa2.6 a_3 + (\spb1.2\spa2.6+\spb1.3\spa3.6) a_4
                           - \spb1.5\spa5.6 a_5 \Bigr)\ . \cr}
\eqn\Feynparam
$$

Now we can employ formula~(VII.8) of ref.~[\use\IntegralsLong],
for $n=6$, to express the two-parameter (``reduced'') hexagon
integrals arising from the third term in~(\Feynparam),
$\hat I_6[a_ia_j]$, in terms of scalar
hexagon and pentagon integrals in $6-2\e$ dimensions,
$\hat I_6^{D=6-2\e}$ and $\hat I_5^{D=6-2\e\ (\ell)}$,
and the desired scalar box integrals, $\hat I_4^{(\ell,p)}$,
$$
\eqalign{
  \hat I_6[a_ia_j]\ &=\ {\eta_{ij}\Det_6 + 2\e \gamma_i\gamma_j
    \over 2N_6\Det_6}\ \hat I_6^{D=6-2\e} \cr
&\quad + {2\e\over 4N_6^2} \sum_{\ell=1}^6 \left[
   \eta_{i\ell}\gamma_j + \eta_{j\ell}\gamma_i
 - {\eta_{i\ell}\eta_{j\ell}\gamma_\ell\over\eta_{\ell\ell}}
 - {\gamma_i\gamma_j\gamma_\ell\over\Det_6} \right]
      \hat I_5^{D=6-2\e\ (\ell)}\cr
&\quad + {1\over 4N_6^2} \sum_{\ell,p=1}^6 \left[
   {\eta_{ip}\eta_{j\ell}\eta_{\ell\ell}
    - \eta_{i\ell}\eta_{j\ell}\eta_{\ell p}
   \over\eta_{\ell\ell}} \right] \hat I_4^{(\ell,p)}\ . \cr }
\eqn\twoparam
$$
The definitions of the reduced integrals $\hat I_n$ and
the kinematical quantities $\Det_6$, $\gamma_i$, $\eta_{ij}$ and
$N_6$ appearing in eq.~(\use\twoparam) are given in
ref.~[\use\IntegralsLong].  The index $\ell$ on the ``daughter''
integral $\hat I_5^{D=6-2\e\ (\ell)}$ means that the propagator
between external legs $(\ell-1)\ $(mod 6) and $\ell$ should be
omitted; that propagator and the one between $(p-1)\ $(mod 6) and $p$
should be omitted for $\hat I_4^{(\ell,p)}$.  (In this formula,
four-dimensional external kinematics should be taken only at the
very end of the calculation, or one encounters unphysical
singularities; this point is discussed in ref.~[\use\IntegralsLong].)

The first two terms on the right-hand side
of eq.~(\use\twoparam) can be dropped:  the first term
cancels (to $\Ord(\e)$) against a $(6-2\e)$-dimensional hexagon
integral from integrating the first term in~(\Feynparam);
the second term is $\Ord(\e)$.
Combining eqs.~(\Feynparam) and (\twoparam), the cut hexagon
integral is expressed in terms of scalar box integrals
$I_4^{(\ell,p)}$ as,
$$
\eqalign{
  C_{123}\ &=\ {i \over (4\pi)^{2-\e}}
     {{\rm Im} \atop t_{123}>0} \Biggl[
     c^{(2,3)} \, I_4^{(2,3)} + c^{(5,6)} \, I_4^{(5,6)} \cr
 &\quad
   + c^{(2,5)} \, I_4^{(2,5)} + c^{(3,6)} \, I_4^{(3,6)}
   + c^{(2,6)} \, I_4^{(2,6)} + c^{(3,5)} \, I_4^{(3,5)}
   \Biggr]\ , \cr}
\eqn\hexbox
$$
where the box coefficients are
$$
  c^{(\ell,p)}\ =\ { (t_{123})^3 \over \spa1.2\spa2.3\spb4.5\spb5.6 }
   \times \sum_{i,j=1}^6 v^i \, (M^{(\ell,p)})_{ij} \, w^j
\eqn\clp
$$
with
$$
  (M^{(\ell,p)})_{ij}\ \equiv\
  {\alpha_\ell \alpha_p\over 4N_6^2}
  \left[  {\eta_{ip}\eta_{j\ell}\eta_{\ell\ell}
    - \eta_{i\ell}\eta_{j\ell}\eta_{\ell p}
   \over\eta_{\ell\ell}}\ + \ (\ell\lr p)\ \right]
\eqn\Mlp
$$
and
$$
\eqalign{
  v^i\ &\equiv \bigl(
   (\spb3.5\spa5.4+\spb3.6\spa6.4)\alpha_1,- \spb3.2\spa2.4\alpha_2,
  0,0,0,\spb3.5\spa5.4\alpha_6\bigr)\, , \cr
  w^j\ &\equiv \bigl(0,0,\spb1.2\spa2.6\alpha_3,
   (\spb1.2\spa2.6+\spb1.3\spa3.6)\alpha_4,-\spb1.5\spa5.6\alpha_5,
   0\bigr)\, .\cr}
\anoneqn
$$
(The kinematical quantities $\alpha_i$ are also defined in
ref.~[\IntegralsLong].)
Of the $6\cdot5/2=15$ box integrals $I_4^{(\ell,p)}$, only
the six appearing in eq.~(\hexbox) have cuts in the $t_{123}$ channel.

It is convenient to rewrite eq.~(\hexbox) in terms of the scalar
box $F$ functions defined in equation~(\use\boxes) of
Appendix~\use\IntegralsAppendix,
$$
\eqalign{
  C_{123}\ =\ -2 \, i \, \cg {{\rm Im} \atop t_{123}>0} \Biggl[
     &{ c^{(2,3)} \over s_{45} s_{56} } \, \Fone{1}
   + { c^{(5,6)} \over s_{12} s_{23} } \, \Fone{4} \cr
   + &{ c^{(2,5)} \over t_{612} t_{123} - s_{12} s_{45} } \, \Feasy{2;1}
   + { c^{(3,6)} \over t_{123} t_{234} - s_{23} s_{56} } \, \Feasy{2;2}
   \cr
   + &{ c^{(2,6)} \over s_{34} t_{123} } \, \Fhard{2;5}
   + { c^{(3,5)} \over s_{61} t_{123} } \, \Fhard{2;2} \Biggr]\ .\cr}
\eqn\newhexbox
$$
Upon simplifying the rather messy form~(\use\clp) for the
quantities~$c^{(\ell,p)}$, we find that the coefficients of the
``easy-two-mass'' box functions $\Feasy{2;1}$, $\Feasy{2;2}$ vanish,
and the others are all equal:
$$
  C_{123}\ =\
  \cg B_0\ \times\ {{\rm Im}\atop t_{123}>0}
            \left[ \Fone{1} + \Fone{4}
                 + \Fhard{2;2} + \Fhard{2;5} \right]\ .
\eqn\Canswer
$$
where the coefficient $B_0$ is defined as
$$
\eqalign{
  B_0\ &\equiv\ -2 \, i \,
   \ { c^{(2,3)} \over s_{45} s_{56} }\,, \cr}
\anoneqn$$ and is given by
$$\eqalign{
B_0\ &=\ i \, {(\spb1.2\spa2.4+\spb1.3\spa3.4)
  \, (\spb3.1\spa1.6+\spb3.2\spa2.6) \, (t_{123})^3
  \over \spa1.2\spa2.3\spb4.5\spb5.6\
  (t_{123}t_{345}-s_{12}s_{45}) \, (t_{123}t_{234}-s_{23}s_{56}) }
 \ . \cr}
\eqn\Bdefanswer
$$
In deriving this expression, we used the vanishing of the Gram determinant
for any subsystem of five vectors, after integral reductions.

Recall the form of the $(+{}+{}+{}-{}-{}-)$ tree
amplitude from refs.~[\use\MPX,\use\ManganoReview],
$$
\eqalign{
 A_6^\tree(1^+,2^+,3^+,4^-,5^-,6^-)\ &=\
  i \, \left[
   {\beta^2 \over t_{234}s_{23}s_{34}s_{56}s_{61}}
 + {\gamma^2 \over t_{345}s_{34}s_{45}s_{61}s_{12}}
 + {\beta\gamma t_{123} \over s_{12}s_{23}s_{34}s_{45}s_{56}s_{61}}
  \right] \, , \cr
 \beta\ &\equiv\
   \spb2.3\spa5.6\ (\spb1.2\spa2.4+\spb1.3\spa3.4)\, , \cr
 \gamma\ &\equiv\
   \spb1.2\spa4.5\ (\spb3.1\spa1.6+\spb3.2\spa2.6) \, . \cr}
\eqn\treepppmmm
$$
The coefficient $B_0$ is related to the third term in this expression:
their ratio is a real quantity, and can be written
without complex spinor products, as a simple
rational function in $s_{i,i+1}$ and $t_{i,i+1,i+2}$.

Having simplified the coefficient $B_0$, we turn to
the linear combination of box functions appearing in
eq.~(\use\Canswer).   Denoting this combination by $\Wsix{1}$,
and its cyclic permutations by $\Wsix{i}$, we have
explicitly,
$$
\eqalign{
  \Wsix{i}\ &\equiv\ \Fone{i} + \Fone{i+3}
                 + \Fhard{2;i+1} + \Fhard{2;i+4} \cr
     \ &=\ -{1\over2\e^2} \sum_{j=1}^6
         \left( { \mu^2 \over -s_{j,j+1} } \right)^\e
 \ -\ \ln\left({-t_{i,i+1,i+2} \over -s_{i,i+1}}\right)
      \ln\left({-t_{i,i+1,i+2} \over -s_{i+1,i+2}}\right)\cr
 &\quad
  \ -\ \ln\left({-t_{i,i+1,i+2} \over -s_{i+3,i+4}}\right)
      \ln\left({-t_{i,i+1,i+2} \over -s_{i+4,i+5}}\right)
 \ +\ \ln\left({-t_{i,i+1,i+2} \over -s_{i+2,i+3}}\right)
      \ln\left({-t_{i,i+1,i+2} \over -s_{i+5,i}}\right)\cr
 &\quad
 \ +\ {1\over 2}\ln\left({-s_{i,i+1} \over -s_{i+3,i+4}}\right)
         \ln\left({-s_{i+1,i+2} \over -s_{i+4,i+5}}\right)
 \ +\ {1\over 2}\ln\left({-s_{i-1,i} \over -s_{i,i+1}}\right)
         \ln\left({-s_{i+1,i+2} \over -s_{i+2,i+3}}\right)\cr
  &\quad
 \ +\ {1\over 2}\ln\left({-s_{i+2,i+3} \over -s_{i+3,i+4}}\right)
         \ln\left({-s_{i+4,i+5} \over -s_{i+5,i}}\right)
 \ +\ {\pi^2\over3}\ . \cr}
\eqn\Wdef
$$
It is amusing that all dilogarithms cancel out of $\Wsix{i}$.
Also note that $\Wsix{i}$ is invariant under a cyclic permutation
by three units; hence there are only three independent objects,
$\Wsix1$, $\Wsix2$, and $\Wsix3$.


\subsection{ Remaining $t_{i,i+1,i+2}$ cuts of
 $A_{6;1}^{N=4}(1^+,2^+,3^+,4^-,5^-,6^-)$,
 and the full amplitude}

Fortunately, no other integrals have to be performed explicitly to
get the other two $(+{}+{}+{}-{}-{}-)$ $t_{i,i+1,i+2}$ cuts, nor for the
$t_{i,i+1,i+2}$ cuts of the other two non-MHV six-gluon amplitudes.
They can all be related to the cut $C_{123}$ defined in
equation~(\nonmhvcut), and thereby to $B_0$ given in
equation~(\Bdefanswer).

First consider the $t_{234}$ cut of
$A_{6;1}^{N=4} (1^+,2^+,3^+,4^-,5^-,6^-)$.
There are two contributions to this cut,
one where the intermediate helicities are the same,
so only gluons contribute,
and one where the helicities of the intermediate states are opposite,
so one sums over the entire $N=4$ multiplet.
The first contribution is given by
$$
\eqalign{
 C_{234}^{(a)}\ &\equiv\ i\int \dlips(-\ell_1,\ell_2)
   \ A^{\tree}_5((-\ell_1)^-,2^+,3^+,4^-,\ell_2^-)
   \ A^{\tree}_5((-\ell_2)^+,5^-,6^-,1^+,\ell_1^+) \cr
  &=\  -{i (\spb2.3\spa5.6)^4
     \over t_{234}\ \spb2.3\spb3.4\spa5.6\spa6.1 }
 \int \dlips(-\ell_1,\ell_2)
   { 1 \over \spb{\ell_1}.{2} \spb{4}.{\ell_2}
          \spa{\ell_2}.{5} \spa{1}.{\ell_1} } \cr
  &=\  \left({ \spb2.3\spa5.6 \over t_{234} } \right)^4
  \times [C_{123}^{\dagger}]\vert_{j\to j+1} \, , \cr
}\eqn\nonmhvcuta
$$
\def\cc{\dagger}
where $\cc$ means complex conjugating spinor products,
$\spa{i}.{j}\leftrightarrow\spb{j}.i$
(without complex conjugating factors of $i$), and where
the subscript $j\to j+1$ means applying a cyclic permutation of
the six momenta $k_i$, $\{1,2,3,4,5,6\}\rightarrow \{2,3,4,5,6,1\}$.

\def\rhot{\tilde\rho}
The second contribution is given in terms of the contribution
of scalar intermediate states,
multiplied by a correction factor $\rhot^2$ similar to that used in
evaluating $N=4$ MHV cuts~[\use\SusyFour],
$$
\eqalign{
  \rhot^2\ &\equiv\
   \left({ \spb{1}.{\ell_1}\spa{\ell_1}.{4}
    \over \spb{1}.{\ell_2}\spa{\ell_2}.{4} }\right)^2
 -4\left({ \spb{1}.{\ell_1}\spa{\ell_1}.{4}
    \over \spb{1}.{\ell_2}\spa{\ell_2}.{4} }\right)
 +6
 -4\left({ \spb{1}.{\ell_1}\spa{\ell_1}.{4}
    \over \spb{1}.{\ell_2}\spa{\ell_2}.{4} }\right)^{-1}
 + \left({ \spb{1}.{\ell_1}\spa{\ell_1}.{4}
    \over \spb{1}.{\ell_2}\spa{\ell_2}.{4} }\right)^{-2} \cr
 &=\ { \bigl(\langle1^+| (\lsl_1-\lsl_2) |4^+\rangle\bigr)^4
   \over {\spa{\ell_1}.4}^2 {\spa{\ell_2}.4}^2
         {\spb{\ell_1}.1}^2 {\spb{\ell_2}.1}^2 } \cr
 &=\ { \bigl(\langle1^+| (\ksl_2+\ksl_3) |4^+\rangle\bigr)^4
   \over {\spa{\ell_1}.4}^2 {\spa{\ell_2}.4}^2
         {\spb{\ell_1}.1}^2 {\spb{\ell_2}.1}^2 }\ , \cr
}\eqn\newrhofactor
$$
$$
\eqalign{
 C_{234}^{(b)}\ &\equiv\ i \int \dlips(-\ell_1,\ell_2)
   \ A^{{\rm tree}}_5((-\ell_1),2^+,3^+,4^-,\ell_2)
   \ A^{{\rm tree}}_5((-\ell_2),5^-,6^-,1^+,\ell_1)
   \times \rhot^2 \cr
  &=\ -{ i\bigl(\langle1^+| (\ksl_2+\ksl_3) |4^+\rangle\bigr)^4
      \over t_{234}\ \spa2.3\spa3.4\spb5.6\spb6.1 }
 \int \dlips(-\ell_1,\ell_2)
   { 1 \over \spa{\ell_1}.{2} \spa{4}.{\ell_2}
          \spb{\ell_2}.{5} \spb{1}.{\ell_1} } \cr
  &=\  \left({ \langle1^+| (\ksl_2+\ksl_3) |4^+\rangle
               \over t_{234} } \right)^4
  \times [C_{123}]\vert_{j\to j+1}\; ,  \cr
}\eqn\nonmhvcutb
$$
where the legs carrying momenta $\ell_1$ and $\ell_2$ are intermediate
scalars.

Adding up these two contributions, and noticing that the remaining
$t_{345}$ cut is related by a reflection symmetry, we can now write
down the full amplitude:
$$
A_{6;1}^{N=4}(1^+,2^+,3^+,4^-,5^-,6^-)\ =\
 \cg\ \LB B_1\,\Wsix1+B_2\,\Wsix2+B_3\,\Wsix3\RB,
\eqn\pppmmmloop
$$
where
$$
\eqalign{
  B_1\ &=\ B_0 \ , \cr
  B_2\ &=\ \left({ \langle1^+| (\ksl_2+\ksl_3) |4^+\rangle
         \over t_{234} } \right)^4  \ B_0 \vert_{j\to j+1}
       + \left({ \spb2.3\spa5.6 \over t_{234} } \right)^4
            \ B_0^\cc \vert_{j\to j+1}\ , \cr
  B_3\ &=\ \left({ \langle3^+| (\ksl_1+\ksl_2) |6^+\rangle
         \over t_{345} } \right)^4  \ B_0 \vert_{j\to j-1}
       + \left({ \spb1.2\spa4.5 \over t_{345} } \right)^4
            \ B_0^\cc \vert_{j\to j-1}\ . \cr}
\eqn\pppmmmdef
$$

There are several consistency checks one may apply to equation~(\pppmmmloop).
The pole terms in $\e$ represent infrared divergences,
which should have a universal form, proportional to the
corresponding tree
amplitude~[\use\GG,\use\KunsztSingular],
$$
A_{6;1}^{N=4}(1,2,3,4,5,6)|_{\rm singular}\ =\
 \cg\ \left[ -{1\over\e^2} \sum_{j=1}^6
         \left( { \mu^2 \over -s_{j,j+1} } \right)^\e \right]
 A_{6}^{\tree}(1,2,3,4,5,6) \ .
\eqn\sixptsingular
$$
This behavior for the result~(\pppmmmloop) can be verified
using the explicit form~(\Wdef) for $\Wsix{i}$, and the
relation
$$
  B_1+B_2+B_3 =\ 2 \, A_6^\tree(1^+,2^+,3^+,4^-,5^-,6^-) \ ,
\eqn\ABCtreesum
$$
which we have verified numerically; this verifies the overall sign in
eq.~(\pppmmmloop). In addition, we have checked
numerically that the result in eq.~(\use\pppmmmloop) is equal to that
given by a direct calculation using the string-based methods of
ref.~[\use\Long].

In appendix~\use\CollinearLimitCheck, we verify
that these amplitudes have the expected behavior
in the limit that two neighboring (in the sense of color-ordering) momenta
become collinear, thereby isolating a two-particle singular invariant
($s_{i,i+1}\rightarrow0$).
One might also wonder how loop amplitudes
factorize in multi-particle channels, i.e. as $t_i^{[r]} \to 0$
for $r>2$.
It is easy to see from the supersymmetry Ward identities that
supersymmetric MHV amplitudes cannot have multi-particle poles,
so the first amplitudes for which this question arises are in fact
non-MHV supersymmetric amplitudes (and also the non-supersymmetric
amplitudes $A_{n;1}^{[0]}(1^-,2^+,\ldots,n^+)$ constructed by
Mahlon~[\use\MahlonB], which we will not study here).
In the $(+{}+{}+{}-{}-{}-)$ amplitude, one expects and finds poles
in the three-particle channels, $t_{234}\to0$ and $t_{345}\to0$.
However, naive factorization does not hold in these
limits, due to the presence
in $\Wsix{i}$ of logarithms of kinematic invariants which get
caught ``across'' the pole.  For example, as $t_{234}\to0$, the invariants
$s_{12}$, $s_{45}$, $t_{123}$ and $t_{345}$ each have at
least one argument on each side of the pole, i.e. at least one argument
belonging to $\{k_2,k_3,k_4\}$ and one to $\{k_5,k_6,k_1\}$.
This lack of naive factorization is present even at the level of
the universal singular terms which come from soft virtual gluon exchange.
One might suspect that naive factorization holds
for the $N=1$ chiral contribution and for the scalar contribution,
since these contributions do not have virtual gluons.
We have examined the three-particle poles of the
$(+{}+{}+{}-{}-{}-)$ $N=1$ chiral contribution, calculated from its cuts,
and have found that a simple factorization does hold there.


\subsection{ The remaining $N=4$ non-MHV amplitudes }

The analysis for the remaining two $N=4$ supersymmetric non-MHV
six-gluon amplitudes is identical to that presented for $(+{}+{}+{}-{}-{}-)$,
so we merely quote the results, in terms of the function $B_0$ given in
eq.~(\Bdefanswer):
$$
A_{6;1}^{N=4}(1^+,2^+,3^-,4^+,5^-,6^-)\ =\
 \cg\ \left[ D_1\,\Wsix1 + D_2\,\Wsix2 + D_3\,\Wsix3 \right],
\eqn\ppmpmmloop
$$
where
$$
\eqalign{
  D_1\ &=\ \left({ \langle4^+| (\ksl_1+\ksl_2) |3^+\rangle
         \over t_{123} } \right)^4    \ B_0
       + \left({ \spb1.2\spa5.6 \over t_{123} } \right)^4
            \ B_0^\cc \ , \cr
  D_2\ &=\ \left({ \langle1^+| (\ksl_2+\ksl_4) |3^+\rangle
         \over t_{234} } \right)^4    \ B_0 \vert_{j\to j+1}
       + \left({ \spb2.4\spa5.6 \over t_{234} } \right)^4
            \ B_0^\cc \vert_{j\to j+1}\ , \cr
  D_3\ &=\ \left({ \langle4^+| (\ksl_1+\ksl_2) |6^+\rangle
         \over t_{345} } \right)^4    \ B_0 \vert_{j\to j-1}
       + \left({ \spb1.2\spa3.5 \over t_{345} } \right)^4
            \ B_0^\cc \vert_{j\to j-1}\ , \cr}
\eqn\DEFdef
$$
and
$$
A_{6;1}^{N=4}(1^+,2^-,3^+,4^-,5^+,6^-)\ =\
 \cg\ \left[ G_1\,\Wsix1 + G_2\,\Wsix2 + G_3\,\Wsix3 \right],
\eqn\pmpmpmloop
$$
where
$$
\eqalign{
  G_1\ &=\ \left({ \langle5^+| (\ksl_4+\ksl_6) |2^+\rangle
         \over t_{123} } \right)^4    \ B_0
       + \left({ \spb1.3\spa4.6 \over t_{123} } \right)^4
            \ B_0^\cc \ , \cr
  G_2\ &=\ \left({ \langle3^+| (\ksl_2+\ksl_4) |6^+\rangle
 \over t_{234} } \right)^4    \ B_0^\cc \vert_{j\to j+1}
       + \left({ \spb5.1\spa2.4 \over t_{234} } \right)^4
            \ B_0 \vert_{j\to j+1}\ , \cr
  G_3\ &=\ \left({ \langle1^+| (\ksl_6+\ksl_2) |4^+\rangle
         \over t_{345} } \right)^4    \ B_0^\cc \vert_{j\to j-1}
       + \left({ \spb3.5\spa6.2 \over t_{345} } \right)^4
            \ B_0 \vert_{j\to j-1}\ . \cr}
\eqn\GHKdef
$$

We have checked (numerically) the relations required for the pole
terms in $\e$ to be correctly given by~(\sixptsingular):
$$
\eqalign{
  D_1 +D_2+D_3 &=\ 2 \, A_6^\tree(1^+,2^+,3^-,4^+,5^-,6^-)\, , \cr
  G_1+G_2+G_3 &=\ 2 \, A_6^\tree(1^+,2^-,3^+,4^-,5^+,6^-)\,. \cr}
\eqn\DEFGHKtreesum
$$
We have taken the tree amplitudes from
refs.~[\use\MPX,\use\ManganoReview].


\section{Non-Supersymmetric Theories.}
\tagsection\NonSusySection

The cut uniqueness result may also be applied to non-supersymmetric
amplitudes, allowing one to trade harder calculations for easier ones.
The supersymmetric decomposition of the $n$-gluon amplitudes in
eq.~(\use\TotalampB) provides one such example.
Instead of computing QCD amplitudes with gluons and quarks in the loop
we compute amplitudes with an $N=4$ multiplet, $N=1$ chiral multiplet
and scalar in the loop.
The supersymmetric amplitudes are cut-constructible,
leaving the easier-to-compute scalar loop contribution to be
obtained by other means (see below).

One can apply the same technique to amplitudes with external
fermions.  Consider the color-ordered partial amplitudes
with two massless external quarks and $n-2$ external gluons.
An example of a one-particle irreducible diagram contributing
to such an amplitude is depicted in {}\fig\NonSusyExampleFigure{a}.
Such an $m$-point graph has a loop-momentum polynomial
of maximum degree $m-1$, as discussed in section~\use\SusySection.
If one adds to this diagram an identical diagram but with the gluon
replaced by a scalar (depicted in fig.\use\NonSusyExampleFigure{b}),
suitably adjusts the scalar-fermion Yukawa coupling,
and works in background-field Feynman gauge~[\use\Background],
then the leading loop-momentum terms cancel.
(From the $\gamma$-matrix algebra the diagram with a gluon in the loop
has a minus sign relative to the case of a scalar in the loop.)
Thus the sum is cut-constructible, and so (after calculating the
cuts for the sum) one can again replace a more difficult calculation
with a gluon in the loop by a simpler calculation with a scalar in the
loop.

\LoadFigure\NonSusyExampleFigure{\baselineskip 13 pt
\narrower\ninerm Diagrams with identical leading
loop momentum behavior; the wavy lines are gluons, the straight lines
are fermions and the dashed lines are scalars.}{\epsfysize 1.5truein}
{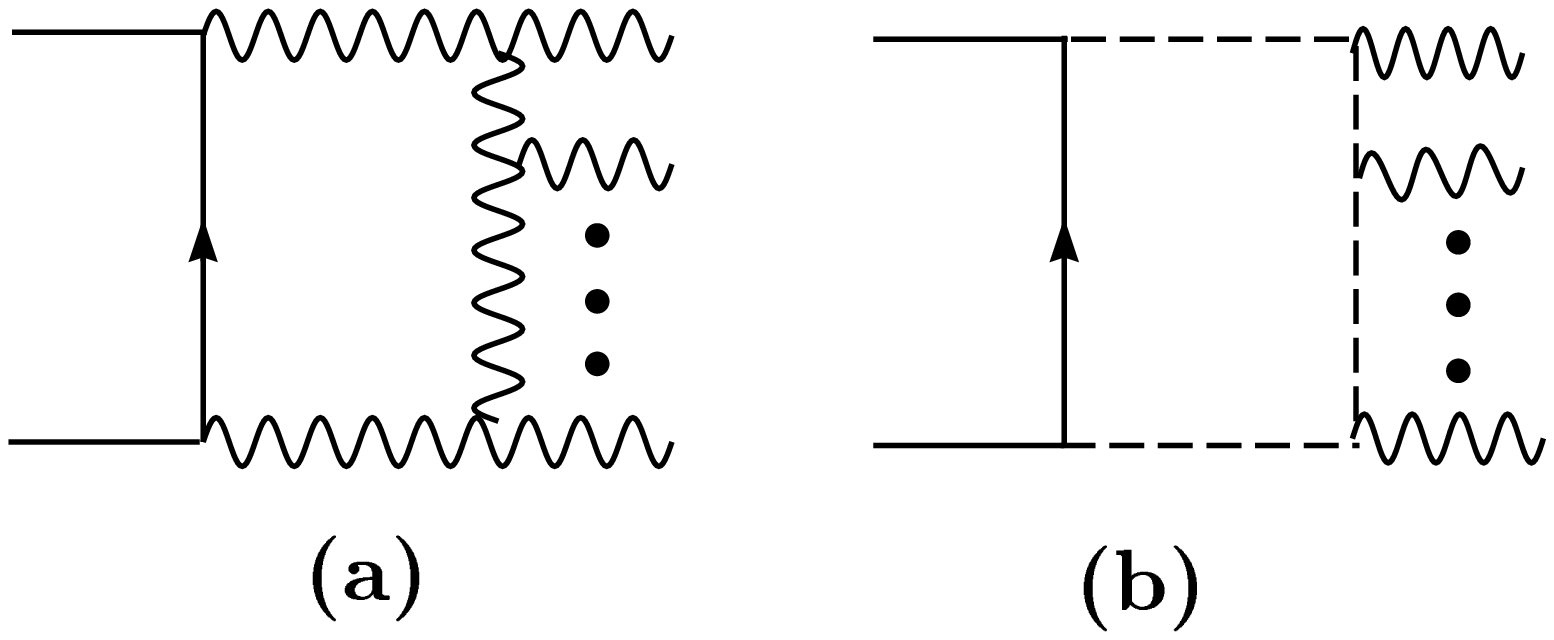}

Amplitudes with four or more external fermions can be treated
similarly.  Indeed, from section~\use\SusySection\ we know that the
only diagrams requiring cancellation of the leading loop-momentum
behavior are those where fermions are ``pinched off'' onto external
trees, such that the one-particle irreducible part of the
diagram has either zero or two external fermion lines.

Although the scalar loop contribution to $n$-gluon amplitudes
is not cut-constructible
(since the $m$-point loop integrals contain $m$ powers of loop momentum)
we can still fix all terms containing cuts.  As an explicit example, by
following the same procedure as for the $N=1$ chiral multiplet,
we obtain the scalar loop contribution for
two negative helicity adjacent legs ($i=1$ and $j=2$),
\def\Lt{\mathop{\rm L}\nolimits_2}
$$
\eqalign{
A_{1\, 2}^{[0]} & (1, \ldots, n)\ =\
 {1\over 3} A_{1\, 2}^{N=1\ \rm chiral} (1, \ldots, n) \cr
 &\quad - {\cg A_{1 \, 2}^{\rm tree}(1,\ldots, n)  \over 3}
   {1 \over  (\tn{2}1)^3 }
  \sum_{m=4}^{n-1} {\Lt \L -\tn{m-2}2/(-\tn{m-1}2)\R\over (\tn{m-1}2)^3 }
  \cr
&\quad \times
 \Bigl(
(\tr_+[\Slash{k_1}\Slash{k_2}\Slash{k_{m}}\Slash{q_{m,1}} ])^2 \,
\tr_+[\Slash{k_1}\Slash{k_2}\Slash{q_{m,1}}\Slash{k_{m}} ]
 -
\tr_+[\Slash{k_1}\Slash{k_2}\Slash{k_{m}}\Slash{q_{m,1}} ] \,
(\tr_+[\Slash{k_1}\Slash{k_2}\Slash{q_{m,1}}\Slash{k_{m}} ])^2 \Bigr)
  \cr
\null & \hskip 3 cm
+  {\rm polynomials.} \cr}
\eqn\ScalarLogs
$$
Although the cut analysis is quite similar to the $N=1$ chiral case,
here we cannot conclude that there are no missing polynomial
(rational function) terms.
Indeed, the known five-gluon amplitude with a scalar in the loop
[\use\FiveGluon] contains polynomials not reproduced by this formula.
Furthermore, if we ignore missing polynomials, this formula has
incorrect factorization properties; in particular, there should
be poles in multi-particle channels ($\tn{r}i \sim 0$ for $r\geq3$),
which are lacking from the non-polynomial terms in~(\ScalarLogs).
The appearance of multi-particle poles significantly
complicates the structure of the polynomial terms, but these pieces
should still be amenable to recursive [\use\MahlonA,\use\MahlonB] or
collinear bootstrap [\use\AllPlus,\use\SusyFour] techniques.

\section{Conclusions}
\tagsection\ConclusionSection

In a previous paper we demonstrated that $N=4$ super-Yang-Mills
amplitudes are constructible solely from their cuts and we explicitly
computed $n$-point one-loop amplitudes with maximal helicity
violation.  In this paper we extended the `cut-constructibility' to
include all massless amplitudes which satisfy a power-counting criterion.
In particular, $N=1$ supersymmetric gauge theory amplitudes satisfy
this criterion, when using string-based [\use\Long,\StringBased] or background
field diagrams [\use\Background,\use\Mapping]. Once the power-counting
criterion has been checked, the one-loop amplitudes can be obtained directly
from tree amplitudes using the cuts.

It is convenient to calculate cuts in terms of the imaginary parts of
one-loop integrals that would have been encountered in a direct
calculation.  This makes it straightforward to write down an analytic
expression with the correct cuts in all channels, avoiding the need to
perform dispersion integrals to reconstruct the full amplitude.  Once
this expression is written in terms of the integral functions
appearing in a direct calculation, it is fixed uniquely so long as
the power-counting criterion is satisfied.

Using this method, we calculated all maximally helicity-violating
one-loop $n$-gluon amplitudes in $N=1$ super-QCD as well as the $N=4$
six-gluon helicity amplitudes for all helicity configurations that
were not presented in ref.~[\use\SusyFour].  The $N=4$ and $N=1$
amplitudes form two of three components of a QCD $n$-gluon amplitude
[\use\FiveGluon,\use\Tasi],
with the third component the scalar loop contribution.  The scalar
loop contributions contain polynomial pieces which must be obtained by
other means, such as string-based~[\use\StringBased],
collinear [\use\AllPlus], or recursive [\use\MahlonA,\use\MahlonB]
techniques.  The replacement of
more difficult calculations with easier ones can also be carried out
for amplitudes with external fermions.

Gauge theory loop calculations are difficult to perform because of the
algebraic complexity which in most cases disappears near the end of a
calculation, with relatively simple final results. By fusing together tree
amplitudes we are utilizing simplifications already performed at
tree level.  When applicable, the technique presented here is
remarkably simple relative to conventional techniques, as there is no
explosion in the size of intermediate expressions.  We expect this
method to have wide applications and to be useful in the continuing
effort to calculate the gauge theory loop amplitudes required by
experiment.

\vskip0.3in
\par\noindent
{\bf Acknowledgements}
\vskip0.1in

Z.B.\ thanks G. Chalmers for discussions on multi-particle
factorization. We are grateful for the support
of NATO Collaborative Research Grants CRG--921322 (L.D. and D.A.K.)
and CRG--910285 (Z.B. and D.C.D).


\appendix{Integrals}
\tagappendix\IntegralsAppendix
\def\Fn{n}
%


In this appendix, we list the integral functions which occur,
after performing a Passarino-Veltman reduction, in
explicit calculations of amplitudes satisfying the power-counting
criterion given in section~\use\UnitarityProofSection, namely
that all $m$-point integrals ($m>2$) contain at most $m-2$ powers of the loop
momentum in the numerator of the integrand, and that all two-point
integrals contain at most a single power of the loop momentum.
(That is, loop integrals with $m$ or $m-1$ powers of the loop momentum
are excluded for $m>2$, as is the tensor bubble integral.)
We consider only purely massless processes; thus while the integral
functions appearing may have off-shell external legs (which we shall
refer to as massive legs, since $K^2\neq 0$), internal legs are
strictly massless.  As discussed in
Section~\use\ReviewSection, a Passarino-Veltman reduction may be used
to re-express any integral satisfying the power-counting criterion
(and thus any amplitude all of whose loop integrals satisfy the criterion),
in terms of
\item{(a)} scalar boxes, triangles and bubbles; and,
\item{(b)} triangles with one or two external massive (off-shell) legs
and  with a single power of loop momentum
in the numerator, and bubbles with a single power of the loop momentum
in the numerator.

\noindent
However, as we shall see the
integrals in category (b) are all  linear combinations (with coefficients
taken to be rational functions of the momentum invariants) of
integrals in category (a).

\vskip .2 cm
\noindent
\subappendix{Bubble Integrals}

The bubble integral is defined by
$$
I_{2}[P(p^\mu)]
= -i \L4\pi\R^{2-\e} \,\int {d^{4-2\e}p\over \L2\pi\R^{4-2\e}}
\;{P(p^\mu)\over p^2 (p-K)^2 }\; ,
\eqn\GeneralFdefnB
$$
where $P(p^\mu)$ is a polynomial in the loop momentum $p^\mu$ and
$K = \sum_{l = i}^{i+r-1} k_l$ is the total momentum flowing out of
one side,
$r$ being the number of external legs clustered on one side of the
bubble starting at leg $i$, as depicted in fig.~\use\BubblesFigure .
It is straightforward to evaluate the integrals of this type belonging
to the cut-constructible class (scalar and vector),
$$
\eqalign{
I_{2:r;i} &\equiv I_2[1] = {\rg \over \eps (1-2\eps)}(-t_i^{[r]})^{-\eps} \cr
& = \rg \Bigl( {1\over \eps} - \ln(-t_i^{[r]}) + 2 \Bigr) + \Ord(\eps)\,,\cr
I_{2}[p^\mu] &= {K^\mu\over 2}\, I_{2:r;i}\,,\cr
}\eqn\bubbles
$$
where $\rg$ was defined in equation~(\use\Prefactor).
Observe that the two terms not containing logarithms are linked to the
logarithm before the expansion in $\eps$; furthermore,
one cannot cancel the logarithms in any linear combination of
the scalar and vector integrals without also cancelling the cut-free
parts.   For the degenerate
case with $r=1$, the standard prescription in
dimensional regularization is to take $I_2$ to vanish for massless
external legs; this is interpreted as a cancellation of ultra-violet and
infrared divergences [\use\QCDReview].

\vskip .2 cm
\noindent
\subappendix{Triangle Integrals}

The triangle integrals are defined by
$$
I_3[P(p^\mu)] = i \L4\pi\R^{2-\e} \,\int {d^{4-2\e}p\over \L2\pi\R^{4-2\e}}
\;{P(p^\mu) \over p^2 \L p-K_1\R^2 \L p+K_3\R^2}\; ,
\eqn\GeneralFdefnT
$$
where $P(p^\mu)$ is again a polynomial of the loop momentum and
the external momentum arguments $K_{1\ldots3}$ in
equation~(\use\GeneralFdefnT) are sums of external momenta $k_i$ that
are the arguments of the $n$-point amplitude.
There are three types of scalar triangle integrals that can appear in
an $n$-point calculation depending on how many legs are massive
(off-shell), as depicted in fig.~\use\TrianglesFigure .

The one-mass triangle in fig.~\use\TrianglesFigure{a}
depends only on the momentum invariant of the massive leg,
$t_{i+2}^{[n-2]} = t_{i}^{[2]}$,
$$
I_{3;i}^{1\rm m} = {\rg\over\e^2} (-t_{i}^{[2]})^{-1-\eps} \ .
\eqn\OneMassTriangle
$$
This integral does not contain any finite polynomials in the
region $\tn2i < 0$.

The next integral function shown in fig.~\use\TrianglesFigure{b} is
the two-mass triangle integral,
$$
I_{3:r;i}^{2 \rm m}(-t_i^{[r]},-t_{i+r}^{[n-r-1]}) = {\rg\over\e^2}
{(-t_i^{[r]})^{-\eps}-(-t_{i+r}^{[n-r-1]})^{-\eps}\over
(-t_i^{[r]})-(-t_{i+r}^{[n-r-1]}) }\ .
\eqn\TwoMassTriangle
$$
Note that the functions in
eqs.~(\use\OneMassTriangle) and (\use\TwoMassTriangle)
are linear combinations of the set of functions
$$
G(-t_i^{[r]})= {(-t_i^{[r]})^{-\eps}  \over\e^2} \; .
\eqn\TrianglesBasis
$$
Conversely, the functions $G(-t_i^{[r]})$
can also be written as linear combinations
of the one and two-mass triangle scalar integrals. For example,
$$
( t_{i}^{[2]}-t_{i+2}^{[n-3]})
{ I_{3:2;i}^{2 \rm m} \over \rg }
-t_{i}^{[2]} { I_{3;i}^{1\rm m} \over \rg }
=G(-t_{i+2}^{[n-3]})
=G(-t_{i-1}^{[3]}) \; .
\anoneqn
$$

The final scalar triangle is the three-mass integral function depicted in
fig.~\use\TrianglesFigure{c}.  The evaluation of this integral is more
involved,
and can be obtained from refs.~[\use\ThreeMassTriangle,\use\IntegralsLong]
$$
I_{3:r,r';i}^{3 \rm m}
=\ {i\over \sqrt{\Delta_3}}  \sum_{j=1}^3
  \left[ \Li_2\left(-\left({1+i\delta_j \over 1-i\delta_j}\right)\right)
       - \Li_2\left(-\left({1-i\delta_j \over 1+i\delta_j}\right)\right)
  \right]\ + \ \Ord(\e),
\eqn\deqfourtrians
$$
where
$$
\eqalign{
\delta_1 & = { t_i^{[r]} - t_{i+r}^{[r']} - t_{i+r+r'}^{[n-r-r']} \over
\sqrt{\Delta_3}} \; , \cr
\delta_2 & = {-t_i^{[r]} + t_{i+r}^{[r']} - t_{i+r+r'}^{[n-r-r']} \over
\sqrt{\Delta_3}} \; , \cr
\delta_3 & = {-t_i^{[r]} - t_{i+r}^{[r']} + t_{i+r+r'}^{[n-r-r']} \over
\sqrt{\Delta_3}}\; ,  \cr}
\anoneqn$$
and
$$
\Delta_3\equiv -(t_i^{[r]})^2 - (t_{i+r}^{[r']})^2 - (t_{i+r+r'}^{[n-r-r']})^2
+ 2 t_i^{[r]} t_{i+r}^{[r']} + 2t_{i+r}^{[r']} t_{i+r+r'}^{[n-r-r']}
+ 2t_{i+r+r'}^{[n-r-r']} t_i^{[r]}\ .
\anoneqn$$
There are three different ways of labeling any three-mass
triangle
$I_{3:r,r';i}^{3 \rm m} = I_{3:r',n-r-r';i+r}^{3 \rm m} =
I_{3:n-r-r',r;i+r+r'}^{3 \rm m}$;
we only need to keep the distinct cases.

We must also consider the single and double mass triangles that have
a single power of the loop momenta in their numerator.
The single mass case with the same
kinematics as in fig.~\use\TrianglesFigure\ gives,
$$
\eqalign{
I_{3;i}^{1\rm m}[p^\mu]
& =
\rg K_1^\mu {(-t_i^{[2]})^{-1-\eps} \over \eps^2 (1 - 2 \eps)}
-\rg (K_1^\mu + K_2^\mu){(-t_i^{[2]})^{-1-\eps} \over \eps
(1 - 2 \eps)}
\; , \cr}
\anoneqn
$$
where $K_1 = k_i$, $K_2 = k_{i+1}$ and $K_3 = k_{i+2} + \cdots + k_{i+n-1}$.
The second term is proportional to
$I_{2:2;i}$ (defined in eq.~(\use\bubbles)). Since
$$
{ 1 \over \eps^2 (1 - 2 \eps)}
={ 1 \over \eps^2} +{ 2 \over \eps (1 - 2 \eps)} \; ,
\anoneqn
$$
the first term is a linear combination of $I_2(t_i^{[2]}) \equiv
I_{2:2;i}$ and
$I_3^{\rm 1 m } (t_i^{[2]}) \equiv I_{3;i}^{1\rm m}$.
Hence $I_{3;i}^{1\rm m}[p^\mu]$ can be
excluded from the set of integral functions.
The two-mass linear triangle is
$$
\eqalign{
I_{3:r;i}^{\rm 2m}[p^\mu]
& = \rg K_1^\mu
{(-t_i^{[r]})^{-\eps} - (-t_{i+r}^{[n-r-1]})^{-\eps}
 \over \eps (1 - 2 \eps) (t_i^{[r]} - t_{i+r}^{[n-r-1]}) } \cr
\null & \hskip .5 cm -
\rg (K_1^\mu + K_2^\mu) \biggl\{
{t_i^{[r]} ((-t_i^{[r]})^{-\eps} - (- t_{i+r}^{[n-r-1]})^{-\eps})
 \over \eps^2 (1-2\eps) (t_i^{[r]} - t_{i+r}^{[n-r-1]})^2 }
+ {(- t_{i+r}^{[n-r-1]})^{-\eps} \over \eps (1 - 2 \eps)
  (t_i^{[r]} - t_{i+r}^{[n-r-1]}) } \biggr\}\; ,  \cr}
\anoneqn
$$
where $K_1 = k_i + \cdots +k_{i+r-1}$, $K_2 = k_{i+r} + \cdots + k_{i+n-2}$,
and $K_3 = k_{i+n-1}$. By the same reasoning this is also
a linear combination of the scalar integrals.

Finally, we can simplify the three-mass linear triangle using
equation~(42) of ref.~[\use\IntegralsShort] since none of the
coefficients are singular in this case, and the three-mass linear
triangle can also be re-expressed as a sum of scalar triangle and
bubble integrals.


\vskip .2 cm
\noindent
\subappendix{Box Integrals}

The scalar box integrals considered here have vanishing internal masses,
but may have from zero to four nonvanishing external
masses.  Again by external masses we mean off-shell legs with $K^2 \not =
0$.  These integrals are defined and given in ref.~[\use\IntegralsLong]
(the four-mass box was computed by
Denner, Nierste, and Scharf~[\use\FourMassBox]) and are shown
in fig.~\use\BoxesFigure\
and in fig.~\FourBasisFigure(a).

The scalar box integral is
$$
I_4 = -i \L4\pi\R^{2-\e} \,\int {d^{4-2\e}p\over \L2\pi\R^{4-2\e}}
\;{1\over p^2 \L p-K_1\R^2 \L p-K_1-K_2\R^2 \L p+K_4\R^2}\;.
\eqn\GeneralFdefnBX
$$
The external momentum arguments $K_{1\ldots4}$ in
equation~(\use\GeneralFdefnBX) are sums of external momenta $k_i$ that
are the arguments of the $n$-point amplitude.

The no-mass box is depicted in fig.~\FourBasisFigure(a) and is through
${\cal O}(\e^0)$
$$
\eqalign{
I^{0\rm m}_4 [1]\ & =\ \rg \, {1\over s t}
\biggl\{
{2 \over \eps^2} \Bigl[ ( -s)^{-\eps}+ (-t)^{-\eps} \Bigr]
- \ln^2\L {-s \over - t} \R - \pi^2 \biggr\} \ , \cr}
\eqn\ZeroMassBox
$$
where $s = (k_1 + k_2)^2$ and $t = (k_2 + k_3)^2$ are the usual
Mandelstam variables.  This function appears only in four-point amplitudes
with massless particles.  In the proof, presented in
Section~\use\UnitarityProofSection , that the cuts uniquely determine
the amplitude, this box appears only as a special case.

With the labeling of legs shown in fig.~\use\BoxesFigure\ (that is,
expressing the functions in terms of the invariants $\tn{r}i$ of
the $n$-point amplitude), the scalar box integrals $I_4$ expanded
through order ${\cal O}(\e^0)$ for the different cases reduce to
\defeqn\Fboxes
$$
\eqalign{
  I_{4:i}^{1{\rm m}} &=\ { -2 \rg  \over \tn{2}{i-3} \tn{2}{i-2} }
 \biggl\{
 -{1\over\e^2} \Bigl[ (-\tn{2}{i-3})^{-\e} +
(-\tn{2}{i-2} )^{-\e} - (-\tn{n-3}{i})^{-\e} \Bigr] \cr
 &\ + \Li_2\left(1-{\tn{n-3}{i} \over \tn{2}{i-3}}\right)
  \ + \ \Li_2\left(1-{\tn{n-3}{i} \over \tn{2}{i-2}}\right)
  \ +{1\over 2} \ln^2\left({ \tn{2}{i-3} \over \tn{2}{i-2}}\right)\
+\ {\pi^2\over6} \biggr\} \ ,
\cr}
\eqno({\rm\Fboxes{a}})
$$
$$
\eqalign{
 I_{4:r;i}^{2{\rm m}e}
&=\
{-2 \rg
     \over \tn{r+1}{i-1}\tn{r+1}{i} -\tn{r}{i}\tn{n-r-2}{i+r+1} }
\biggl\{
 - {1\over\e^2} \Bigl[ (-\tn{r+1}{i-1})^{-\e} + (-\tn{r+1}{i})^{-\e}
              - (-\tn{r}{i} )^{-\e} - (-\tn{n-r-2}{i+r+1} )^{-\e} \Bigr] \cr
&\ +\ \Li_2\left(1-{\tn{r}{i} \over \tn{r+1}{i-1} }\right)
 \ +\ \Li_2\left(1-{\tn{r}{i} \over \tn{r+1}{i}}\right)
 \ +\ \Li_2\left(1-{\tn{n-r-2}{i+r+1} \over \tn{r+1}{i-1} }\right)
\cr
&\
 \ +\ \Li_2\left(1-{\tn{n-r-2}{i+r+1} \over \tn{r+1}{i}}\right)
-\ \Li_2\left(1-{\tn{r}{i} \tn{n-r-2}{i+r+1}
\over \tn{r+1}{i-1} \tn{r+1}{i}}\right)
   \ +\ {1\over 2} \ln^2\left({\tn{r+1}{i-1} \over \tn{r+1}{i}}\right)
\biggr\} \ ,
\cr }
\eqno({\rm\Fboxes{b}})
$$
$$
\eqalign{ \hskip -1.4 truecm
  I_{4:r;i}^{2{\rm m}h}
&=\ { -2 \rg  \over \tn{2}{i-2} \tn{r+1}{i-1} }
\biggl\{
 -{1\over\e^2} \Bigl[ (- \tn{2}{i-2})^{-\e} + (-\tn{r+1}{i-1})^{-\e}
              - (-\tn{r}{i} )^{-\e} - (-\tn{n-r-2}{i+r})^{-\e} \Bigr]
\cr &
  \ -\ {1\over2\e^2}
    { (-\tn{r}{i})^{-\e}(-\tn{n-r-2}{i+r})^{-\e}
     \over (- \tn{2}{i-2})^{-\e} }
  \ +\ {1\over 2} \ln^2\left({ \tn{2}{i-2}\over \tn{r+1}{i-1} }\right)
\cr &
  \ +\ \Li_2\left(1-{ \tn{r}{i} \over \tn{r+1}{i-1}}\right)
  \ +\ \Li_2\left(1-{\tn{n-r-2}{i+r}\over \tn{r+1}{i-1} }\right)
  \biggr\}  \ ,
\cr}
\eqno({\rm\Fboxes{c}})
$$
$$
\eqalign{ \hskip -1.4 truecm
  I_{4:r,r',i}^{3{\rm m}}
&=\ { -2 \rg
     \over \tn{r+1}{i-1} \tn{r+r'}i -\tn{r}{i} \tn{n-r-r'-1}{i+r+r'} }
\biggl\{
\cr &
 -{1\over\e^2} \Bigl[ (-\tn{r+1}{i-1} )^{-\e} + (-\tn{r+r'}{i})^{-\e}
     - (-\tn{r}{i})^{-\e}
     - (-\tn{r'}{i+r} )^{-\e}
     - (-\tn{n-r-r'-1}{i+r+r'} )^{-\e} \Bigr] \cr
  &  \ -\ {1\over2\e^2}
   { (-\tn{r}{i})^{-\e}(-\tn{r'}{i+r} )^{-\e} \over(-\tn{r+r'}{i})^{-\e} }
  \ -\ {1\over2\e^2}
    {(-\tn{r'}{i+r})^{-\e}(-\tn{n-r-r'-1}{i+r+r'})^{-\e}
           \over (-\tn{r+1}{i-1})^{-\e} }
      \ +\ {1\over2}\ln^2\left({\tn{r+1}{i-1} \over \tn{r+r'}i}\right)
\cr
  &\ +\ \Li_2\left(1-{\tn{r}{i}\over \tn{r+1}{i-1} }\right)
   \ +\ \Li_2\left(1-{\tn{n-r-r'-1}{i+r+r'} \over \tn{r+r'}i}\right)
  \ -\  \Li_2
\left(1-{\tn{r}{i} \tn{n-r-r'-1}{i+r+r'}\over \tn{r+1}{i-1}\tn{r+r'}i }\right)
\biggr\}\ ,
 \cr }
\eqno({\rm\Fboxes{d}})
$$
$$
\eqalign{
I_{4: r, r', r'', i}^{4{\rm m}} = &
{-\rg  \over  t_i^{[r+ r']}\; t_{i+r}^{[r'+r'']}\;\rho}
\biggl\{ - \Li_2\left(\hf(1-\lambda_1+\lambda_2+\rho)\right)
  \ +\ \Li_2\left(\hf(1-\lambda_1+\lambda_2-\rho)\right) \cr
 &\ -\ \Li_2\left(
   \textstyle-{1\over2\lambda_1}(1-\lambda_1-\lambda_2-\rho)\right)
 \ +\ \Li_2\left(\textstyle-{1\over2\lambda_1}(1-\lambda_1-
    \lambda_2+\rho)\right) \cr
  &\ -\ {1\over2}\ln\left({\lambda_1\over\lambda_2^2}\right)
   \ln\left({ 1+\lambda_1-\lambda_2+\rho \over 1+\lambda_1
        -\lambda_2-\rho }\right) \biggr\} \ ,
   \cr}
\eqno({\rm\Fboxes{e}})
$$
where
$$
 \rho\ \equiv\ \sqrt{1 - 2\lambda_1 - 2\lambda_2
+ \lambda_1^2 - 2\lambda_1\lambda_2 + \lambda_2^2}\ ,
\eqn\rdefinition
$$
and
$$
\lambda_1 = {t_i^{[r]} \; t_{i+r +r'}^{[r'']} \over t_i^{[r+ r']} \;
t_{i+r}^{[r'+r'']} } \; , \hskip 1.5 cm
\lambda_2 = {t_{i+r}^{[r']}\; t_{i+r+r'+r''}^{[n-r-r'-r'']} \over
t_i^{[r+ r']}\; t_{i+r}^{[r'+r'']} }\ .
\anoneqn
$$
There are different ways of labeling $I_{4: r, r', r'', i}^{4{\rm m}}$
and $I_{4:r;i}^{2{\rm m}e}$; once again we need only keep the distinct
cases.


For the explicit six-point amplitudes presented in
section~\use\NonMHVSection, it is convenient to define
scalar box functions $F$ in which prefactors have been removed
from the scalar box integrals $I_4$:
\defeqn\boxes
$$
\eqalignno{
  I_{4:i}^{1{\rm m}} &=\ -2 \rg {\Fone{i} \over \tn{2}{i-3} \tn{2}{i-2} }
        \, , \hskip 4 cm & ({\rm\boxes{a}}) \cr
 I_{4:r;i}^{2{\rm m}e}
&=\ -2 \rg {\Feasy{r;i}
      \over \tn{r+1}{i-1}\tn{r+1}{i} -\tn{r}{i}\tn{n-r-2}{i+r+1} }\,, &
     ({\rm\boxes{b}}) \cr
  I_{4:r;i}^{2{\rm m}h}
&=\ -2 \rg {\Fhard{r;i} \over \tn{2}{i-2} \tn{r+1}{i-1} } \,, &
 ({\rm\boxes{c}}) \cr
  I_{4:r,r',i}^{3{\rm m}}
&=\ -2 \rg {\Fthree{r,r';i}
     \over \tn{r+1}{i-1} \tn{r+r'}i -\tn{r}{i} \tn{n-r-r'-1}{i+r+r'} }\,,
 & ({\rm\boxes{d}}) \cr
I_{4: r, r', r'', i}^{4{\rm m}} & =
-2 {\Ffour{r, r', r'';i}\over t_i^{[r+ r']}\; t_{i+r}^{[r'+r'']}\;\rho}\, .
 & ({\rm\boxes{e}})   \cr}
$$


\appendix{Functions Used in the $N=1$ Supersymmetric Amplitudes}
\tagappendix\FunctionAppendix

Here we define the functions appearing in the $N=1$
supersymmetric MHV amplitudes.
In addition to the definitions we demonstrate
that these functions, with the appropriate
arguments, are linear combinations of the integral functions
of Appendix~\use\IntegralsAppendix .
$$
\eqalign{
\Kz ( s)\ =&\
\Bigl( {\rm -ln} (-s ) + 2 + {1\over\eps} \Bigr)+{\cal O}(\e)
=  {1 \over \eps(1-2\eps)  } ({-s})^{-\eps}  \; ,  \cr
\Lz ( r)\ =&\ { {\rm ln} (r) \over 1-r } \; ,
\cr
\Mz ( s_1, s_2 ; m_1^2, m_2^2   )\ =&\
\Li_2 \Bigl( 1 -{m_1^2 m_2^2 \over s_1 s_2 } \Bigr)
-\Li_2 \Bigl( 1 -{m_1^2\over s_1} \Bigr)
-\Li_2 \Bigl( 1 -{m_1^2\over s_2} \Bigr)
-\Li_2 \Bigl( 1 -{m_2^2 \over s_1 } \Bigr)
\cr
&\ -\Li_2 \Bigl( 1 -{m_2^2 \over s_2} \Bigr)
- {1\over2} \ln^2 \Bigl( {s_1\over s_2} \Bigr)\ .
\cr}
\anoneqn
$$
The function $\Kz(s)$ is simply proportional
to the scalar bubble function
$$
\Kz(t_i^{[r]})\ =\ {I_{2:r;i}\over \rg}\ .
\anoneqn
$$
In the amplitudes it is the combination
$$
{ \Lz\L(-t_i^{[r]})/(-t_i^{[r+1]})\R
\over
t_i^{[r+1]} }
\anoneqn
$$
which appears.
This has several representations; it can be
expressed as a linear combination of bubble functions,
$$
\eqalign{
{ \Lz\L(-t_i^{[r]})/(-t_i^{[r+1]})\R
\over
t_i^{[r+1]} }
&=
-{ \ln(-t_i^{[r]})-\ln(-t_i^{[r+1]}) \over
t_i^{[r+1]}-t_i^{[r]} }
\cr
&=
{ 1 \over t_i^{[r+1]}-t_i^{[r]}  }
\Bigl(
I_{2:r;i} -I_{2:r+1;i}
\Bigr)\,.
\cr}
\anoneqn
$$
Secondly it is a Feynman parameter integral,
for a two mass triangle integral,
$$
{ \Lz\L(-t_i^{[r]})/(-t_i^{[r+1]})\R
\over
t_i^{[r+1]} }
= { 1 \over \rg } I^{2m}_{3:r,i} [a_2]
\anoneqn
$$
where $t_i^{[r]}$ and $\tn{n-r-1}{i+r}=\tn{r+1}i$ are the
momentum invariants of the massive legs
and
$a_2$ is the Feynman parameter for the leg between the two massive
external legs.
We mention this represention as it arises naturally
when one carries out the calculation of the $N=1$ chiral multiplet
amplitude in a manner analogous to the $N=4$ calculation
in ref.~[\use\SusyFour].

The $\Mz$ is a linear combination of the scalar `easy two-mass' box
integral
(with the masses on diagonally opposite legs) and certain triangle
integrals.
As discussed in Appendix~\use\IntegralsAppendix\
the function
$$
G(s) \equiv {1\over \eps^2} ({-s})^{-\eps}
\anoneqn
$$
is a linear combination of integral functions.
Using this function we have
$$
\eqalign{
\Mz ( \tn{r+1}{i-1}, \tn{r+1}{i} ; \tn{r}{i} , \tn{n-r-2}{i+r+1}  )
\ = &\
{\tn{r+1}{i-1}\tn{r+1}{i} -\tn{r}{i}\tn{n-r-2}{i+r+1}
     \over 2 \rg}  I_{4:r;i}^{2{\rm m}e}
\cr
&\
+G(\tn{r}{i})+G(\tn{n-r-2}{i+r+1})
-G(\tn{r+1}{i-1})-G(\tn{r+1}{i}) \; .
\cr}
\anoneqn
$$
It is in fact precisely the combination that yields the
$D=6$ `easy two-mass' box integral~[\use\IntegralsShort]
multiplied by its denominator and a constant factor.
The $D=6$ box integrals arise here from the
reduction of box integrals with insertions of loop momentum polynomials
to scalar box integrals and triangle integrals.  It provides a slightly
preferable representation to the $D=4$ box integral because it is
manifestly free of soft divergences, as should be the case for contributions
of internal $N=1$ matter supermultiplets.  (Only internal gluons can give rise
to soft divergences.)

In the case $r=n-3$ the two mass box reduces to a single mass
box, because the momentum invariant giving rise to the mass of
the second massive leg is zero,
$\tn{n-r-2}{i+r+1}=0$. This corresponds to
setting one of the last two arguments of $\Mz$ to zero whence,
$$
\eqalign{
\Mz ( s_1 ,s_2; m_1^2, 0   ) &=
-\Li_2 \Bigl( 1 -{m_1^2\over s_1} \Bigr)
-\Li_2 \Bigl( 1 -{m_1^2\over s_2} \Bigr)
-{1\over2} \ln^2 \Bigl( {s_1\over s_2} \Bigr)
-{\pi^2 \over 6 }\ ,
\cr}
\anoneqn
$$
so that
$$
\eqalign{
\Mz (  \tn{n-2}{i-1} , \tn{n-2}{i} ;\tn{n-3}{i} , 0 )
\ = &\ { \tn{n-2}{i}\tn{n-2}{i-1} \over 2 \rg} I_{4:i}^{1{\rm m}}
\cr
&\
+G(\tn{r}{i})-G(\tn{r+1}{i-1})-G(\tn{r+1}{i})\ ,
\cr}
\anoneqn
$$
and the function corresponds to a single mass box integral,
plus $G$ functions with various arguments.

As we have expressed the functions $\Kz$, $\Lz$ and $\Mz$ in terms
of the set of integral functions, any
cut constructible amplitude
which is written in terms of these functions
and has the correct cuts, automatically has the correct polynomial terms.


\appendix{Collinear Limits}
\tagappendix\CollinearLimitCheck

In this appendix, we verify that the
$N=1$ MHV $n$-point amplitudes and (one of) the $N=4$ non-MHV
six-point amplitudes have the expected collinear limits.
In general, one expects~[\use\AllPlus,\use\SusyFour]
the collinear limits of the leading-color
one-loop partial amplitudes to have the following form,
$$
\eqalign{
A_{n;1}^{\rm loop}\ \mathop{\longrightarrow}^{a \parallel b}\
\sum_{\lambda=\pm}  \biggl(
  \Split^{\rm tree}_{-\lambda}(a^{\lambda_a},b^{\lambda_b})\,
&
      A_{n-1;1}^{\rm loop}(\ldots(a+b)^\lambda\ldots)
\cr
&  +\Split^{\rm loop}_{-\lambda}(a^{\lambda_a},b^{\lambda_b})\,
      A_{n-1}^{\rm tree}(\ldots(a+b)^\lambda\ldots) \biggr) ,
\cr}
\eqn\loopsplit
$$
where $\Split^{\rm tree}$ and $\Split^{\rm loop}$ are universal
tree and loop ``splitting amplitudes'', given in the appendix of
ref.~[\use\SusyFour], and the two collinear legs are adjacent in
the color ordering.

The $N=1$ chiral multiplet contribution to the loop splitting
amplitudes for gluons vanishes, while in
an $N=4$ supersymmetric theory, the loop splitting function
is proportional to the tree splitting function,
$$
 \Split^{\rm loop}_{-\lambda}(a^{\lambda_a},b^{\lambda_b})
   \ =\ {c_\Gamma }
  \times \Split^{\rm tree}_{-\lambda}(a^{\lambda_a},b^{\lambda_b})
  \times r_S^{\rm SUSY}(z,s_{ab}),
\eqn\rSdef
$$
where $s_{ab}=(k_a+k_b)^2$, and
$$
\eqalign{
 r_S^{\rm SUSY}(z,s)
  \ &=
\ - {1\over\e^2}\L{\mu^2\over z(1-z)(-s)}\R^{\e}
        + 2 \ln z\,\ln(1-z) - {\pi^2\over6} \cr}
\eqn\rSanswer
$$
is independent of the helicities.

\vskip .2 cm
\noindent
\subappendix{$N=1$ MHV Collinear Limits}

We begin with the $N=1$ MHV amplitudes appearing in
eq.~(\use\NonadjacentMinusAmplitude).
In all collinear limits the net loop splitting amplitude for
an $N=1$ chiral multiplet vanishes, so that defining
$$
V^{N=1\,\rm chiral}_n\ =\ {A_{1j}^{N=1\,\rm chiral} (1,\ldots,n)
                        \over \cg { A_{1j}^{\rm tree}   (1,\ldots,n)}} \; ,
\anoneqn$$
the expected behavior of $V_n$ in the collinear limits is simply
\def\vnone{V^{N=1\,\rm chiral}}
$$
\vnone_n\ \rightarrow\ \vnone_{n-1}.
\eqn\VcollinearLimit
$$
Denote the collinear legs by $c$ and $c+1$, and the momentum fraction
within the fused momentum $k_P$ by $z$ (so that $k_c = z k_P$ and
$k_{c+1} = (1-z) k_P$).
Then relabel the fused leg $P\to c$, and shift the labels
of legs $c+2,c+3,\ldots,n$ down by one, in order to recover
the standard labeling $1,2,\ldots,n-1$ for $(n-1)$-point kinematics.
For $2<j<n$, so that the two negative helicities are not adjacent,
and using the reflection symmetry of the $N=1$ MHV amplitudes,
we need only consider two cases, $1<c,c+1<j$ and $c=1$.

In the case $1<c,c+1<j$, we note that
$$
b^j_{c,m_2} \rightarrow b^j_{c,m_2-1},\hskip 1cm
b^j_{c+1,m_2} \rightarrow b^j_{c,m_2-1} \; ,
\anoneqn$$
and also that
$$\eqalign{
\Mz\L \tn{m_2-c}{c+1}, \tn{m_2-c}{c}; \right.
    & \left. \tn{m_2-c-1}{c+1}, \tn{n+c-m_2-1}{m_2+1}\R
+\Mz\L \tn{m_2-c-1}{c+2}, \tn{m_2-c-1}{c+1};
     \tn{m_2-c-2}{c+2}, \tn{n+c-m_2}{m_2+1}\R\cr
&\rightarrow
\Mz\L t_{c+1}^{[m_2-c-1]}, t_{c}^{[m_2-c-1]};
     t_{c+1}^{[m_2-c-2]}, t_{m_2}^{[n+c-m_2-1]} \R \; , \cr
}\anoneqn$$
where the $t$s on the right-hand side refer to the daughter kinematics.
Other $\Mz$s reduce trivially to corresponding ones for the daughter
kinematics.  As a result, the $\Mz$ terms satisfy
equation~(\use\VcollinearLimit) on their own.  Furthermore,
$$\eqalign{
c^j_{m,c} &\rightarrow 0\,,\cr
c^j_{m,c+1} &\rightarrow c^j_{m,c}\,,\cr
c^j_{c,a} &\rightarrow z c^j_{c,a-1}\,,\cr
c^j_{c+1,a} &\rightarrow (1-z) c^j_{c,a-1}\,,\cr
}\anoneqn$$
all other coefficients reducing in a trivial manner to the
corresponding ones for the reduced kinematics.
For the generic case $2<c<j-2$, the pair of $\Lz$s whose coefficients are
$c^j_{c,a}$ and $c^j_{c+1,a}$ combine to produce the appropriate
$\Lz$ for $V_{n-1}$,
$$
\eqalign{
z {\Lz\L \tn{a-c}{c+1}\big/\tn{a-c+1}c\R\over \tn{a-c+1}c}
+
(1-z) {\Lz\L \tn{a-c-1}{c+2}\big/\tn{a-c}{c+1}\R\over \tn{a-c}{c+1}}
&
\rightarrow {\Lz\L \tn{a-c-1}{c+1}\big/\tn{a-c}c\R\over \tn{a-c}c} \; .
\cr
}\anoneqn
$$
The remaining $\Lz$s as well as the $\Kz$s
reduce trivially to those appearing in $V_{n-1}$.  In the boundary
cases $c=2$ and $c+1=j-1$, an $\Lz$ combines with a $\Kz$ to produce
the desired $\Kz$.  In the case $c=2$, for example, there is no
$c^j_{c,n}$ term in the double sum; instead,
$$\eqalign{
c^j_{3,n} {\Lz\L \tn{n-3}4\big/\tn{n-2}3\R\over\tn{n-2}3}
+{c^j_{2,n}\over \tn21} \Kz\bigl(\tn21\bigr)
&\rightarrow {c^j_{2,n-1}\over \tn21} \Kz\bigl(\tn21\bigr).
}\anoneqn$$

In the second case, $c=1$.  Here, $b^j_{c+1,m_2}$ and
$c^j_{c+1,a}$ (as well as $c^j_{m,c}$) vanish in the collinear limit,
while again $c^j_{m,c+1}\rightarrow c^j_{m,c}$, so that all terms except
the $c^j_{3,n}$, $c^j_{2,n}$, $c^j_{n,2}$,
and $c^j_{n,1}$ terms reduce trivially to the
corresponding terms in $V_{n-1}$.  For the remaining terms, note
that
$$
c^j_{3,n} \rightarrow c^j_{2,n-1}\,,\qquad
{c^j_{2,n}\over \tn21} \rightarrow 1\,,\qquad
c^j_{n,2} \rightarrow c^j_{n-1,1}\,,\qquad
c^j_{n,1} \rightarrow 0\,.
\anoneqn
$$
Combining these equations with equation~(\use\cRelations),
we can see that
the singular $\ln(-\tn21)=\ln(-P^2)$ terms of the collinear limit cancel;
furthermore, the remaining parts of the $\Lz$ functions in the
$c^j_{3,n}$ and $c^j_{n,2}$ terms reproduce the $\Kz$ functions
with $c^j_{2,n-1}$ and $c^j_{n-1,1}$ coefficients respectively.

Finally we examine the collinear limit $1\parallel 2$ in the
amplitude $A_{12}$ with adjacent negative-helicity gluons.
In this case all four of the terms on the right-hand-side of~(\loopsplit)
vanish due to the supersymmetry Ward identity~(\use\susyward),
so the limit should be nonsingular, and it is so, by
virtue of the factors of $\spa1.2$ in the tree amplitude.
This completes the explicit verification of the
collinear limits of the $N=1$ MHV amplitudes.

\vskip .2 cm
\noindent
\subappendix{$N=4$ Non-MHV Collinear Limits}

Verification of the behavior~(\loopsplit) for the six-point
non-MHV $N=4$ amplitudes hinges on the collinear behavior
of the coefficient $B_0$ and the functions $\Wsix{i}$.
When two like-helicity gluons become collinear, $B_0$ has a
singularity.
For example, in the limit $5\parallel6$ (with $k_5+k_6=k_P$),
$$\eqalign{
  B_0\ &\mathop{\longrightarrow}^{5\parallel6}\ \Split^\tree_{+}(5^-,6^-)
    \times A^\tree_{5}(1^+,2^+,3^+,4^-,P^-)\,,\cr
  B_0^\cc\ &\mathop{\longrightarrow}^{5\parallel6}\ \Split^\tree_{-}(5^+,6^+)
    \times A^\tree_{5}(1^-,2^-,3^-,4^+,P^+)\,,\cr
}\eqn\Blimit
$$
where
$$
\Split^\tree_{+}(5^-,6^-)\ =\ -{1\over\sqrt{z(1-z)}\spb5.6}\ ,\hskip1cm
\Split^\tree_{-}(5^+,6^+)\ =\ {1\over\sqrt{z(1-z)}\spa5.6}\ .
\eqn\splitfivesix
$$
The singular behavior of $B_0$ in the limits $1\parallel2$, $2\parallel3$ and
$4
   \parallel5$
is related to eq.~(\Blimit) by symmetries.
In the limit that two opposite-helicity gluons become collinear
($3\parallel4$ or $6\parallel1$), $B_0$ is nonsingular.
The only result needed regarding the $\Wsix{i}$ is that
the sum of two of the three $\Wsix{i}$'s has a simple collinear
limit in two of the six collinear channels:
$$
\eqalign{
 \Wsix3\ +\ \Wsix1\ &\mathop{\longrightarrow}^{s \to 0}\
   V_5^g\ +\ r_S^{\rm SUSY}(z,s),
      \qquad\quad s\ =\ s_{45},s_{12}, \cr
 \Wsix1\ +\ \Wsix2\ &\mathop{\longrightarrow}^{s \to 0}\
   V_5^g\ +\ r_S^{\rm SUSY}(z,s),
      \qquad\quad s\ =\ s_{23},s_{56}, \cr
 \Wsix2\ +\ \Wsix3\ &\mathop{\longrightarrow}^{s \to 0}\
   V_5^g\ +\ r_S^{\rm SUSY}(z,s),
      \qquad\quad s\ =\ s_{34},s_{61}, \cr}
\eqn\Wcollinear
$$
where $V_5^g$ is the universal, cyclicly symmetric function appearing
in the $N=4$ supersymmetric five-gluon amplitudes,
$$
V_5^g =
\sum_{i=1}^{5} -{ 1 \over \eps^2 } \Bigl(
{ \mu^2  \over -s_{i,i+1} } \Bigr)^{\eps}
           +\sum_{i=1}^5 \ln\L{-s_{i,i+1} \over-s_{i+1,i+2} }\R\,
                     \ln\L{-s_{i+2,i+3}\over-s_{i-2,i-1}}\R+{5\over6}\pi^2
 \; ,
\eqn\fiveptVg
$$
which should be evaluated in the appropriate five-point kinematics.

Up to symmetries, there are only two inequivalent collinear limits
for the $(+{}+{}+{}-{}-{}-)$ amplitude,
say $5^- \parallel 6^-$ and $6^-\parallel 1^+$.
Here we will check the latter limit, which is more intricate in that
both helicities of the intermediate gluon $P$ enter.
In the limit $6\parallel 1$, $B_0$ is nonsingular.
The singular behavior of $B_2$ is
$$
\hskip -15pt\eqalign{
  B_2\ &\mathop{\longrightarrow}^{6 \parallel 1}
  \left({ (1-z)^{1/2} \spb{P}.{5} \spa{5}.{4} \over s_{5P} }\right)^4
  \LP\LB \Split^\tree_{+}(5^-,6^-)
    \times A^\tree_{5}(1^+,2^+,3^+,4^-,P^-) \RB\right\vert_{j\to j+1} \cr
&\quad
+ \left({ z^{1/2} \spb2.3 \spa{5}.{P} \over s_{5P} }\right)^4
  \LP\LB \Split^\tree_{-}(5^+,6^+)
    \times A^{\tree}_{5}(1^-,2^-,3^-,4^+,P^+) \RB\right\vert_{j\to j+1} \cr
  &=\ (1-z)^2 \Split^\tree_{+}(6^-,1^-) \times
  { {\spa{4}.{5}}^4 \over {\spa{5}.{P}}^4 }
           A^\tree_{5}(2^+,3^+,4^+,5^-,P^-)  \cr
&\quad
  + z^2 \Split^\tree_{-}(6^+,1^+) \times
  { {\spb2.3}^4 \over {\spb{5}.{P}}^4 }
           A^{\tree}_{5}(2^-,3^-,4^-,5^+,P^+)  \cr
  &=\ \Split^\tree_{-}(6^-,1^+) \times A^\tree_5(2^+,3^+,4^-,5^-,P^+) \cr
 & \hskip 2 cm
 \ +\ \Split^\tree_{+}(6^-,1^+) \times A^\tree_5(2^+,3^+,4^-,5^-,P^-).
\cr}
\eqn\hardAlimit
$$
Using symmetries, the singular behavior of $B_3$ is identical
to~(\hardAlimit).
Using eq.~(\hardAlimit), the third line of~(\Wcollinear),
and the form~(\use\NeqFourMHV) of the (MHV) five-gluon amplitudes,
the expected behavior~(\loopsplit) for $6\parallel1$ is verified.

\listrefs

\listfigs
\end